\documentclass[preprint,12pt]{elsarticle}

\usepackage{graphicx}
\usepackage{cases}
\usepackage{amssymb}
\usepackage{amsthm}

\usepackage{color}
\definecolor{rouge}{rgb}{1,0,0}
\definecolor{bleu}{rgb}{0,0,1}
\definecolor{vert}{rgb}{0,0.5,0}

\journal{Journal of Computational Physics}

\begin{document}

\begin{frontmatter}

\title{Numerical modeling of nonlinear acoustic waves in a tube connected with Helmholtz resonators}

\author[LMA]{Bruno Lombard}\corref{cor1}
\ead{lombard@lma.cnrs-mrs.fr}
\author[ENSTA]{Jean-Fran\c{c}ois Mercier}
\ead{jean-francois.mercier@ensta.fr}
\cortext[cor1]{Corresponding author. Tel.: +33 491 16 44 13.}
\address[LMA]{LMA, CNRS UPR 7051, Centrale Marseille, Aix-Marseille Univ, F-13402 Marseille Cedex 20, France}
\address[ENSTA]{POEMS, CNRS UMR 7231 CNRS-INRIA-ENSTA, 91762 Palaiseau, France}

\begin{abstract}
Acoustic wave propagation in a one-dimensional waveguide connected with Helmholtz resonators is studied numerically. Finite amplitude waves and viscous boundary layers are considered. The model consists of two coupled evolution equations: a nonlinear PDE describing nonlinear acoustic waves, and a linear ODE describing the oscillations in the Helmholtz resonators. The thermal and viscous losses in the tube and in the necks of the resonators are modeled by fractional derivatives. A diffusive representation is followed: the convolution kernels are replaced by a finite number of memory variables that satisfy local ordinary differential equations. A splitting method is then applied to the evolution equations: their propagative part is solved using a standard TVD scheme for hyperbolic equations, whereas their diffusive part is solved exactly. Various strategies are examined to compute the coefficients of the diffusive representation; finally, an optimization method is preferred to the usual quadrature rules. The numerical model is validated by comparisons with exact solutions. The properties of the full nonlinear solutions are investigated numerically. In particular, the existence of acoustic solitary waves is confirmed.
\end{abstract}

\begin{keyword}
nonlinear acoustics \sep solitons \sep Burgers equation \sep fractional derivatives \sep diffusive representation \sep time splitting \sep shock-capturing schemes
\end{keyword}

\end{frontmatter}



\section{Introduction}\label{SecIntro}

Propagation of linear acoustic waves in lattices has already been the subject of a large body of theoretical and experimental works. Floquet-Bloch band gaps are present in ordered lattices \cite{Brillouin56}, whereas localization occurs in disordered cases \cite{Richoux06}. Nonlinearities, when they are present, are usually considered at discrete points \cite{Richoux07}.

\begin{figure}[htbp]
\begin{center}
\begin{tabular}{c}
\includegraphics[scale=0.25]{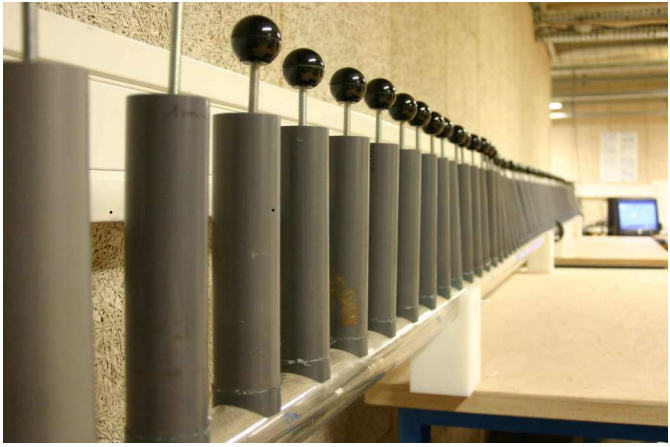} 
\end{tabular}
\caption{Tube connected with Helmholtz resonators (courtesy given by O. Richoux).}
\end{center}
\label{FigPhoto} 
\end{figure}

The propagation of nonlinear acoustic waves in lattices was addressed by Sugimoto and his coauthors in a series of theoretical and experimental studies, whose original purpose was the reduction of shock waves generated by high-speed train into a tunnel \cite{Sugimoto91,Sugimoto92,Sugimoto96,Sugimoto99,Sugimoto04}. The configuration under study was made up of a tube connected with Helmholtz resonators (figure \ref{FigPhoto}). These resonators induce dispersion that competes with the nonlinear effects and may prevent from the emergence of shocks. 

More fundamental questions are also raised, concerning well-known nonlinear waves called solitons \cite{Toda89,Tao09}. Those are stable solitary waves that maintain their shape while traveling at constant speed. Solitons are caused by the cancellation of nonlinear and dispersive effects in the medium. Many models have soliton solutions, for instance the Korteweg-de Vries equation, the nonlinear Schr\"{o}dinger equation, and the sine-Gordon equation. In acoustics, dissipation is largely greater than dispersion, so that it was commonly thought that it was impossible to generate those solitary waves. Thanks to the Sugimoto's configuration, it was shown that acoustic solitons can exist and propagate in place of shock waves \cite{Sugimoto04}.  

The model proposed by Sugimoto involves two coupled equations: a nonlinear Partial Differential Equation (PDE) PDE describing the propagation of finite amplitude acoustic waves in the tube, and a linear Ordinary Differential Equation (ODE) describing the oscillations in the Helmholtz resonators. The dissipative effects in the tube and in the necks of the resonators are modeled by fractional derivatives \cite{Matignon08}, that introduce convolution products. A good numerical modeling relies on the following three specifications:
\begin{itemize}
\item accurate computation of nonlinear non-smooth waves;
\item efficient computation of fractional derivatives, without storing the previous values of the solutions;
\item stable algorithm under a Courant-Friedrichs-Lewy (CFL) condition, whatever the physical parameters and the amplitude of the waves are.
\end{itemize}
The first specification, based for example on shock-capturing schemes, has been well-known for decades \cite{LeVeque92}. The second specification is much less standard. A diffusive representation of fractional operators is used instead of a discretization of convolution operators \cite{Yuan02,Lu05,Galucio06,Diethelm08,Deu10,Birk10}. In doing so, the fractional derivatives are replaced by a set of memory variables that satisfy local-in-time linear differential equations. Determining the quadrature coefficients of the diffusive representation is a crucial issue for the accuracy and the efficiency of the method. Various strategies, usually based on orthogonal polynomials, have been proposed in the literature. We propose here a more efficient strategy, where the coefficients are optimized with respect to the dispersion relation of Sugimoto's model. Lastly, the stability specification requires an adequate coupling between the PDE and the ODE. A naive coupling between these equations usually ensures an increase of discrete energy. On the contrary, we obtain here a stable scheme under the optimal CFL condition.

The paper is organized as follows. The Sugimoto's model is presented in section \ref{SecPhys}. Dispersion analysis in the linear case is developed, and degeneracy towards Korteweg-de Vries equations is specified. The diffusive representation of fractional derivatives, leading to a first-order system of PDE, is described in section \ref{SecMath}. The numerical methods are detailed in section \ref{SecScheme}: a splitting procedure to ensure an optimal CFL condition and to take advantage of efficient methods; a Total Variation Diminishing (TVD) scheme for the advection-Burgers PDE; and an exact integration of the diffusive part. The determination of the weights and nodes of the diffusive representation is discussed in section \ref{SecCoeff}. Numerical experiments are proposed in section \ref{SecExp}. Four tests are presented, concerning successively nonlinear acoustic waves in the tube, oscillations in the resonators, and the coupling (linear and nonlinear) between the two subsystems. The numerical simulations reveal acoustic solitary waves. A conclusion is drawn and future directions of research are outlined in section \ref{SecConclusion}.


\section{Physical modeling}\label{SecPhys}

\subsection{Notations}\label{SecPhysNota}

\begin{figure}[htbp]
\begin{center}
\begin{tabular}{c}
\includegraphics[scale=0.68]{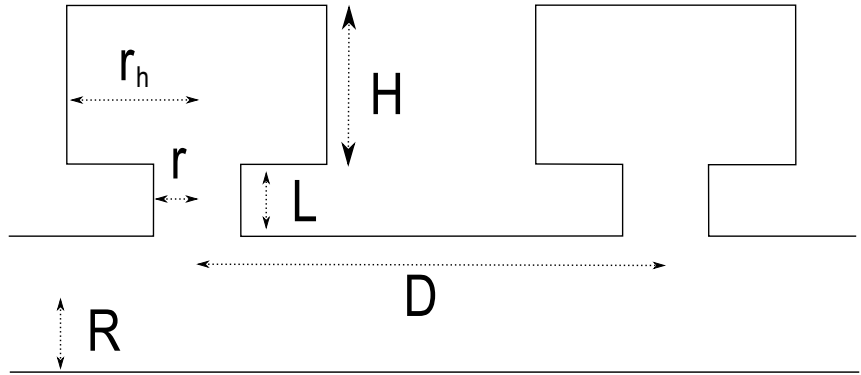} 
\end{tabular}
\caption{Sketch of the guide connected with Helmholtz resonators.}
\end{center}
\label{FigCrobar} 
\end{figure}

The configuration under study is made up of an air-filled tube connected with Helmholtz resonators (figure 2). The cylindrical resonators are uniformly distributed along the tube. The geometrical parameters are the radius of the guide $R$; the axial spacing between resonators $D$; the radius of the neck $r$; the length of the neck $L$; the radius of the cavity $r_h$; and the height of the cavity $H$, which may vary depending on the resonator. Hence, the cross-sectional area of the guide, $A$, is $\pi\,R^2$ and that of the neck, $B$, $\pi\,r^2$, the volume of each resonator, $V$, is $\pi\,r_h^2\,H$, and the reduced radius is:
\begin{equation}
R^*=\frac{\textstyle R}{\textstyle \displaystyle 1-\frac{\textstyle R}{\textstyle 2\,D}\,\frac{\textstyle B}{\textstyle A}}=\frac{\textstyle R}{\textstyle \displaystyle 1-\frac{\textstyle r^2}{\textstyle 2\,D\,R}}.
\end{equation}
The physical parameters are the ratio of specific heats at constant pressure and volume $\gamma$; the pressure at equilibrium $p_0$; the density at equilibrium $\rho_0$; the Prandtl number Pr; the kinematic viscosity $\nu$; and the ratio of shear and bulk viscosities $\mu_v/\mu$; hence the linear sound speed, $a_0$, the sound diffusivity, $\nu_d$, the dissipation in the boundary layer, $C$, the natural angular frequency of the resonator, $\omega_0$ and the natural angular frequency of the tube coupled with the resonator, $\omega_1$, are:
\begin{equation}
\begin{array}{l}
\displaystyle
a_0=\sqrt{\frac{\textstyle \gamma\,p_0}{\textstyle \rho_0}}, \qquad
\nu_d=\nu\left(\frac{\textstyle 4}{\textstyle 3}+\frac{\textstyle \mu_v}{\textstyle \mu}+\frac{\textstyle \gamma-1}{\textstyle \mbox{Pr}}\right), \qquad
C=1+\frac{\textstyle \gamma-1}{\textstyle \sqrt{\mbox{Pr}}},\\
\\
\displaystyle
\omega_0=a_0\,\sqrt{\frac{\textstyle B}{\textstyle L\,V}}=a_0\,\frac{\textstyle r}{\textstyle r_h}\frac{\textstyle 1}{\textstyle \sqrt{L\,H}},\qquad \omega_1=\omega_0\sqrt{1+\frac{\textstyle V}{\textstyle 2\,A\,D}}.
\end{array}
\label{Omega0}
\end{equation}
Under a one-dimensional assumption, the variables are the axial velocity of the gas $u$ and the excess pressure in the cavity $p$ (denoted $p^{'}_2=p_2-p_0$ in the papers by Sugimoto, $p_2$ being the pressure in the cavity of the resonators). The wavelength of the initial disturbance is $\lambda$; hence the acoustic Mach number, $M$, the parameter of nonlinearity, $\varepsilon$, the characteristic angular frequency, $\omega$, the excess pressure in the tube (in the linear theory), $p^{'}$, and the intensity of sound (in dB), $I$, are:
\begin{equation}
\begin{array}{l}
\displaystyle
M=\frac{\textstyle u}{\textstyle a_0}, \qquad 
\varepsilon=\frac{\textstyle \gamma+1}{\textstyle 2}\,M,\qquad
\omega=\frac{\textstyle 2\,\pi\,a_0}{\textstyle \lambda}, \\
[10pt]
\displaystyle
\frac{\textstyle p^{'}}{\textstyle p_0}=\gamma\frac{\textstyle u}{\textstyle a_0},\qquad
I=20\,\log\left(\frac{\textstyle 2\,\gamma}{\textstyle \gamma+1}\,\frac{\textstyle p_0}{\textstyle p_{ref}}\,\varepsilon\right),
\end{array}
\label{IdB}
\end{equation}
where $p_{ref}=2\,10^{-5}$ Pa.


\subsection{Governing equations}\label{SecPhysEdp}

The main assumptions underlying Sugimoto's model are \cite{Sugimoto92}:
\begin{itemize}
\item low-frequency ($\omega<\omega^*=\frac{1.84\,a_0}{R}$), so that only the plane mode propagates and the 1D approximation is valid \cite{Chaigne08};
\item weak acoustic nonlinearity in the tube ($\varepsilon \ll 1$) \cite{Hamilton98};
\item continuous distribution of resonators ($\lambda\gg D$);
\item no interactions between neighboring resonators ($\frac{V}{A\,D} = \left( \frac{r_h}{R} \right)^2 \left( \frac{H}{D} \right) \ll 1$); 
\item linear response of the resonators, no turbulence.
\end{itemize}
Under these hypotheses, the right-going simple wave is modeled by the following coupled PDE-ODE system  
\begin{subnumcases}{\label{EDP}}
\displaystyle
\frac{\textstyle \partial u}{\textstyle \partial t}+\frac{\textstyle \partial}{\textstyle \partial x}\left(a \textstyle u +b\,\displaystyle \frac{\textstyle u^2}{\textstyle 2}\right)=c\,\frac{\textstyle \partial^{-1/2}}{\textstyle \partial t^{-1/2}}\frac{\textstyle \partial u}{\textstyle \partial x}+d\,\frac{\textstyle \partial^2 u}{\textstyle \partial x^2}-e\,\frac{\textstyle \partial p}{\textstyle \partial t},\label{EDP1}\\
[8pt]
\displaystyle
\frac{\textstyle \partial^2 p}{\textstyle \partial t^2}+f\,\frac{\textstyle \partial^{3/2} p}{\textstyle \partial t^{3/2}}+g\,p=h\,u,\label{EDP2}
\end{subnumcases}
with the parameters 
\begin{equation}
\begin{array}{l}
\displaystyle
a=a_0,\qquad b=\frac{\textstyle \gamma+1}{\textstyle 2},\qquad c=\frac{\textstyle C\,a_0 \sqrt{\nu}}{\textstyle R^*},\qquad d=\frac{\textstyle \nu_d}{\textstyle 2},\\
[8pt]
\displaystyle
e=\frac{\textstyle V}{\textstyle 2\,\rho_0\,a_0\,A\,D},\qquad f=\frac{\textstyle 2\,\sqrt{\nu}}{\textstyle r},\qquad g=\omega_0^2,\qquad h=\omega_0^2\,\frac{\textstyle \gamma\,p_0}{\textstyle a_0}.
\end{array}
\label{NotationsEDP}
\end{equation}
PDE (\ref{EDP1}) models nonlinear acoustic waves in the tube (coefficients $a$ and $b$). Viscous and thermal losses in the boundary layer of the tube are introduced by the coefficient $c$ \cite{Chester64}. The diffusivity of sound in the tube is also modeled by the coefficient $d$. ODE (\ref{EDP2}) models the air oscillation in the neck of the resonators (coefficients $f$ and $g$) \cite{Monkewitz85a,Monkewitz85b}. The coupling between the two equations is done by the coefficients $e$ and $h$. If the resonators are suppressed ($H\rightarrow 0$ and thus $V\rightarrow 0$), then the coefficient $e\rightarrow 0$: no coupling occurs, and the classical Chester's equation is recovered \cite{Menguy00}.

Fractional operators of order -1/2 and 3/2 are involved in the system (\ref{EDP}). These operators model the viscous and thermal losses in the tube and in the resonators. These losses are respectively proportional to $1/(i\,\omega)^{1/2}$ and $(i\,\omega)^{3/2}$ in the frequency domain. In (\ref{EDP1}), the Riemann-Liouville fractional integral of order 1/2 of a function $w(t)$ is defined by
\begin{equation}
\frac{\textstyle \partial^{-1/2}}{\textstyle \partial t^{-1/2}}w(t)=\frac{\textstyle H(t)}{\textstyle \sqrt{\pi\,t}}*w=\frac{\textstyle 1}{\textstyle \sqrt{\pi}}\int_0^t(t-\tau)^{-1/2}w(\tau)\,d\tau,
\label{RiemannLiouville}
\end{equation}
where * is the convolution product in time, and $H(t)$ is the Heaviside step function \cite{Matignon08}. The fractional derivative of order 3/2 in (\ref{EDP2}) is obtained by differentiating (\ref{RiemannLiouville}) twice with respect to $t$. 


\subsection{Dispersion analysis}\label{SecPhysDisp}

In this section we present the dispersion analysis of (\ref{EDP}) in the linear case $b=0$. The results obtained will be useful to adjust the terms in the diffusive representation of the fractional derivatives (section \ref{SecCoeff}).

\begin{figure}[htbp]
\begin{center}
\begin{tabular}{cc}
$\nu=0$ & $\nu\neq 0$\\
\includegraphics[scale=0.31]{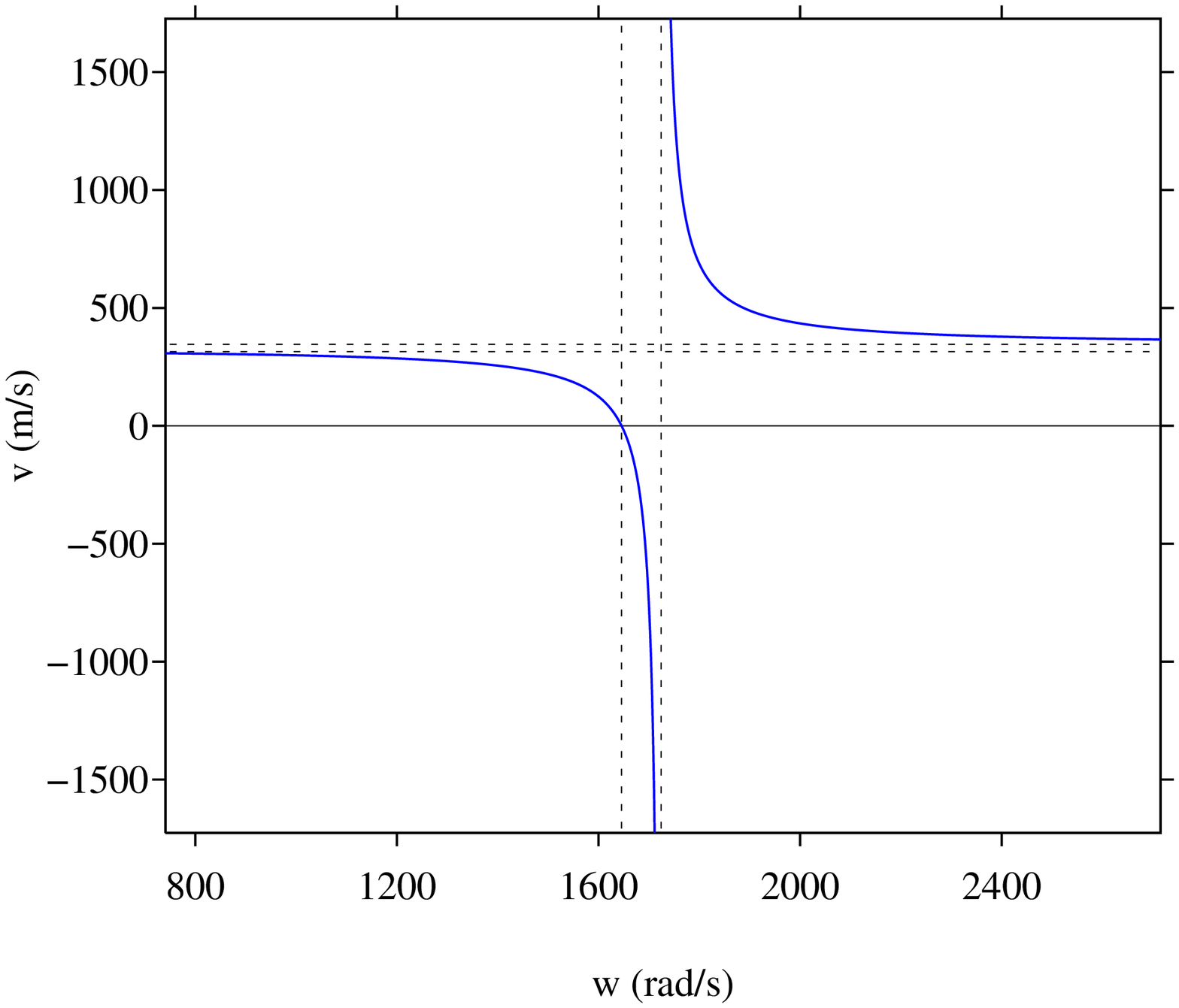} &
\includegraphics[scale=0.31]{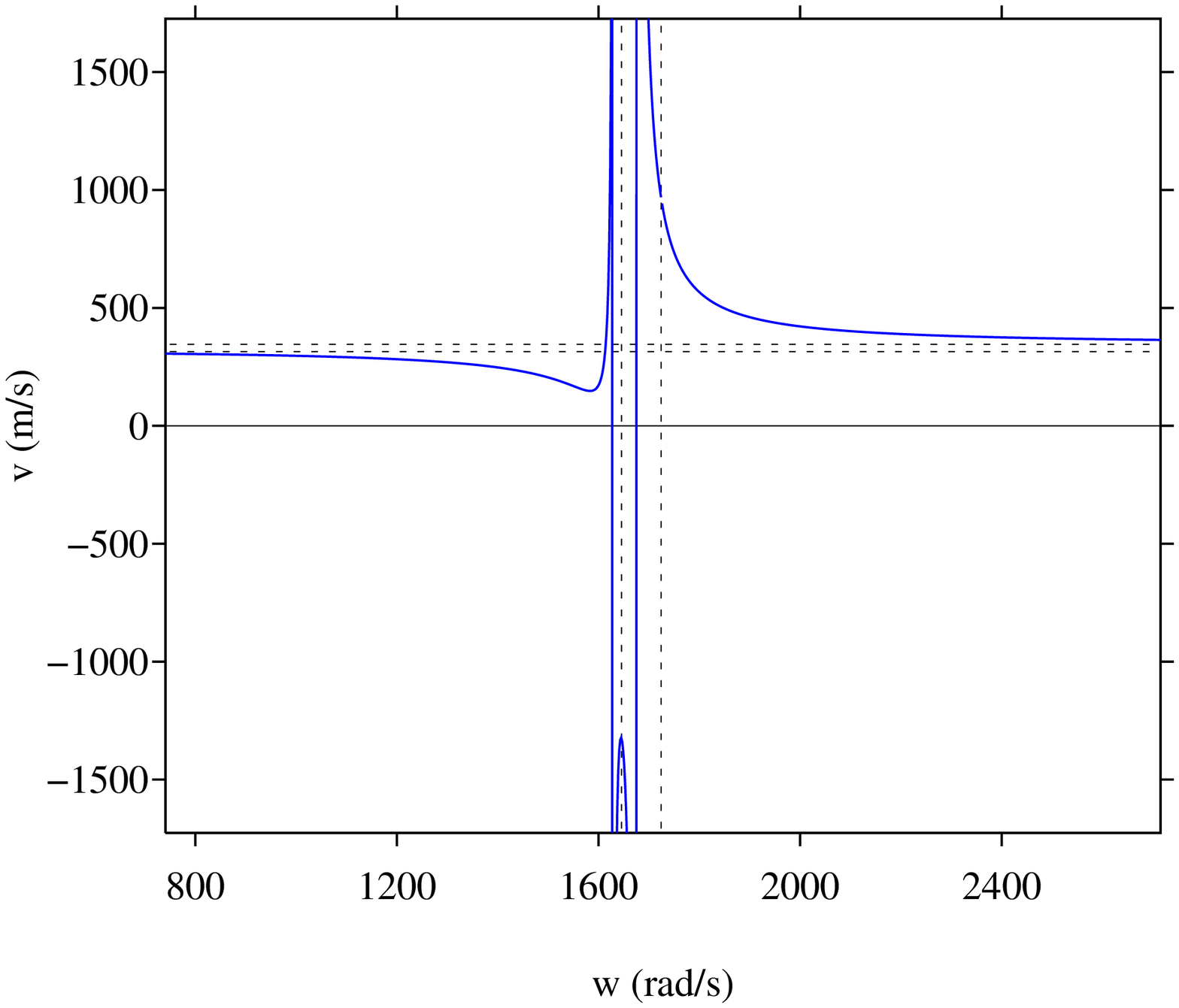} 
\end{tabular}
\vspace{-0.5cm}
\caption{phase velocity in the linear regime, with resonators ($V\neq0 \Leftrightarrow e\neq 0$). Inviscid ($\nu=0$) and viscous ($\nu\neq0$) case; see table \ref{TabParam} for the parameters. The vertical dotted lines denote $\omega_0$ and $\omega_1$. The horizontal dotted lines denote $\overline{\upsilon}<a_0$ and $a_0$ (\ref{PropertyVphase}).} 
\end{center}
\label{FigCelCou} 
\end{figure}

Let us define the Fourier transforms in time and space
\begin{equation}
\widehat{w}(\omega)=\int_{-\infty}^{+\infty}w(t)\,e^{-i\,\omega\,t}\,dt,\hspace{1cm}
\widehat{w}(k)=\int_{-\infty}^{+\infty}w(x)\,e^{i\,kx}\,dx,\hspace{1cm}
\label{Fourier}
\end{equation}
where $\omega$ is the angular frequency and $k$ is the wavenumber. Applying these transforms to (\ref{EDP}) provides a system of two linear equations whose determinant must be null, which yields the dispersion relation between $\omega$ and $k$. Defining the symbol of the half-order integral (\ref{RiemannLiouville})
\begin{equation}
\chi(\omega)=\frac{\textstyle 1}{\textstyle \left(i\,\omega\right)^{1/2}},
\label{ChiSugi}
\end{equation}
and setting the coefficients
\begin{equation}
\left\{
\begin{array}{l}
\displaystyle
{\cal D}_2(\omega)=i \,d [ g-\omega^2 (1+ f \,\chi) ],\\
[8pt]
\displaystyle
{\cal D}_1(\omega)=(a -c\,\chi ) [ g-\omega^2(1+ f\,\chi) ],\\
[8pt]
\displaystyle
{\cal D}_0(\omega)= \omega [ \omega^2(1+ f\,\chi)-(g+e\,h) ],
\end{array}
\right.
\label{CoeffDisp}
\end{equation}
the dispersion relation takes the form
\begin{equation}
{\cal D}(k,\,\omega)={\cal D}_2(\omega)\,k^2+{\cal D}_1(\omega)\,k+{\cal D}_0(\omega)=0.
\label{Dispersion}
\end{equation}
\begin{figure}[htbp]
\begin{center}
\begin{tabular}{c}
\includegraphics[scale=0.31]{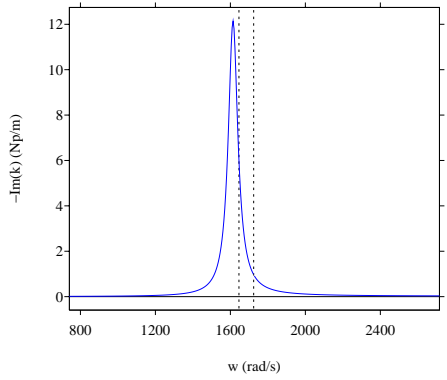} 
\end{tabular}
\vspace{-0.5cm}
\caption{attenuation in the linear regime, with resonators ($V\neq0 \Leftrightarrow e\neq 0$). Viscous case ($\nu\neq0$); see table \ref{TabParam} for the parameters. The vertical dotted lines denote $\omega_0$ and $\omega_1$.} 
\end{center}
\label{FigAttCou} 
\end{figure}
Let us now describe this dispersion relation. First, we will discuss the general case in which the guide is coupled with resonators ($e\neq 0$). Neglecting the diffusivity of sound ($d=0$), one obtains $k=-{\cal D}_0(\omega)\,/\,{\cal D}_1(\omega)$. Otherwise, (\ref{Dispersion}) has two roots $k_1$ and $k_2$, and the root $k$ with minimal modulus is selected, hence the phase velocity $\upsilon=\omega\,/\,\Re\mbox{e}(k)$ and the attenuation $-\Im\mbox{m}(k)$. In the inviscid case ($c=f=0$ adding to $d=0$), the explicit expressions are obtained
\begin{equation}
\upsilon(\omega)=a_0\,\frac{\textstyle \omega^2-\omega_0^2}{\textstyle \omega^2-\omega_1^2} \quad \mbox{and} \quad \Im\mbox{m}(k)=0,
\label{Vphase}
\end{equation}
where we use (\ref{Omega0}) and (\ref{NotationsEDP}). The basic properties of (\ref{Vphase}) are deduced:
\begin{equation}
\left\{
\begin{array}{l}
\displaystyle
\overline{\upsilon}=\upsilon(0)=\frac{\textstyle a_0}{\textstyle \displaystyle 1+\frac{\textstyle V}{\textstyle 2\,A\,D}}<a_0,\hspace{1cm} \upsilon(\omega_0)=0,\\
[18pt]
\displaystyle
\lim_{\omega\rightarrow \omega_1^{\pm}}\upsilon(\omega)=\mp \infty,\hspace{1cm} \lim_{\omega \rightarrow +\infty}\upsilon(\omega)=a_0,\\
[10pt]
\displaystyle
\upsilon^{'}(\omega)=-2\,a_0\,\omega_0 ^2\,\left(\frac{\textstyle V}{\textstyle 2\,A\,D}\right)\frac{\textstyle \omega}{\textstyle \left(\omega^2-\omega_1^2\right)^2}<0.
\end{array}
\right.
\label{PropertyVphase}
\end{equation}
Hypothetically, $\frac{V}{A\,D}\ll 1$ and hence $\omega_1$ is close to $\omega_0$ (\ref{Omega0}). Far from $\omega_0$ and $\omega_1$, the curve of $\upsilon$ is quite flat. In the viscous case ($c\neq 0$ and $f\neq 0$), the phase velocity does not vanish anymore at $\omega_0$. Two vertical asymptotes of $\upsilon$ occur near $\omega_0$ and $\omega_1$. When $\omega\rightarrow +\infty$, the horizontal asymptote of $\upsilon$ is still $a_0$. On the other hand, the maximum attenuation occurs near $\omega_0$. These properties are illustrated in figures 3 and 4.

\begin{figure}[htbp]
\begin{center}
\begin{tabular}{cc}
phase velocity & attenuation\\
\includegraphics[scale=0.31]{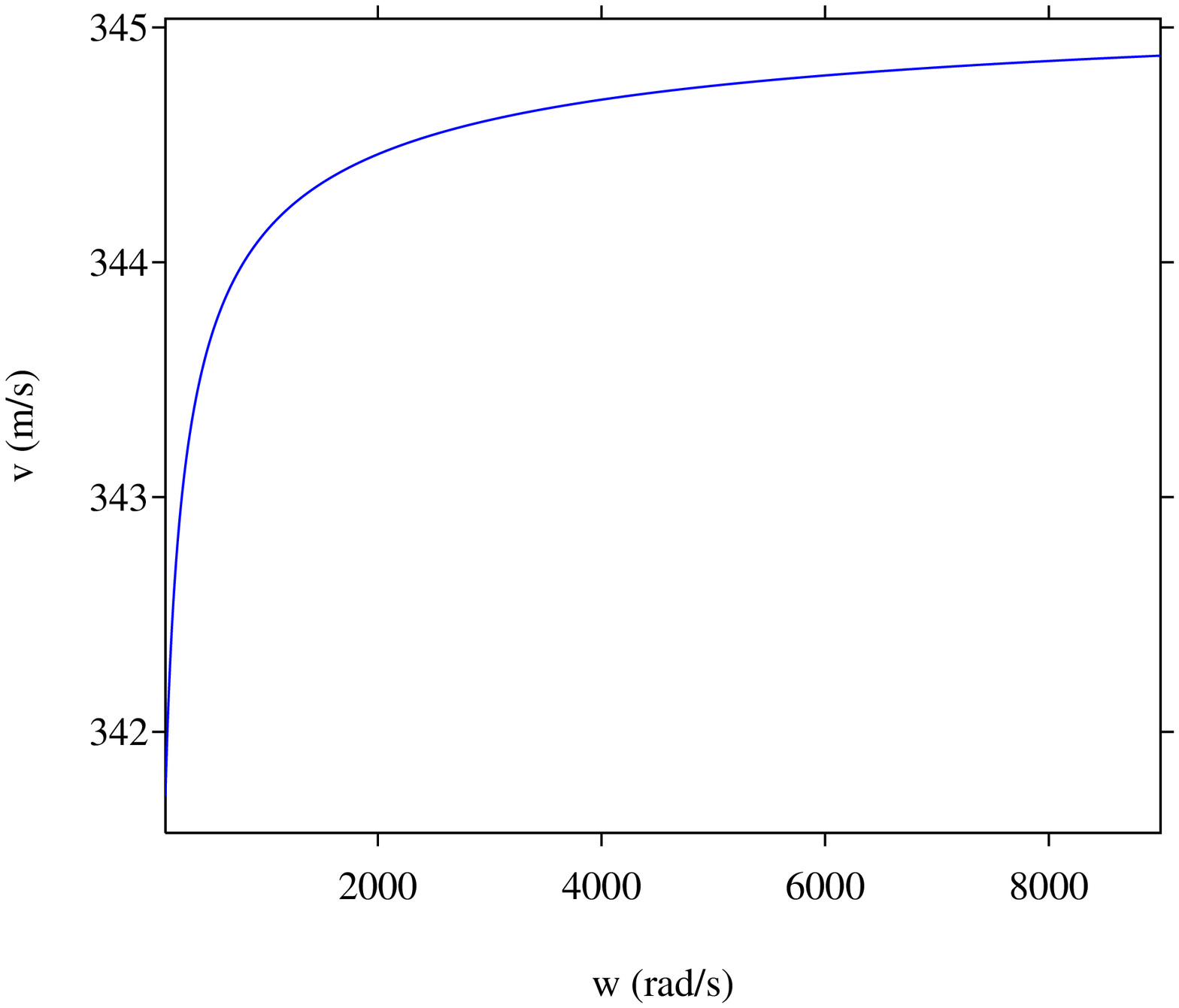}&
\includegraphics[scale=0.31]{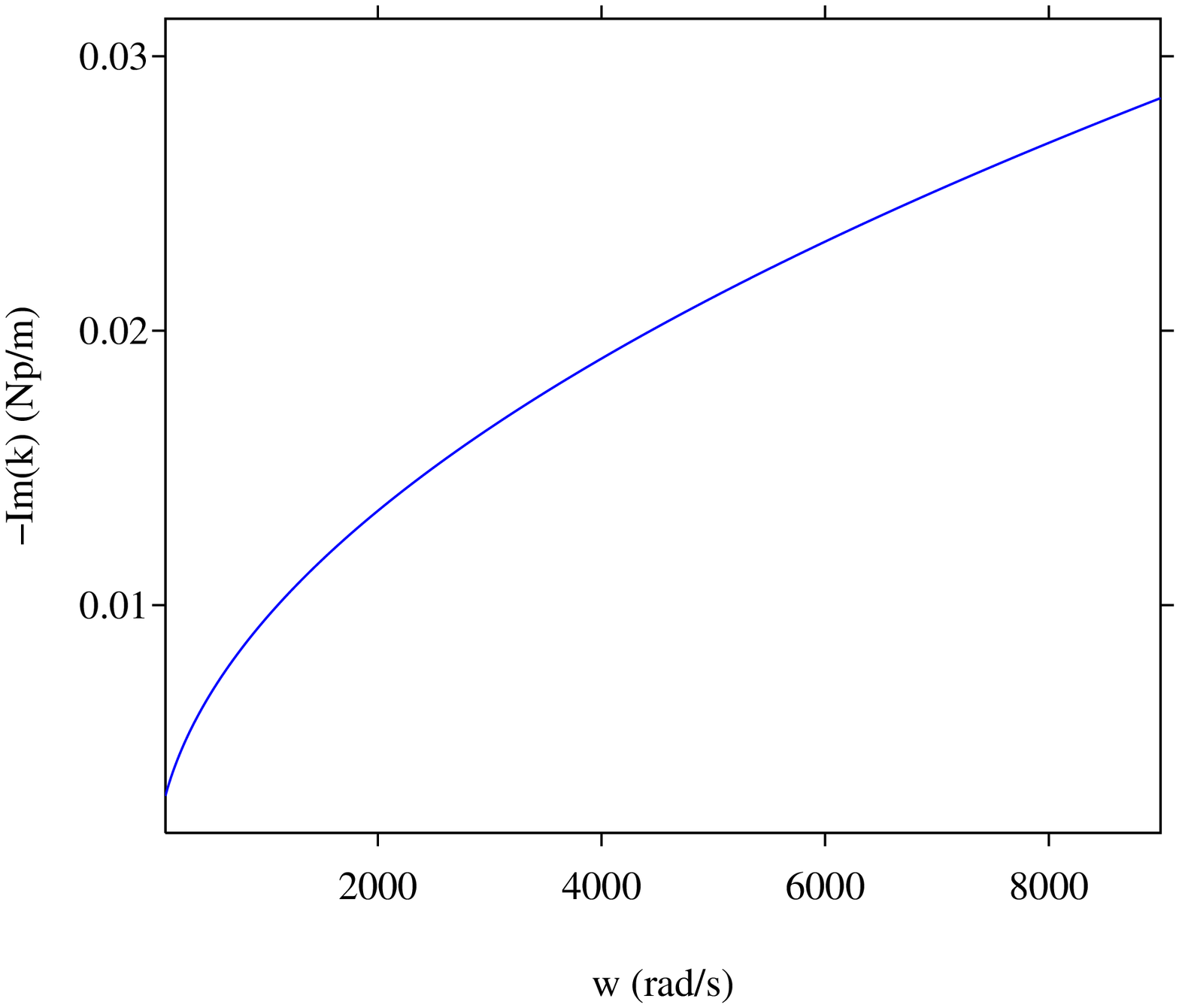} 
\end{tabular}
\vspace{-0.5cm}
\caption{phase velocity and attenuation of the tube without resonators ($V=0\Leftrightarrow e=0$). Viscous case $(\nu \neq 0$); see table \ref{TabParam} for the parameters.} 
\end{center}
\label{FigCelAttDecou} 
\end{figure}

In the limit-case without resonators ($e=0$), the linear dispersion relation (\ref{Dispersion}) simplifies to
\begin{equation}
[ g-\omega^2(1+ f\,\chi) ] [i \,d\,k^2 + k (a -c\,\chi ) - \omega] =0. 
\end{equation}
In this equation, we focus on the part corresponding to (\ref{EDP2}), which leads to
\begin{equation}
i\,d\,k^2+\left(a-\frac{\textstyle c}{\textstyle \left(i\,\omega\right)^{1/2}}\right)\,k-\omega=0.
\label{DispersionGuide}
\end{equation}
Neglecting the diffusivity of sound  ($d=0$) leads to the following phase velocity and attenuation
\begin{equation}
\upsilon=\frac{\textstyle a^2\,\omega-a\,c\sqrt{2\,\omega}+c^2}{\textstyle \displaystyle a\,\omega-c\sqrt{\omega/2}},\hspace{1cm}
\alpha=\frac{\textstyle c}{\textstyle \sqrt{2}}\,\frac{\textstyle \omega^{3/2}}{\textstyle a^2\,\omega-a\,c\,\sqrt{2\,\omega}+c^2}.
\label{CelAttGuide}
\end{equation}
In the inviscid case ($c=d=0$), the phase velocity is equal to $a$ and the attenuation is null. Otherwise, the following properties are deduced:
\begin{equation}
\left\{
\begin{array}{l}
\displaystyle
\upsilon(\omega)\mathop{\sim}\limits_{0} - c \sqrt{\frac{2}{\omega}},\hspace{1cm}\lim_{\omega\rightarrow+\infty}\upsilon(\omega)=a_0,\\
[8pt]
\displaystyle
\alpha(0)=0,\hspace{1.6cm} \alpha(\omega)\mathop{\sim}\limits_{+\infty} \frac{c}{a^2} \sqrt{\frac{\omega}{2}}.
\end{array}
\right.
\label{PropertyGuide}
\end{equation}
These properties are illustrated in figure 5.


\subsection{Regimes of propagation}\label{SecPhysRegime}

An analysis of wave regimes is presented in \cite{Sugimoto92}. Under the hypothesis of weak nonlinearity, $\partial u/\partial x$ in (\ref{EDP1}) is replaced by $-(1/a_0) \, \partial u/\partial t$ in the terms with coefficients $b$, $c$ and $d$. The resulting system is written in the $(T,\,X)$ coordinates, where $T$ is a non-dimensional retarded time, $X$ is a non-dimensional slow space variable and $\varepsilon$ is defined in (\ref{IdB}): 
\begin{equation}
T=\omega\left(t-\frac{\textstyle x}{\textstyle a_0}\right),\qquad X=\varepsilon\,\omega\,\frac{\textstyle x}{\textstyle a_0}.
\label{ThetaX}
\end{equation}
The reduced variables $F={\cal O}(1)$ and $G={\cal O}(1)$ are introduced:
\begin{equation}
F=\frac{\textstyle 1}{\textstyle \varepsilon}\,\frac{\textstyle \gamma+1}{\textstyle 2}\,\frac{\textstyle u}{\textstyle a_0}=\frac{\textstyle 1}{\textstyle \varepsilon}\,\frac{\textstyle \gamma+1}{\textstyle 2\,\gamma}\,\frac{\textstyle p^{'}}{\textstyle p_0},\qquad
G=\frac{\textstyle 1}{\textstyle \varepsilon}\,\frac{\textstyle \gamma+1}{\textstyle 2\,\gamma}\,\frac{\textstyle p}{\textstyle p_0}=\frac{\textstyle p}{\textstyle p^{'}}\,F,
\label{FG}
\end{equation}
leading to
\begin{subnumcases}{\label{EDPscale}}
\displaystyle
\frac{\textstyle \partial F}{\textstyle \partial X}-F\,\frac{\textstyle \partial F}{\textstyle \partial T}=-\delta_R\,\frac{\textstyle \partial^{1/2} F}{\textstyle \partial T^{1/2}}+\beta\,\frac{\textstyle \partial^2 F}{\textstyle \partial T^2}-K\,\frac{\textstyle \partial G}{\textstyle \partial T},\label{EDPscale1}\\
[8pt]
\displaystyle
\frac{\textstyle \partial^2 G}{\textstyle \partial T^2}+\delta_r\,\frac{\textstyle \partial^{3/2} G}{\textstyle \partial T^{3/2}}+\Omega\,G=\Omega\,F,\label{EDPscale2}
\end{subnumcases}
with the new sets of parameters
\begin{equation}
\begin{array}{l}
\displaystyle
\delta_R=\frac{\textstyle C}{\textstyle \varepsilon\,R^*}\sqrt{\frac{\textstyle \nu}{\textstyle \omega}},\qquad \beta=\frac{\textstyle \nu_d\,\omega}{\textstyle 2\,\varepsilon\,a_0^2},\qquad K=\frac{\textstyle V}{\textstyle 2 A\,D\,\varepsilon},\\
[12pt]
\displaystyle
\delta_r=\frac{\textstyle 2}{\textstyle r}\sqrt{\frac{\textstyle \nu}{\textstyle \omega}},\qquad \Omega=\left(\frac{\textstyle \omega_0}{\textstyle \omega}\right)^2.
\end{array}
\label{NotationsEDPscale}
\end{equation}
The effect of dissipative terms $\delta_r$, $\beta$ and $\delta_R$ is analyzed in \cite{Sugimoto91,Sugimoto92}; in particular, $\beta$ is negligible. The dynamics of the system is mainly governed by the two parameters $K$ and $\Omega$. $K$ is the ratio of geometrical dispersion terms to the nonlinear terms. It is assumed that the resonators are coupled to the guide ($V\neq 0$), hence $\omega_0 \neq \infty$ and $0<\Omega<+\infty$. Three limit-cases are then obtained in the inviscid case:
\begin{itemize}
\item first, if $K\ll 1$ or $\Omega \ll 1$, one gets the evolution in a tube without resonators, leading to a shock
\begin{equation}
\frac{\textstyle \partial F}{\textstyle \partial X}-F\,\frac{\textstyle \partial F}{\textstyle \partial T}=0;
\label{CasLimit1}
\end{equation}
\item second, if $K \gg 1$, (\ref{EDPscale}) degenerates into a linear dispersive equation
\begin{equation}
\frac{\textstyle \partial F}{\textstyle \partial X}+K\,\frac{\textstyle \partial F}{\textstyle \partial T}+\frac{\textstyle 1}{\textstyle \Omega}\frac{\textstyle \partial^3 F}{\textstyle \partial T^2 \partial X}=0;
\label{CasLimit2}
\end{equation}
\item third and last, if $\Omega \gg 1$, it is possible to obtain the Korteweg-de Vries equation
\begin{equation}
\frac{\textstyle \partial F}{\textstyle \partial X}+K\,\frac{\textstyle \partial F}{\textstyle \partial T}-F\,\frac{\textstyle \partial F}{\textstyle \partial T}=\frac{\textstyle K}{\textstyle \Omega}\,\frac{\textstyle \partial^3 F}{\textstyle \partial T^3}.
\label{CasLimit3}
\end{equation}
\end{itemize}
This case is of particular interest, since it yields to solitons.


\section{Mathematical modeling}\label{SecMath}

\subsection{Diffusive representation of fractional derivatives}\label{SecMathFract}

Here we focus on the fractional terms in (\ref{EDP}). The fractional integral (\ref{RiemannLiouville}) is non local in time, and relies on the full history of $w(t)$. A much more efficient formulation relies on a diffusive representation of this operator, which can be written equivalently \cite{Diethelm08}
\begin{equation}
\frac{\textstyle \partial^{-1/2}}{\textstyle \partial t^{-1/2}}w(t)=\int_0^{+\infty}\phi(\theta,t)\,d\theta,
\label{I12}
\end{equation}
where the diffusive variable $\phi$ defined by
\begin{equation}
\phi(\theta,t)=\frac{\textstyle 2}{\textstyle \pi}\int_0^t e^{-(t-\tau)\,\theta^2}w(\tau)\,d\tau
\label{Phi12}
\end{equation}
satisfies the local-in-time differential equation
\begin{equation}
\left\{
\begin{array}{l}
\displaystyle
\frac{\partial \phi}{\partial t}=-\theta^2\,\phi+\frac{\textstyle 2}{\textstyle \pi}\,w,\\
[8pt]
\phi(\theta,0)=0.
\end{array}
\right.
\label{ODEI12}
\end{equation}
To compute the derivative of order 3/2, we differentiate the fractional integral twice (\ref{RiemannLiouville}). A first differentiation leads to the derivative of order 1/2:
\begin{equation}
\begin{array}{lll}
\displaystyle
\frac{\textstyle \partial^{1/2}}{\textstyle \partial t^{1/2}}w(t) &=& \displaystyle \frac{\textstyle d}{\textstyle dt} \left( \frac{\textstyle \partial^{-1/2}}{\textstyle \partial t^{-1/2}}w(t) \right),\\
[12pt]
&=& \displaystyle \frac{\textstyle d}{\textstyle dt} \left( \frac{\textstyle 1}{\textstyle \sqrt{\pi}} \int_0^t(t-\tau)^{-1/2}w(\tau)\,d\tau \right),\\
[12pt]
&=& \displaystyle \frac{\textstyle 1}{\textstyle \sqrt{\pi}} \int_0^t(t-\tau)^{-1/2}w^{'}(\tau)\,d\tau,
\end{array}
\label{D12int}
\end{equation}
where $w(0)=0$ has been used. In the rest of the paper $w(\cdot)=p(x,\cdot)$ and a zero initial condition $p(x,0)=0$ will always be chosen in the numerical tests. Thus, proceeding as previously for the fractional integral, a diffusive representation is introduced
\begin{equation}
\displaystyle 
\frac{\textstyle \partial^{1/2}}{\textstyle \partial t^{1/2}}w(t) = \displaystyle \int_0^{+\infty}\xi(\theta,t)\,d\theta,
\label{D12}
\end{equation}
where the diffusive variable $\xi$ satisfies
\begin{equation}
\left\{
\begin{array}{l}
\displaystyle
\frac{\partial \xi}{\partial t}=-\theta^2\,\xi+\frac{\textstyle 2}{\textstyle \pi}\,w^{'},\\
[8pt]
\xi(\theta,0)=0.
\end{array}
\right.
\label{ODED12}
\end{equation}
The derivative of order 3/2 is immediately deduced \cite{Deu10}:
\begin{equation}
\begin{array}{lll}
\displaystyle
\frac{\textstyle \partial^{3/2}}{\textstyle \partial t^{3/2}}w(t)&=& \displaystyle \frac{\textstyle d}{\textstyle dt} \left( \frac{\textstyle \partial^{1/2}}{\textstyle \partial t^{1/2}}w(t) \right),\\
[12pt]
&=& \displaystyle \int_0^{+\infty}\frac{\partial \xi}{\partial t}(\theta,t)d\theta,\\
[12pt]
&=& \displaystyle \int_0^{+\infty}\left(-\theta^2\,\xi+\frac{\textstyle 2}{\textstyle \pi}\,w^{'}\right)\,d\theta.
\end{array}
\label{D32}
\end{equation}


\subsection{First-order system}\label{SecMathSystem}

To approximate the integral (\ref{I12}), a quadrature formula on $N$ points is used, with weights $\mu_{\ell}$ and abscissae $\theta_{\ell}$:
\begin{equation}
\frac{\textstyle \partial^{-1/2}}{\textstyle \partial t^{-1/2}}w(t)\simeq\sum_{\ell=1}^N\mu_{\ell}\,\phi(\theta_{\ell},t)=\sum_{\ell=1}^N\mu_{\ell}\,\phi_{\ell}(t),
\label{RDI12}
\end{equation}
where the $\phi_{\ell}$ satisfiy the ODE (\ref{ODEI12}). Similarly, the integral (\ref{D32}) is written
\begin{equation}
\frac{\textstyle \partial^{3/2}}{\textstyle \partial t^{3/2}}w(t)\simeq\sum_{\ell=1}^N\mu_{\ell}\left(-\theta_{\ell}^2\,\xi(\theta_{\ell},t)+\frac{\textstyle 2}{\textstyle \pi}\,w^{'}\right)=\sum_{\ell=1}^N\mu_{\ell}\left(-\theta_{\ell}^2\,\xi_{\ell}+\frac{\textstyle 2}{\textstyle \pi}\,w^{'}\right),
\label{RDD32}
\end{equation}
where the $\xi_{\ell}$ satisfy the ODE (\ref{ODED12}). The determination of weights and nodes $\mu_{\ell}$ and $\theta_{\ell}$ is discussed in section \ref{SecCoeff}. Then equations (\ref{EDP}), (\ref{ODEI12}, (\ref{ODED12})), (\ref{RDI12}) and (\ref{RDD32}) are written as a first-order system
\begin{equation}
\left\{
\begin{array}{l}
\displaystyle
\frac{\textstyle \partial u}{\textstyle \partial t}+\frac{\textstyle \partial}{\textstyle \partial x}\left(a \textstyle u +b\,\displaystyle \frac{\textstyle u^2}{\textstyle 2}\right)=c\sum_{\ell=1}^N\mu_{\ell}\phi_{\ell}+d\,\frac{\textstyle \partial^2 u}{\textstyle\partial x^2}-e\,q,\\
[10pt]
\displaystyle
\frac{\textstyle \partial p}{\textstyle \partial t}=q,\\
[8pt]
\displaystyle
\frac{\textstyle \partial q}{\textstyle \partial t}=h\,u-g\,p-f\,\sum_{\ell=1}^N\mu_{\ell}\left(-\theta_{\ell}^2\,\xi_{\ell}+\frac{\textstyle 2}{\textstyle \pi}\,q\right),\\
[12pt]
\displaystyle
\frac{\textstyle \partial \phi_{\ell}}{\textstyle \partial t}-\frac{\textstyle 2}{\textstyle \pi}\,\frac{\textstyle \partial u}{\textstyle \partial x}=-\theta_{\ell}^2\,\phi_{\ell},\hspace{1cm} \ell=1\cdots N,\\
[12pt]
\displaystyle
\frac{\textstyle \partial \xi_{\ell}}{\textstyle \partial t}=-\theta_{\ell}^2\,\xi_{\ell}+\frac{\textstyle 2}{\textstyle \pi}\,q,\hspace{1.5cm} \ell=1\cdots N,
\end{array}
\right.
\label{SystComplet}
\end{equation}
in an unbounded domain. The initial conditions are ($\ell=1\cdots N$)
\begin{equation}
\begin{array}{l}
\displaystyle
u(x,0)=u_0(x),\quad p(x,0)=p_0(x)\equiv 0, \quad \frac{\partial p}{\partial t}(x,0)=q_0(x)\equiv 0,\\
[8pt]
\displaystyle
\phi_{\ell}(x,0)=0,\quad \quad \xi_{\ell}(x,0)=0.
\end{array}
\label{CondInit}
\end{equation}
Taking the vector of $(3+2\,N)$ unknowns
\begin{equation}
{\bf U}=\left(u,\,p,\,q,\phi_1,\cdots,\,\phi_N,\,\xi_1,\cdots,\,\xi_N\right)^T,
\label{VecU}
\end{equation}
the system (\ref{SystComplet}) can be written in the form
\begin{equation}
\frac{\textstyle \partial}{\textstyle \partial t}{\bf U}+\frac{\textstyle \partial}{\textstyle \partial x}{\cal F}({\bf U})={\bf S\,U}+{\bf G}\,\frac{\textstyle \partial^2}{\textstyle \partial x^2}{\bf U},
\label{SystHyper}
\end{equation}
where ${\cal F}$ is the nonlinear flux function
\begin{equation}
{\cal F}=\left(a\,u+b\,\frac{\textstyle u^2}{\textstyle 2},\,0,\,0,\,-\frac{\textstyle 2}{\textstyle \pi}u,\cdots,\,-\frac{\textstyle 2}{\textstyle \pi}u,\,0,\cdots,\,0\right)^T,
\label{Fnonlin}
\end{equation}
${\bf G}$ is the $(3+2\,N)\times (3+2\,N)$ diagonal matrix $\mbox{diag}(d,\,0,\cdots,\,0)$. ${\bf S}$ is the $(3+2\,N)\times (3+2\,N)$ diffusive matrix, as it contains the diffusive representation:
\begin{equation}
{\bf S}=
\left(
\begin{array}{ccccccccc}
0 & 0 & -e & c\,\mu_1 & \cdots & c\,\mu_N & 0 & \cdots & 0\\
[10pt]
0 & 0 &  1  & 0 & \cdots & 0 & 0 & \cdots & 0\\
h & -g & \displaystyle -\frac{\textstyle 2}{\textstyle \pi}f\sum_{\ell=1}^N\mu_{\ell} & 0 & \cdots & 0 & f\,\mu_1\,\theta_1^2 & \cdots & f\,\mu_N\,\theta_N^2\\
0 & 0 & 0 & -\theta_1^2 &  &  &  &  &  \\
\vdots & \vdots & \vdots & & \ddots & & & \\
0 & 0 & 0 &  &  & -\theta_N^2 & & & \\
0 & 0 & \displaystyle \frac{\textstyle 2}{\textstyle \pi} & & & & -\theta_1^2 & & \\
\vdots & \vdots & \vdots & & & & & \ddots & \\
0 & 0 & \displaystyle \frac{\textstyle 2}{\textstyle \pi} & & & & & & -\theta_N^2
\end{array}
\right).
\label{MatS}
\end{equation}
Three properties are deduced from (\ref{SystHyper}):
\begin{itemize}
\item the eigenvalues of the Jacobian matrix ${\bf J}={\cal F}^{'}$ in (\ref{Fnonlin}) are real: $a+b\,u$, and 0 with multiplicity $2\,N+2$. These eigenvalues do not depend on the coefficients of the diffusive representation;
\item if the viscosity is neglected in the resonators ($f=0$), then the $3+2\,N$ eigenvalues of  ${\bf S}$ are ($\ell=1,\cdots,N$):
\begin{equation}
Sp({\bf S})=\left\{\pm i\,\omega_1,\,0,\,-\theta_{\ell}^2\right\},
\label{SpectreS}
\end{equation}
with $\omega_1=\sqrt{e\,h+g}$; see (\ref{Omega0}) and (\ref{NotationsEDP}). In particular, the real parts in (\ref{SpectreS}) are null or negative, which implies a decrease in energy.
\item a linear dispersion analysis can be performed as in the original model (\ref{EDP}). The formula (\ref{Dispersion}) with coefficients (\ref{CoeffDisp}) still holds, replacing $\chi$ in (\ref{ChiSugi}) by 
\begin{equation}
\tilde{{\chi}}(\omega)=\frac{\textstyle 2}{\textstyle \pi}\sum_{\ell=1}^N\frac{\textstyle \mu_{\ell}}{\textstyle \theta_{\ell}^2+i\,\omega}.
\label{ChiDR}
\end{equation}
\end{itemize}


\section{Numerical scheme}\label{SecScheme}

\subsection{Splitting}\label{SecSchemeSplit}

In order to integrate the system (\ref{SystHyper}), a grid was introduced, with a uniform spatial mesh size $\Delta x$ and a variable time step $\Delta t_n$, which for the sake of simplicity will be noted $\Delta t$. The approximation of the exact solution ${\bf U}(x_j = j\,\Delta\,x, t_n = t_{n-1}+\,\Delta t)$ is denoted by ${\bf U}_j^n$. Unsplit integration of (\ref{SystHyper}) is not optimal, because the time step stability condition involves the spectral radius of ${\bf S}$ which increases with $N$. Moreover, it requires building an adequate scheme for the coupled system.

A more efficient strategy, based on a splitting method, was adopted here. Instead of integrating the original equation (\ref{SystHyper}), a propagative equation 
\begin{equation}
\frac{\textstyle \partial}{\textstyle \partial t}{\bf U}+\frac{\textstyle \partial}{\textstyle \partial x}{\cal F}({\bf U})={\bf G}\,\frac{\textstyle \partial^2}{\textstyle \partial x^2}{\bf U},
\label{SplitPropa}
\end{equation}
and a diffusive equation
\begin{equation}
\frac{\partial}{\partial t}{\bf U}={\bf S}\,{\bf U},
\label{SplitDiffu}
\end{equation}
were considered successively. The discrete operators to solve (\ref{SplitPropa}) and (\ref{SplitDiffu}) were denoted by ${\bf H}_a$ and ${\bf H}_b$, respectively. Strang splitting \cite{LeVeque92,Holden11} was then used between $t_n$ and $t_{n+1}$, solving successively (\ref{SplitPropa}) and (\ref{SplitDiffu}) with adequate time increments:
\begin{equation}
\begin{array}{lllll}
\displaystyle
&\bullet& {\bf U}_{j}^{(1)}&=&{\bf H}_{b}(\frac{\Delta\,t}{2})\,{\bf U}_{j}^{n},\\
[6pt]
\displaystyle
&\bullet& {\bf U}_{j}^{(2)}&=&{\bf H}_{a}(\Delta\,t)\,{\bf U}_{j}^{(1)},\\
[6pt]
\displaystyle
&\bullet& {\bf U}_{j}^{n+1}&=&{\bf H}_{b}(\frac{\Delta\,t}{2})\,{\bf U}_{j}^{(2)}.
\end{array}
\label{AlgoSplitting}
\end{equation}
Provided that ${\bf H}_a$ and ${\bf H}_b$ are second-order accurate and stable operators, the time-marching (\ref{AlgoSplitting}) gave a second-order accurate approximation of the original equation (\ref{SystHyper}).


\subsection{Propagative part of the system}\label{SecSchemePropa}

Equation (\ref{SplitPropa}) is solved by any standard scheme for nonlinear hyperbolic PDE:
\begin{equation}
\begin{array}{l}
\displaystyle
u_j^{n+1}=u_j^n-\frac{\textstyle \Delta t}{\textstyle \Delta x}\left({\cal F}^1_{j+1/2}-{\cal F}^1_{j-1/2}\right)+\frac{\textstyle d\,\Delta t}{\textstyle \Delta x^2}\left(u_{j+1}^n-2\,u_j^n+u_{j-1}^n\right),\\
[12pt]
\displaystyle
\phi_{j,\ell}^{n+1}=\phi_{j,\ell}^n+\frac{\textstyle 1}{\textstyle \pi}\frac{\textstyle \Delta t}{\textstyle \Delta x}\left(u_{j+1}^n-u_{j-1}^n\right), \hspace{1cm} \ell=1,\cdots,N,
\end{array}
\label{TVD}
\end{equation}
where ${\cal F}^1_{j\pm1/2}$ is the numerical flux function of the advection-Burgers equation in (\ref{Fnonlin}). In practice, a second-order TVD scheme with MC-limiter was used in our numerical experiments \cite{LeVeque92}. Stability analysis of (\ref{TVD}) provides the necessary and sufficient condition \cite{Sousa03,Dehghan04}
\begin{equation}
\frac{\textstyle \alpha-\alpha^2}{\textstyle 2}\leq \delta \leq \frac{\textstyle 1-\alpha}{\textstyle 2},
\label{CFL}
\end{equation}
with the adimensionalized parameters $\alpha$ and $\delta$ and the discrete velocity $a_{\max} ^{(n)}$
\begin{equation}
\alpha=a_{\max} ^{(n)} \frac{\textstyle \Delta t}{\textstyle \Delta x},\hspace{1cm} \delta=d\frac{\textstyle \Delta t}{\textstyle \Delta x^2},\hspace{1cm} a_{\max} ^{(n)}=a+b\,\max_j(u_j^n).
\label{Amax}
\end{equation}
Condition (\ref{CFL}) was proven rigorously in the case of the advection equation and the upwind scheme, but numerical experiments indicated that it still holds for the nonlinear advection (modifying $a$ into $a_{\max} ^{(n)}$) and for the TVD scheme. Solving (\ref{CFL})-(\ref{Amax}) gives the condition
\begin{equation}
\alpha \leq \min \left( 1+\frac{\textstyle 1}{\textstyle \mbox{Pe}}, \frac{1}{1+\frac{\textstyle 1}{\textstyle \mbox{Pe}}} \right),
\end{equation}
where $\mbox{Pe}=\alpha/2 \delta=a_{\max} ^{(n)}\,\Delta x/2\,d$ is the discrete P\'eclet number. In our configuration, $\mbox{Pe}\approx10^5$ which leads to the restriction on the time step
\begin{equation}
\frac{a_{\max} ^{(n)} \Delta t}{\Delta x} \leq \left(1+\frac{\textstyle 1}{\textstyle \mbox{Pe}}\right)^{-1} \approx 1-\frac{\textstyle 1}{\textstyle \mbox{Pe}} \approx 1.
\label{Dt}
\end{equation}
Therefore despite the explicit discretization of $d\,\partial^2 u/\partial x^2$, the optimal CFL condition is maintained.\\
 

\subsection{Diffusive part of the system}\label{SecSchemeDiffu} 
 
Since the physical parameters do not vary with time, the diffusive part (\ref{SplitDiffu}) can be solved exactly. This gives
\begin{equation}
{\bf H}_b\left(\frac{\Delta\,t}{2}\right)\,{\bf U}_j = e^{{\bf S}\frac{\Delta\,t}{2}}\,{\bf U}_j.
\label{SplitDiffuExp}
\end{equation}
In the inviscid case $\nu=0$, only the unknowns $p$, $q$ and $u$ are involved, and since $N=0$, the exponential can be computed analytically. Using $\omega_1$ (\ref{Omega0}) and defining $\tau=\Delta t/2$, one obtains:
\begin{equation}
e^{{\bf S}\tau}=\left(
\begin{array}{lll}
\displaystyle \frac{\textstyle 1}{\textstyle \omega_1^2}\left(g+e\,h\,\cos \omega_1\tau\right) & \displaystyle \frac{\textstyle e\,g}{\textstyle \omega_1^2}\left(1-\cos \omega_1\tau\right) & \displaystyle -\frac{\textstyle e}{\textstyle \omega_1}\sin \omega_1\tau\\
[10pt]
\displaystyle \frac{\textstyle h}{\textstyle \omega_1^2}\left(1-\cos \omega_1\tau\right) & \displaystyle \frac{\textstyle 1}{\textstyle \omega_1^2}\left(e\,h+g\,\cos \omega_1\tau\right) & \displaystyle \frac{\textstyle 1}{\textstyle \omega_1}\sin \omega_1\tau\\
[10pt]
\displaystyle \frac{\textstyle h}{\textstyle \omega_1}\sin \omega_1\tau & \displaystyle -\frac{\textstyle g}{\textstyle \omega_1}\sin \omega_1\tau & \cos \omega_1\tau
\end{array}
\right).
\label{ExpoS}
\end{equation}
In the general case $N>0$, the exponential is computed numerically using a $(6,6)$ Pad\'e approximation in the ``scaling and squaring method" \cite{Moler03}. If the physical parameters are constant, the computation is done only once at each time step, leading to a negligible computational cost. Even in the case $N=0$, using the numerical evaluation  of $e^{{\bf S}\tau}$ is twice as fast as computing (\ref{ExpoS}), because of the numerical evaluations of trigonometric functions.

This part of the splitting is unconditionally stable, so that the global stability requirement is (\ref{Dt}) and is not penalized by the diffusive part. In other words, the time step depends only on the advection and on the Burgers coefficient in (\ref{EDP}). In particular, $\Delta t$ does not depend on the fractional parameters $c$ and $f$ or on the coupling parameters $e$ and $h$.


\section{Coefficients of the diffusive representation}\label{SecCoeff}
 
The $2\,N$ coefficients of the diffusive representation $\mu_{\ell}$ and $\theta_{\ell}$ in (\ref{MatS}) have yet to be determined. These coefficients are derived from (\ref{I12}), (\ref{RDI12}) and (\ref{RDD32}), and they are used to approximate improper integrals of the form
\begin{equation}
\int_0^{+\infty}\phi(\theta)\,d\theta\simeq\sum_{\ell=1}^N\mu_{\ell}\,\phi(\theta_{\ell}),
\label{ImpropInt}
\end{equation}
where time $t$ has been omitted for the sake of simplicity. This issue is crucial both for the accuracy of the modeling and for the computational efficiency of the method. Many strategies exist for this purpose. We will begin by recalling three known methods based on orthogonal polynomials, and then we will propose another method based on optimization. To be consistent with the literature, the following notations are introduced: $\alpha$ is the order of the fractional derivative; $\lceil \alpha \rceil$ is the ceiling function that rounds up to the next integer; and lastly
\begin{equation}
\overline{\alpha}=2\,\alpha-2\lceil \alpha \rceil+1.
\end{equation}
Since $\alpha=1/2$ or $3/2$ in (\ref{EDP}), then $\overline{\alpha}\equiv 0$. 


\subsection{Method 1: Gauss-Laguerre}\label{SecCoeff1}

This algorithm is proposed in \cite{Yuan02}. The improper integral (\ref{ImpropInt}) is evaluated with the Gauss-Laguerre quadrature:
\begin{equation}
\int_0^{+\infty}\theta^\gamma\,e^{-\theta}\,\psi(\theta)\,d\theta\simeq\sum_{\ell=1}^N w_{\ell}\,\psi(z_{\ell}),
\end{equation}
where $\gamma$ is a parameter, $w_{\ell}$ are the weights and $z_{\ell}$ are the nodes \cite{NRPAS}. It implies
\begin{equation}
\int_0^{+\infty}\phi(\theta)\,d\theta\simeq \sum_{\ell=1}^N w_{\ell}\,z_{\ell}^{-\gamma}\,e^{z_{\ell}}\,\phi(z_{\ell}),
\label{GL1}
\end{equation}
and consequently the desired coefficients in (\ref{ImpropInt}) are
\begin{equation}
\mu_{\ell}=w_{\ell}\,z_{\ell}^{-\gamma}\,e^{z_{\ell}},\hspace{1cm} \theta_{\ell}=z_{\ell}.
\label{GL2}
\end{equation}
Very slow convergence, with the usual value $\gamma=0$, was observed by many authors. In \cite{Diethelm08}, two problems were identified :
\begin{itemize}
\item $\phi(\theta)\mathop{\sim}\limits_{0}\theta^{\overline{\alpha}}$, and since $\overline{\alpha}\in]-1,1[$, an integrable singularity may occur at 0 if $\overline{\alpha}\neq 0$. Taking $\gamma=\overline{\alpha}$ in (\ref{GL1})-(\ref{GL2}) eliminates this problem;
\item the diffusive variable $\phi$ (\ref{Phi12}) decreases polynomially: $\phi(\theta)\mathop{\sim}\limits_{+\infty}\theta^{\overline{\alpha}-2}\equiv 1/\theta^2$, which is badly represented by Gauss-Laguerre exponential weight. This problem cannot be solved.
\end{itemize}


\subsection{Method 2: Gauss-Jacobi}\label{SecCoeff2}

A more efficient approach has been proposed and analysed in \cite{Diethelm08}. The improper integral (\ref{ImpropInt}) is evaluated with the Gauss-Jacobi quadrature:
\begin{equation}
\begin{array}{lll}
\displaystyle 
\int_0^{+\infty}\phi(\theta)\,d\theta&=&\displaystyle\int_{-1}^{+1}\left(1-z\right)^\gamma\left(1+z\right)^\beta\tilde{\phi}(z)\,dz,\\
[8pt]
&&
\displaystyle \simeq\sum_{\ell=1}^N w_{\ell}\,\tilde{\phi}(z_{\ell}),
\end{array}
\label{GJ1}
\end{equation}
where
\begin{equation}
\tilde{\phi}(z)=\frac{\textstyle 2}{\textstyle \left(1-z\right)^\gamma\left(1+z\right)^{\beta+2}}\,\phi\left(\frac{\textstyle 1-z}{\textstyle 1+z}\right).
\label{GJ2}
\end{equation}
The quadrature coefficients in (\ref{ImpropInt}) are deduced:
\begin{equation}
\mu_{\ell}=w_{\ell}\frac{\textstyle 2}{\textstyle \left(1-z_{\ell}\right)^\gamma\left(1+z_{\ell}\right)^{\beta+2}}, \hspace{1cm}\theta_{\ell}=\frac{\textstyle 1-z_{\ell}}{\textstyle 1+z_{\ell}}.
\label{GJ3}
\end{equation}
In \cite{Diethelm08}, it is proposed to take $\gamma=\overline{\alpha}\equiv 0$ and $\beta=-\overline{\alpha}\equiv 0$ in (\ref{GJ1})-(\ref{GJ3}). The nodes of Gauss-Jacobi quadrature (\ref{GJ3}) cover a much wider interval in the $\theta$-space than those of the Gauss-Laguerre quadrature (\ref{GL2}), which explains qualitatively why the slowly decreasing diffusive variables $\phi$ (\ref{Phi12}) are better approximated.


\subsection{Method 3: modified Gauss-Jacobi}\label{SecCoeff3}

In \cite{Birk10}, an improvement is proposed for method 2 , which consists in widening the range of nodes. Based on a modified Gauss-Jacobi quadrature, the function $\tilde{\phi}$ in (\ref{GJ1}) is now
\begin{equation}
\tilde{z}(\theta)=\frac{\textstyle 4}{\textstyle \left(1-z\right)^{\gamma-1}\left(1+z\right)^{\beta+3}}\,\phi\left(\left(\frac{\textstyle 1-z}{\textstyle 1+z}\right)^2\right).
\label{Birk1}
\end{equation} 
The quadrature coefficients in (\ref{ImpropInt}) are deduced:
\begin{equation}
\mu_{\ell}=w_{\ell}\frac{\textstyle 4}{\textstyle \left(1-z_{\ell}\right)^{\gamma-1}\left(1+z_{\ell}\right)^{\beta+3}}, \hspace{1cm}\theta_{\ell}=\left(\frac{\textstyle 1-z_{\ell}}{\textstyle 1+z_{\ell}}\right)^2.
\label{Birk2}
\end{equation}
In \cite{Birk10}, it is proposed to take $\gamma=2\,\overline{\alpha}+1\equiv 1$, $\beta=-(2\,\overline{\alpha}-1)\equiv 1$ in (\ref{Birk1})-(\ref{Birk2}).


\subsection{Method 4: optimization}\label{SecCoeff4}

Finally, we propose a fourth and last method based on the dispersion relation (\ref{Dispersion}). The original problem (\ref{EDP}) and the first-order system (\ref{SystHyper}) differ only in their symbol $\chi(\omega)$: (\ref{ChiSugi}) in the first case, (\ref{ChiDR}) in the second one. Adjusting them provides a way to estimate $\mu_{\ell}$ and $\theta_{\ell}$. This technique is physically meaningful, and has proven its efficiency in a previous work about poroelastic waves \cite{Blanc13}. Let $Q(\omega)$ be the optimized quantity and $Q_{ref}$ the desired one:
\begin{equation}
\begin{array}{l}
\displaystyle
Q(\omega)=\frac{\textstyle \tilde{{\chi}}(\omega)}{\textstyle \chi(\omega)}=\frac{\textstyle 2}{\textstyle \pi}\sum_{\ell=1}^N\frac{\textstyle \mu_{\ell}}{\textstyle \theta_{\ell}^2+i\,\omega}\left(i\,\omega\right)^{1/2}=\sum_{\ell=1}^N\mu_{\ell}\,q_{\ell}(\omega),\\
\\
\displaystyle
Q_{ref}(\omega)=1.
\end{array}
\label{Qomega}
\end{equation}
We implement a linear optimization procedure \cite{Groby06,Blanc13} in order to minimize the distance between $Q$ and $Q_{ref}$ in the interval $[\omega_{min},\omega_{max}]$ containing the characteristic angular frequency $\omega$ of the initial pulse. The abscissae $\theta_{\ell}$ are chosen so that they are distributed linearly on a logarithmic scale
\begin{equation}
\theta_{\ell}^2=\omega_{min}\left( \frac{\omega_{max}}{\omega_{min}}\right) ^{\frac{\ell-1}{N-1}}\mbox{,}\qquad \ell = 1,...,N.
\end{equation}
The weights $\mu_{\ell}$ are obtained by solving the system
\begin{equation}
\sum \limits _{\ell=1}^{N}\mu_{\ell}\,q_{\ell}(\tilde{\omega}_k) = 1\mbox{,}\qquad k = 1,...,K,
\end{equation}
where the $\tilde{\omega}_k$ are also distributed linearly on a logarithmic scale of $K$ points
\begin{equation}
\tilde{\omega}_k = \omega_{min}\left( \frac{\omega_{max}}{\omega_{min}}\right) ^{\frac{k-1}{K-1}}\mbox{,}\qquad k = 1,...,K.
\label{OmegaK}
\end{equation}
Since the $q_{\ell}(\omega)$ are complex functions, optimization is performed simultaneously on the real and imaginary parts 
\begin{equation}
\left\lbrace 
\begin{array}{ll}
\displaystyle \sum \limits _{\ell=1}^{N}\mu_{\ell}\,\mathbb{R}\mbox{e}(q_{\ell}(\tilde{\omega}_k))  & = 1,\\
[12pt]
\displaystyle \sum \limits _{\ell=1}^{N}\mu_{\ell}\,\mathbb{I}\mbox{m}(q_{\ell}(\tilde{\omega}_k))  & = 0\mbox{,}\qquad k=1,...,K.\\
\end{array}
\right. 
\label{OptiLin}
\end{equation}
A square system is obtained when $2K=N$, whereas $2\,K>N$ yields an overdetermined system, which can be solved by writing normal equations \cite{NRPAS}. Higher accuracy is obtained with $K=N$, so we will make this choice in numerical experiments. The interval of optimisation $[\omega_{min},\,\omega_{max}]$ depends on the configuration under study: 
\begin{itemize}
\item for the coupled system with resonators, the attenuation is bounded (figure 4), and the existence of smooth solitary waves globally maintains the frequency content of the initial disturbance. Consequently, we chose a narrow interval centered around $\omega$, by taking for instance $\omega_{min}=\omega\,/\,2$ and $\omega_{max}=\omega \times 3/2$;
\item for the tube without resonators, shocks are expected. Consequently, higher harmonics are generated, and we proposed to use $\omega_{min}=\omega\,/\,2$ but $\omega_{max}=\omega \times {\cal N}$, where ${\cal N}$ is the number of harmonics of interest in a Fourier decomposition of the wave \cite{Menguy00}. In numerical experiments, we took ${\cal N}=20$.
\end{itemize}  

\begin{figure}[htbp]
\begin{center}
\begin{tabular}{cc}
\includegraphics[scale=0.31]{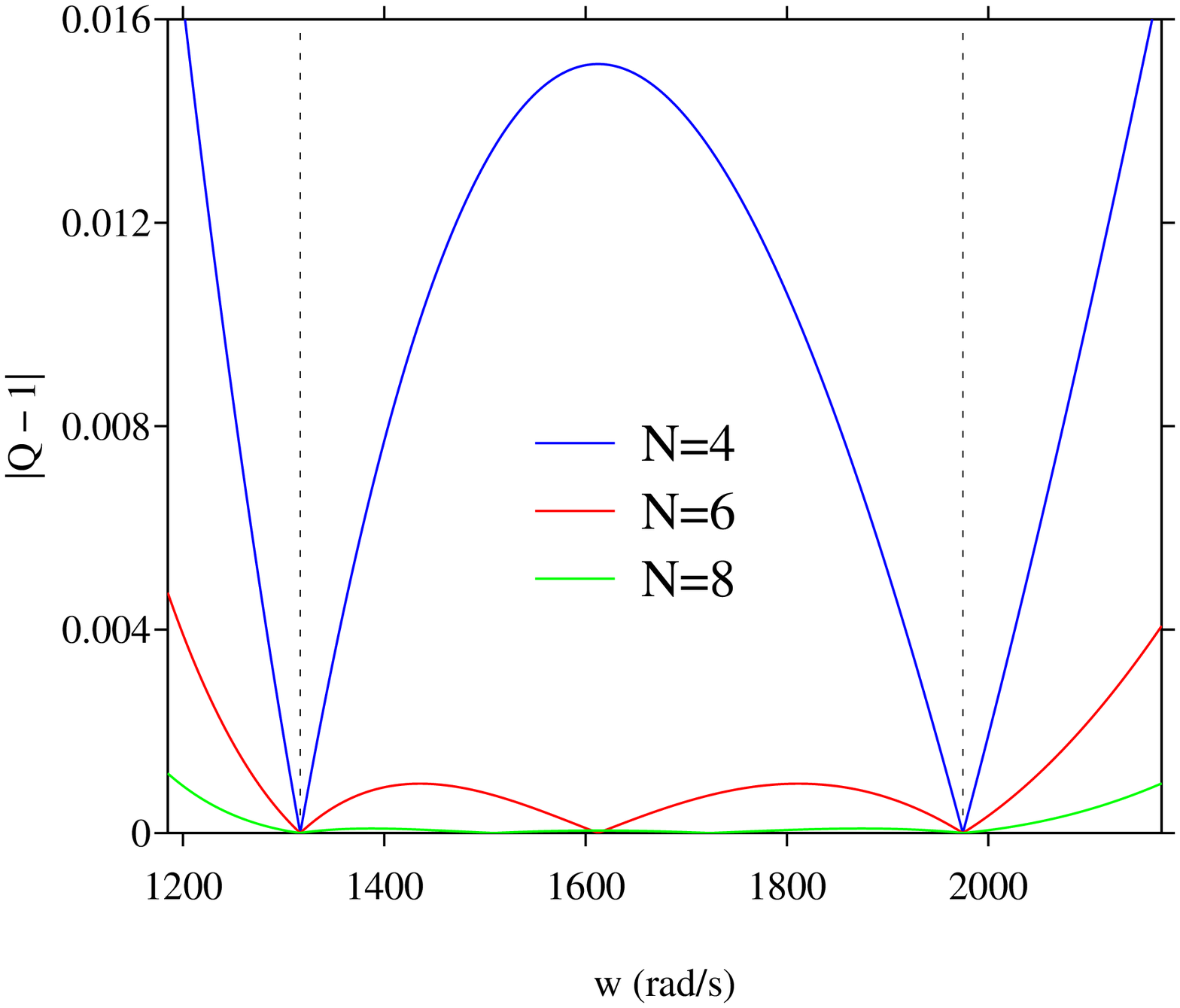} &
\includegraphics[scale=0.31]{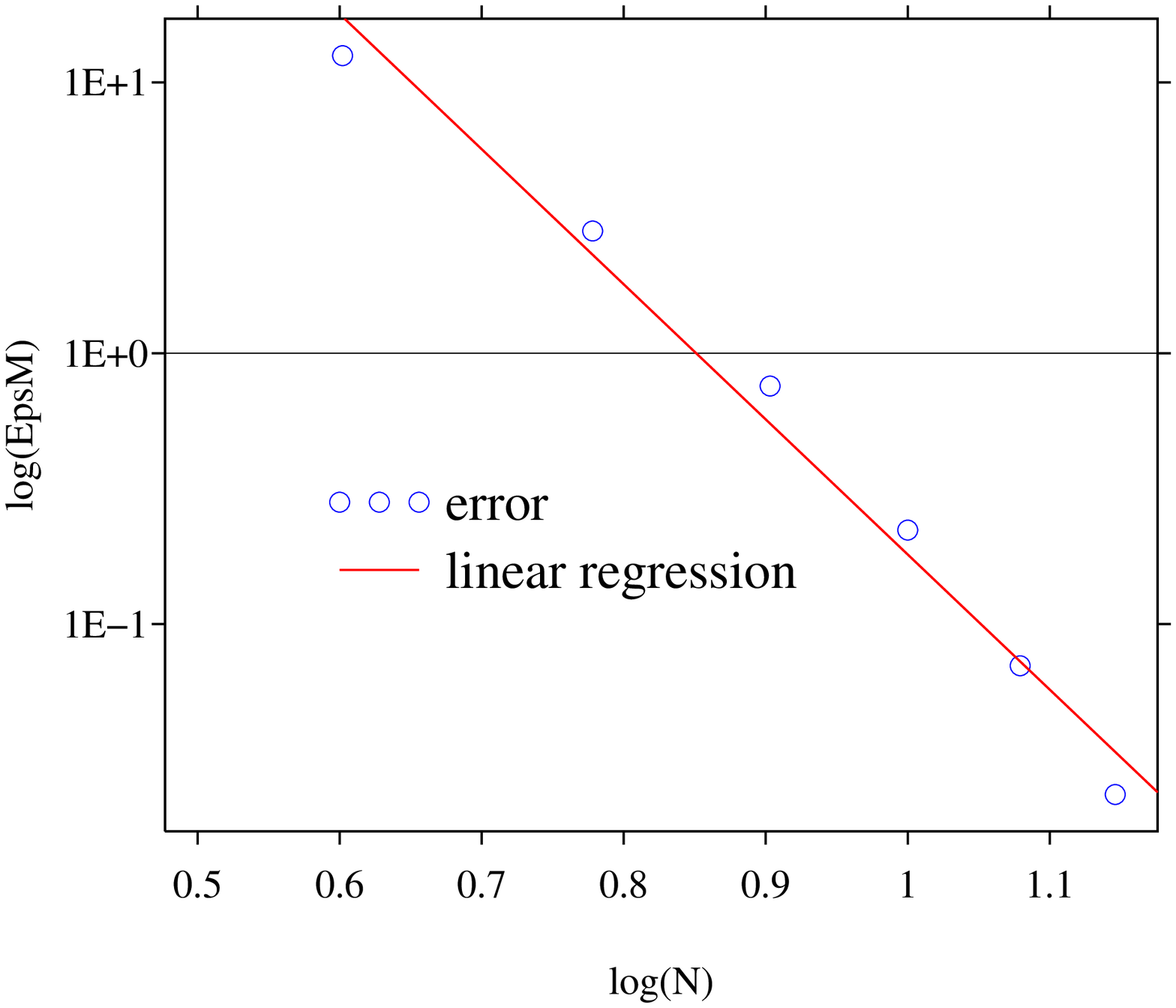}
\end{tabular}
\vspace{-0.5cm}
\caption{error of model due to the optimization procedure, for the coupled system (tube with resonators, $V\neq 0$). Left: $|Q(\omega)-1|$ in (\ref{Qomega}), for various numbers of diffusive variables $N$; the vertical dotted lines denote the upper and lower ranges of optimisation $\omega_{min}$ and $\omega_{max}$. Right: $\varepsilon_m=||Q(\omega)-1||_{L_2}$, in terms of $N$; the slope of the linear regression is -5.} 
\end{center}
\label{FigQm1} 
\end{figure}

Figure 6 illustrates how the number of diffusive variables influences the accuracy of the optimiaation procedure, in the case of the coupled system. In the left part, we show the error $|Q(\omega)-1|$ for various values of $N$. The vertical dotted lines represent the range of optimisation. By construction, the error vanishes at the abscissae $\tilde{\omega}_k$. As expected, the accuracy of the diffusive approximation increases with $N$. The right part of the figure displays the error of model $\varepsilon_m=||Q(\omega)-1||_{L_2}$ on the range of interest, in terms of $N$ and in log-log scale. The measured values are close to a straight line with slope -5, hence one can postulate a power-law $\varepsilon_m\approx \varepsilon_0(1/N)^5$. 


\subsection{Comparison of the methods}\label{SecCoeffsCompare}

\begin{figure}[htbp]
\begin{center}
\begin{tabular}{cc}
phase velocity: coupled & attenuation: coupled\\
\includegraphics[scale=0.31]{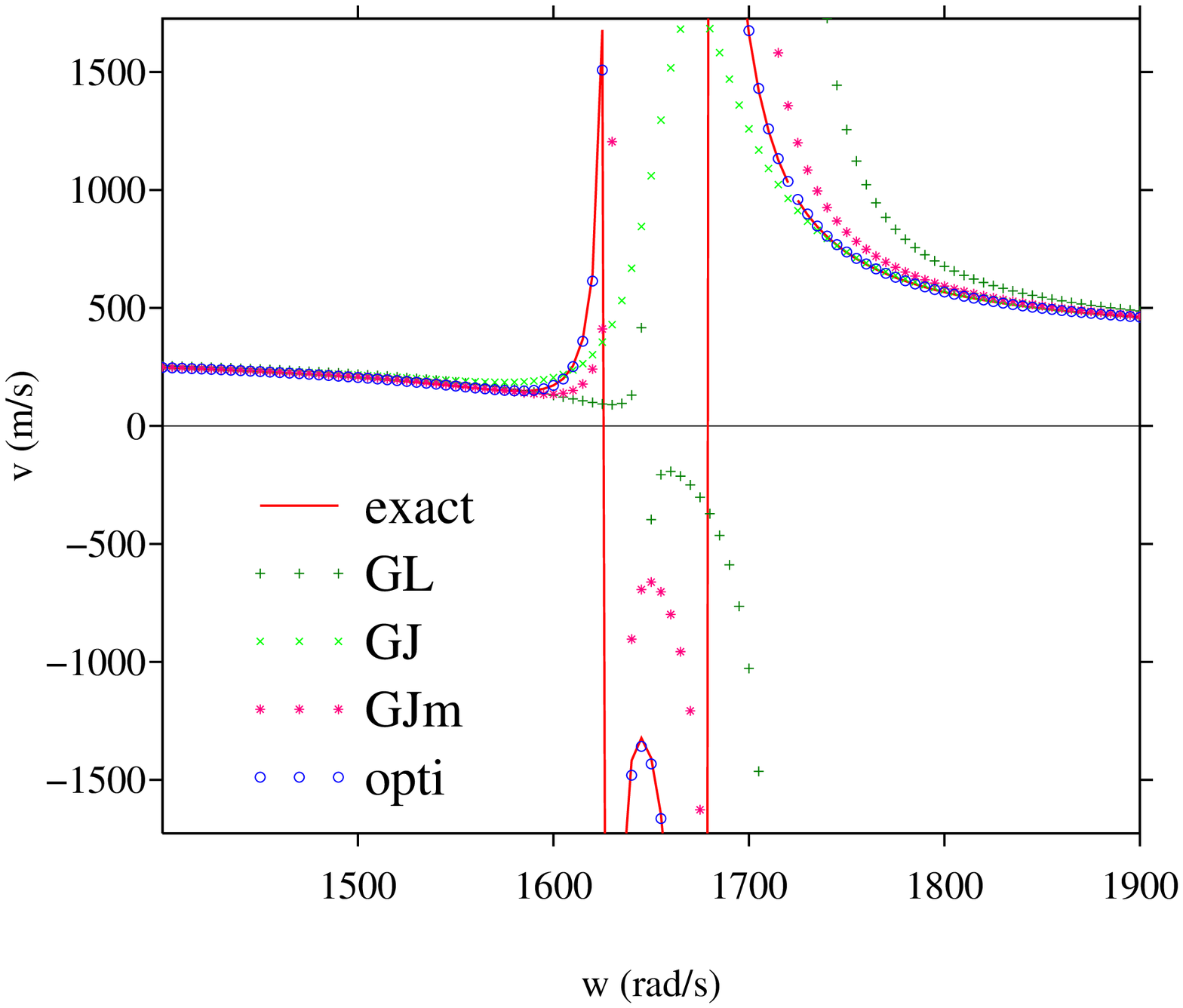} &
\includegraphics[scale=0.31]{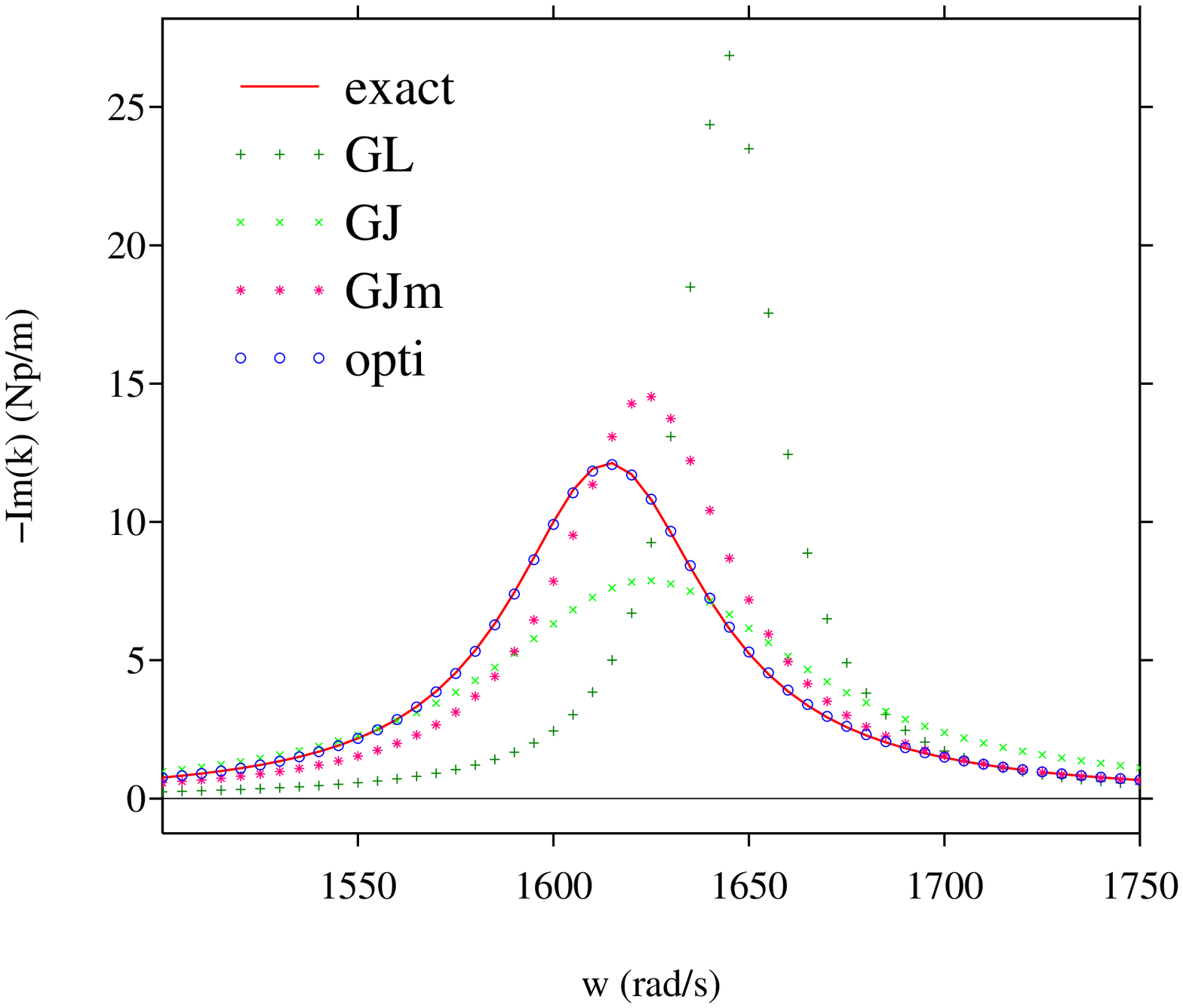} \\
phase velocity: uncoupled & attenuation: uncoupled\\
\includegraphics[scale=0.31]{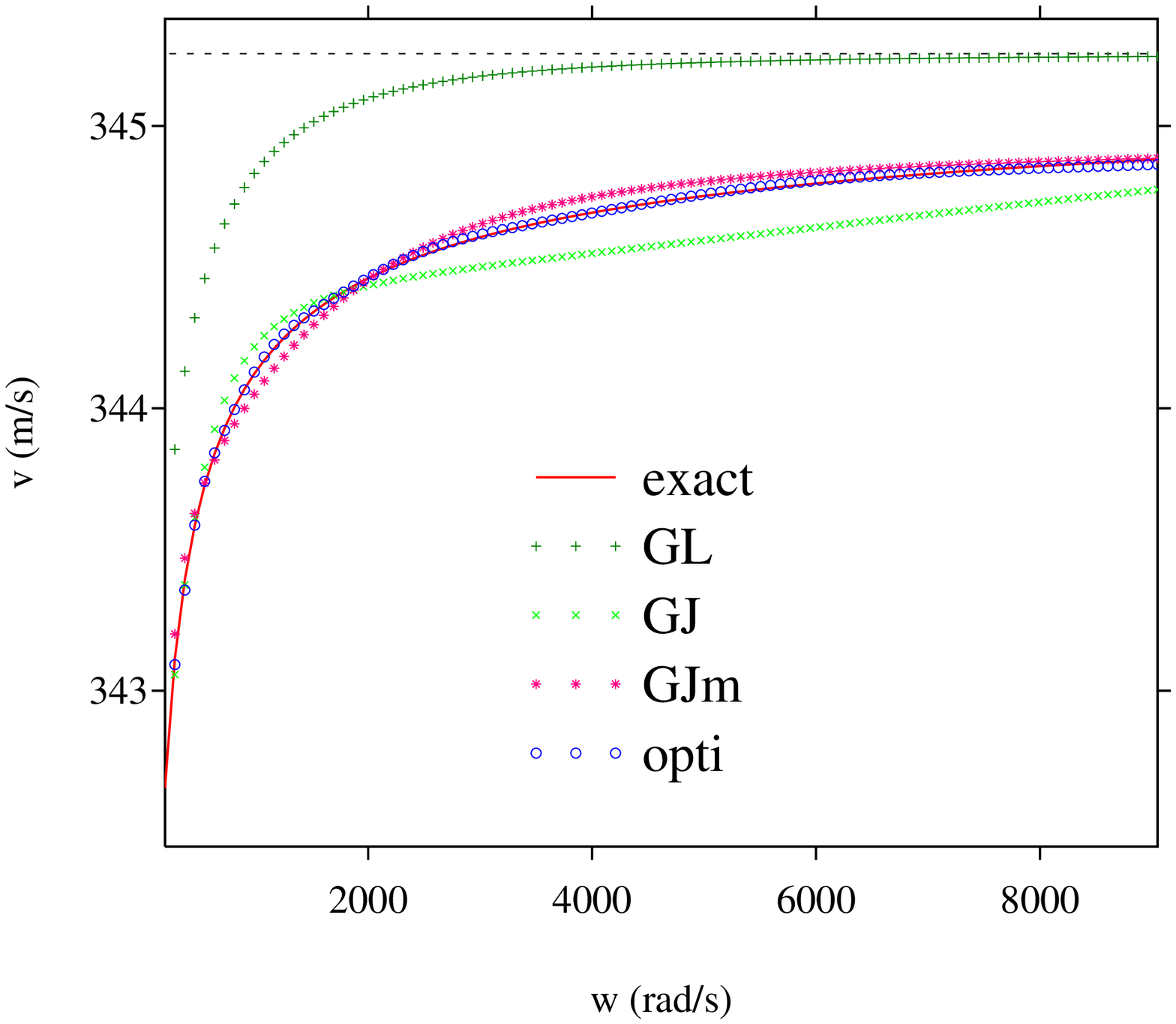} &
\includegraphics[scale=0.31]{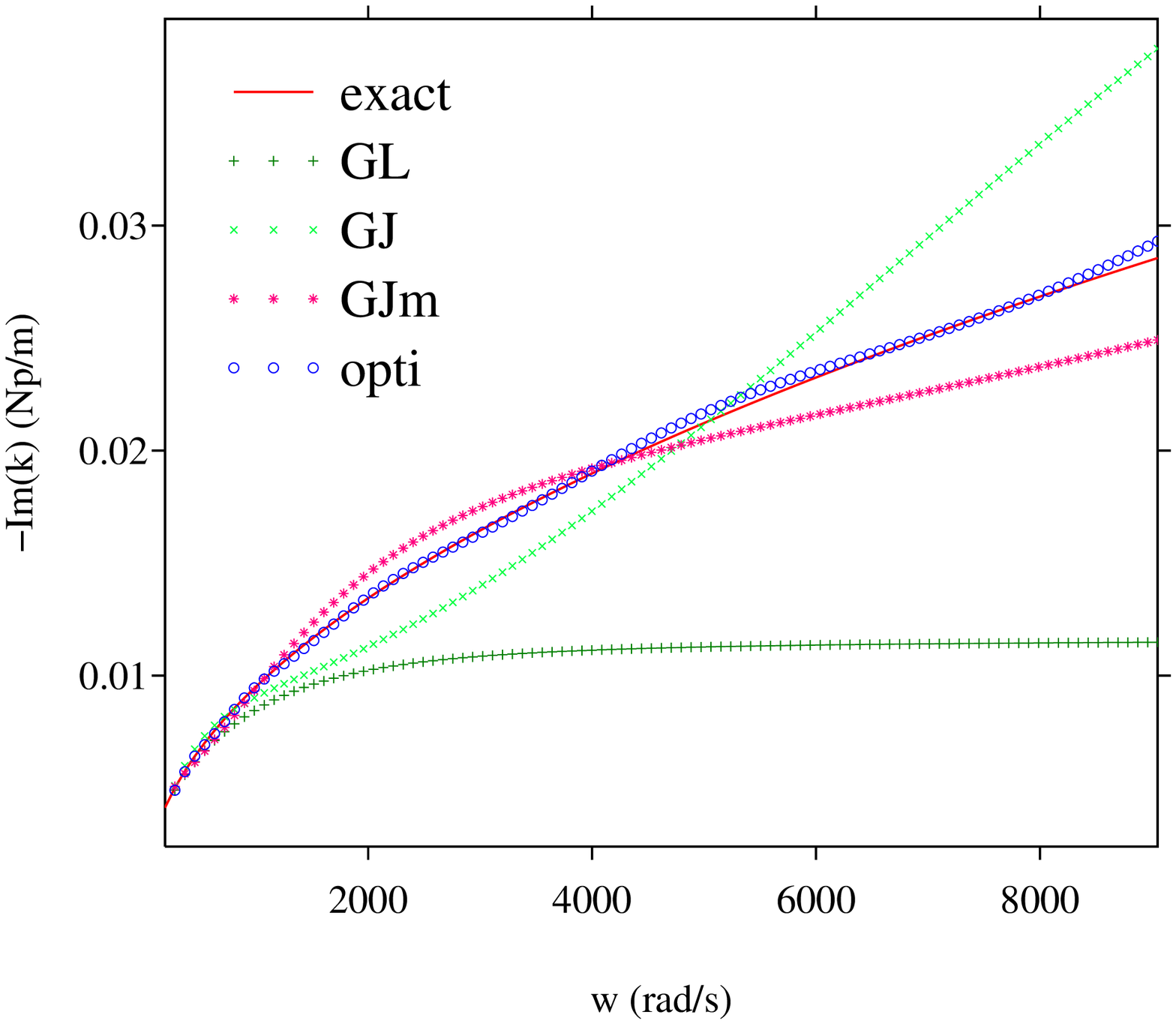}
\end{tabular}
\vspace{-0.5cm}
\caption{phase velocity and attenuation, with viscosity and in the linear regime. Coupled system with resonators (top, $N=6$); uncoupled system without resonators (bottom, $N=12$). Comparisons between the exact fractional model (\ref{EDP}) and the approximate diffusive model (\ref{SystHyper}). GL = Gauss-Laguerre (method 1); GJ = Gauss-Jacobi (method 2); GJm = Gauss-Jacobi modified (method 3); opti=optimisation (method 4).}
\end{center}
\label{FigCelAttMethods} 
\end{figure}

The accuracy of the methods presented along section \ref{SecCoeff} is illustrated in Figure 7. The exact values of the phase velocity and of the attenuation are shown in solid red lines, and correspond to the cases shown in figures 3, 4 and 5. In the case with resonators (top), $N=6$ diffusive variables are used, leading to 15 unknowns in (\ref{VecU}). In the case without resonators (bottom), $N=12$ diffusive variables are used, leading to 27 unknowns in (\ref{VecU}). Increasing accuracy is observed passing successively from Gauss-Laguerre to Gauss-Jacobi, to Gauss-Jacobi modified, and then to optimization. Consequently, the optimization procedure was chosen in the numerical experiments presented in section \ref{SecExp}.


\section{Numerical experiments}\label{SecExp}

\subsection{Configuration}\label{SecExpConfig}

\begin{table}[htbp]
\begin{center}
\begin{small}
\begin{tabular}{|l|l|l|l|l|l|}
\hline
$\gamma$ & $p_0$ (Pa) & $\rho_0$ (kg/m$^{3}$) & $Pr$    & $\nu$ (m$^2$/s) & $\mu_v/\mu$\\
\hline  
1.403    & $ 10^5$    & 1.177                 & 0.708   & $1.57\,10^{-5}$ & 0.60       \\
\hline
\hline
$R$ (m)  & $D$ (m)    & $r$ (m)               & $L$ (m) & $r_h$ (m)       & $H$ (m)    \\
\hline
0.04  & 0.05 & 0.00355 & 0.0356 & 0.0125 & 0.1\\
\hline
\end{tabular}
\end{small} 
\end{center}
\vspace{-0.5cm}
\caption{physical parameters of air at $15\,^{\circ}\mathrm{C}$, and geometrical data from \cite{Sugimoto04}.}
\label{TabParam}
\end{table}

The physical and geometrical parameters are given in table \ref{TabParam}. The physical data correspond to air at $15\,^{\circ}\mathrm{C}$, and the geometrical data are from \cite{Sugimoto04}. $a_0=345.25$ m/s, $C=1.478$ and $\nu_d=3.92\,10^{-5}$ m$^2$/s are obtained from (\ref{Omega0}). The parameters of (\ref{EDP})-(\ref{NotationsEDP}) are deduced. They are given in the upper part of table \ref{TabEdp}. The tube ${\cal L}=80$ m in length is discretized on $N_x=8000$ grid nodes. The maximal CFL number is $1-1/\mbox{Pe}$, and the P\'eclet number is $\mbox{Pe}=1.75\,10^5$ (\ref{Dt}). The CFL number is taken equal to 0.95. Lastly, a set of 10 receivers is put on the computational domain at abscissas $x_r=15+5\,(i-1)$, $i=1\cdots 10$, where the time history of $u$ was recorded at each time step. 

\begin{table}[htbp]
\begin{center}
\begin{small}
\begin{tabular}{|l|l|}
\hline
$a$ (m/s)                   &  345.25          \\
$b$                         &  1.20            \\
$c$ (m/s$^{3/2}$)           &  50.42           \\
$d$ (m$^2$/s)               & $1.96\,10^{-5}$  \\
$e$ (m/s$^3$/Pa)            & $2.40\,10^{-4}$  \\
$f$ (s$^{-1/2}$)            & 2.23             \\
$g$ (s$^{2}$)               & $2.70\,10^6$     \\
$h$ (Pa/m/s)                & $1.10\,10^9$     \\
\hline
$\omega_0$ (rad/s)          & 1645.67          \\
$\omega_1$ (rad/s)          & 1724.16          \\ 
$\overline{\upsilon}$ (m/s) & 314.53           \\
\hline
\end{tabular}
\end{small} 
\end{center}
\vspace{-0.5cm}
\caption{coefficients of the PDE (\ref{EDP}). Lower part: natural angular frequencies (\ref{Omega0}), and low frequency limit of the phase velocity (\ref{Vphase}) in the coupled system.}
\label{TabEdp}
\end{table}

Except in section \ref{SecExpTest2}, computations were initialized by a Gaussian pulse or by a rectangular force pulse on the velocity
\begin{equation}
u_0(x)=
\left\{
\begin{array}{l}
\displaystyle
u_m\,\exp\left(-\displaystyle \left(\frac{\textstyle x-x_0}{\textstyle \sigma}\right)^2\right),\\
[10pt]
\displaystyle
u_m\,\left({\cal H}\left(x-x_0-\frac{\textstyle \lambda}{\textstyle 2}\right)-{\cal H}\left(x-x_0+\frac{\textstyle \lambda}{\textstyle 2}\right)\right),
\end{array}
\right.
\label{U0}
\end{equation}
where ${\cal H}$ is the Heaviside function and $x_0=7$ m. $\lambda$ is the width of the rectangular force pulse, and is also the width of the Gaussian pulse: taking $\sigma=\lambda/2\,\sqrt{\ln 100}$ gives $u_0(x)=u_m/100$ at $x=x_0 \pm \lambda/2$. All the other initial conditions in (\ref{CondInit}) are null.

The key parameters governing the evolution of the system were $K$ and $\Omega$ (\ref{NotationsEDPscale}); see section \ref{SecPhysRegime}. On the one hand, we took $K=0.5$ which ensures a nonlinear regime of propagation, hence the amplitude $u_m\approx 56.12$ m/s (\ref{U0}), the parameter of nonlinearity $\varepsilon=0.19$, the overpressure $p^{'}/p_0=0.22$ and the sound intensity $I=181.1$ dB (\ref{IdB}). On the other hand, we took $\Omega=1$ or $\Omega=16$, yielding theoretically to dispersive waves and solitons, respectively. In the case $\Omega=1$, the linear dispersion analysis predicts a maximal attenuation. In the case $\Omega=16$, the natural frequency is $\omega=411.4$ rad/s and the central wavelength is $\lambda=5.27$ m. The cut-off angular frequency is $\omega^*=15881$ rad/s, so that the 1D approximation is justified (section \ref{SecPhysEdp}).


\subsection{Test 1: nonlinear acoustics}\label{SecExpTest1}

\begin{figure}[htbp]
\begin{center}
\begin{tabular}{cc}
(i) & (ii)\\
\includegraphics[scale=0.31]{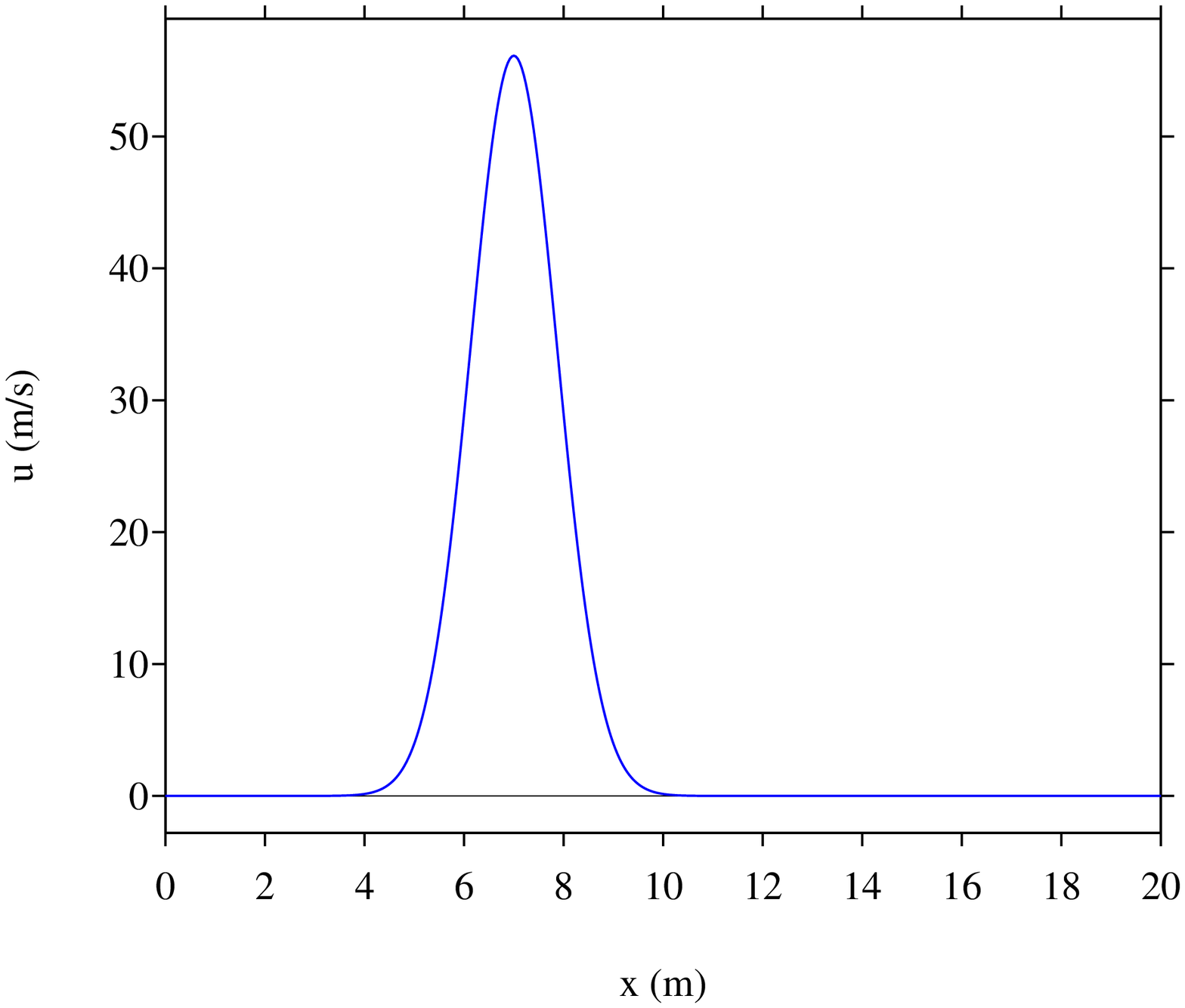} &
\includegraphics[scale=0.31]{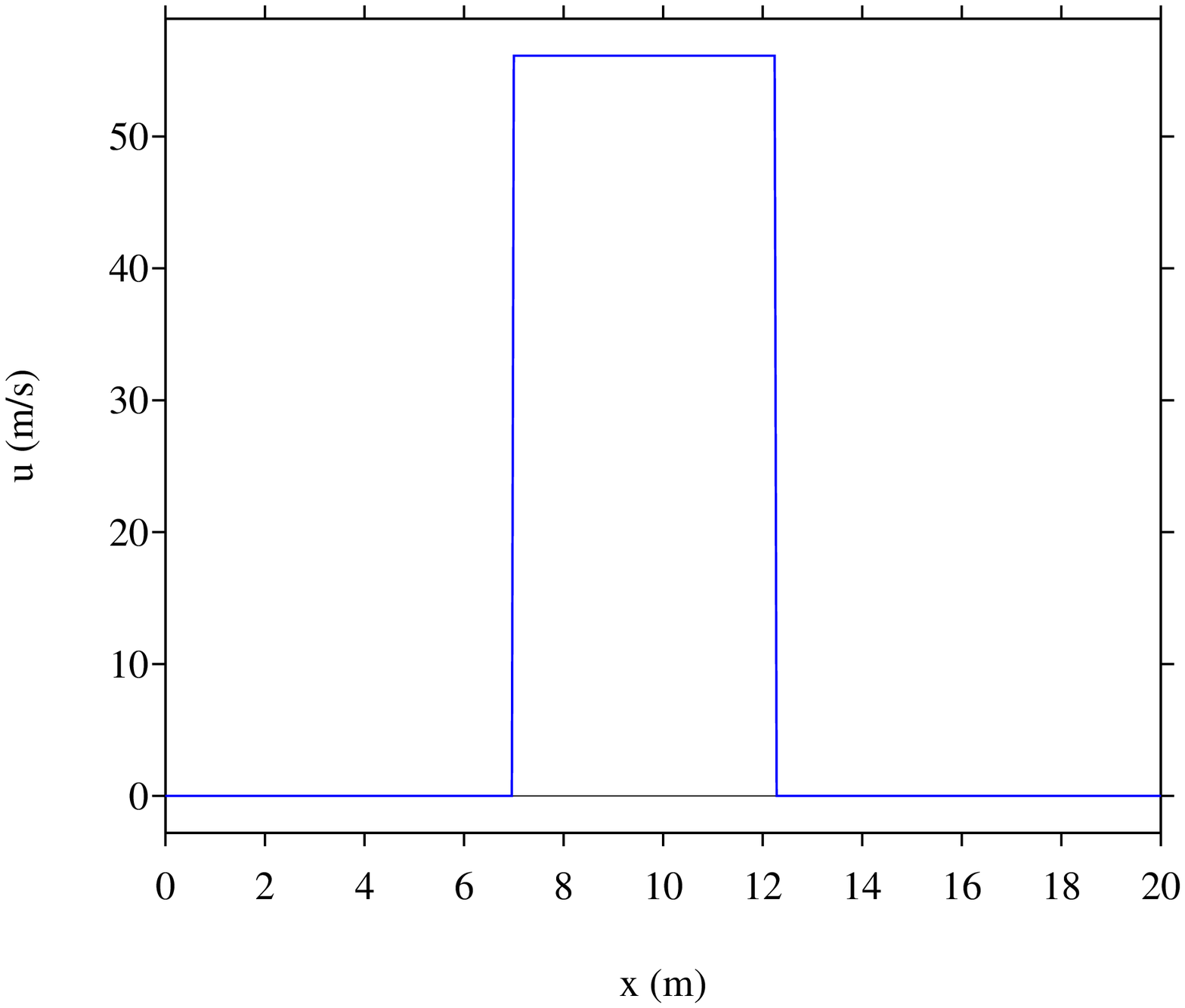} \\
(iii) & (iv)\\
\includegraphics[scale=0.31]{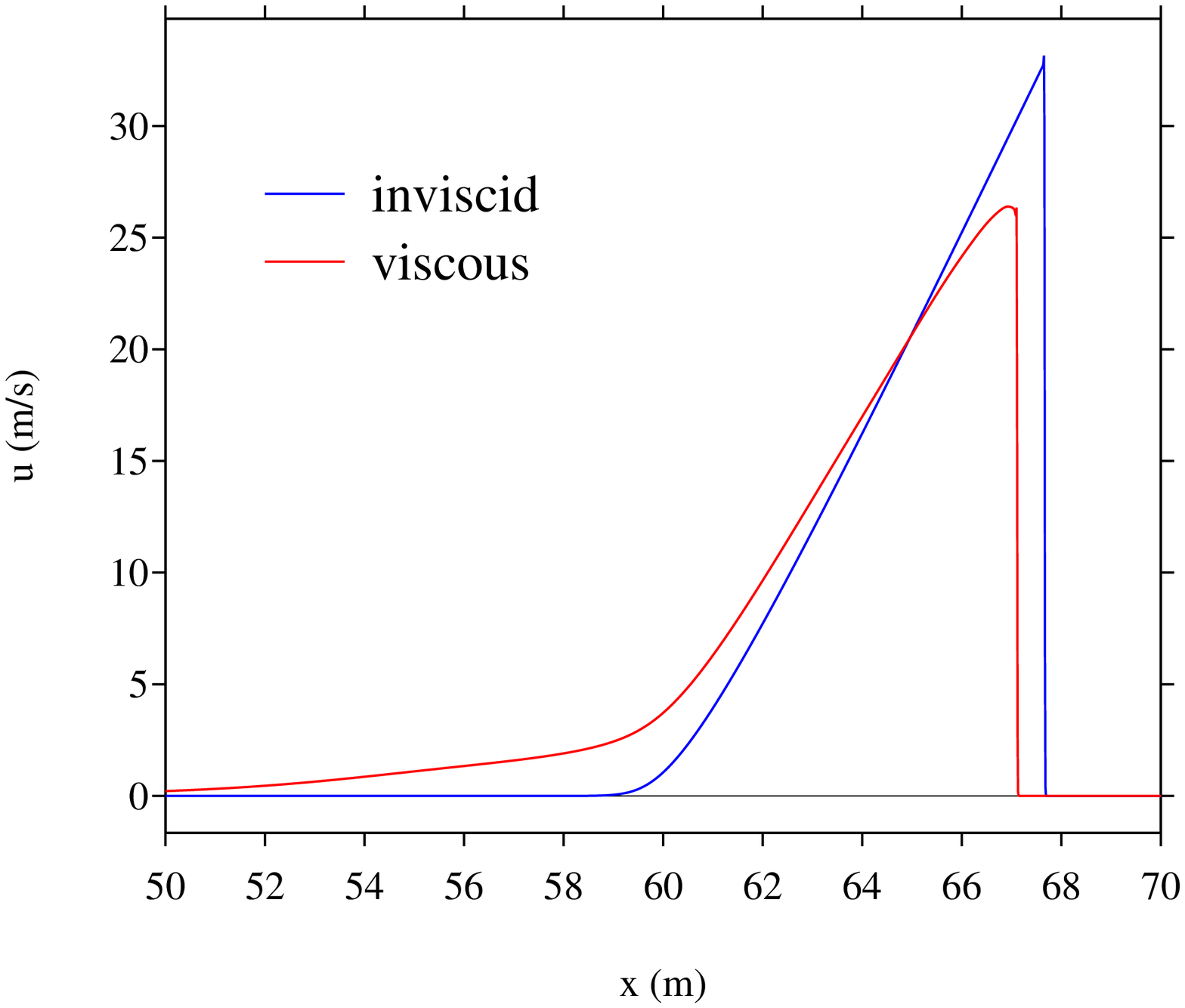} &
\includegraphics[scale=0.31]{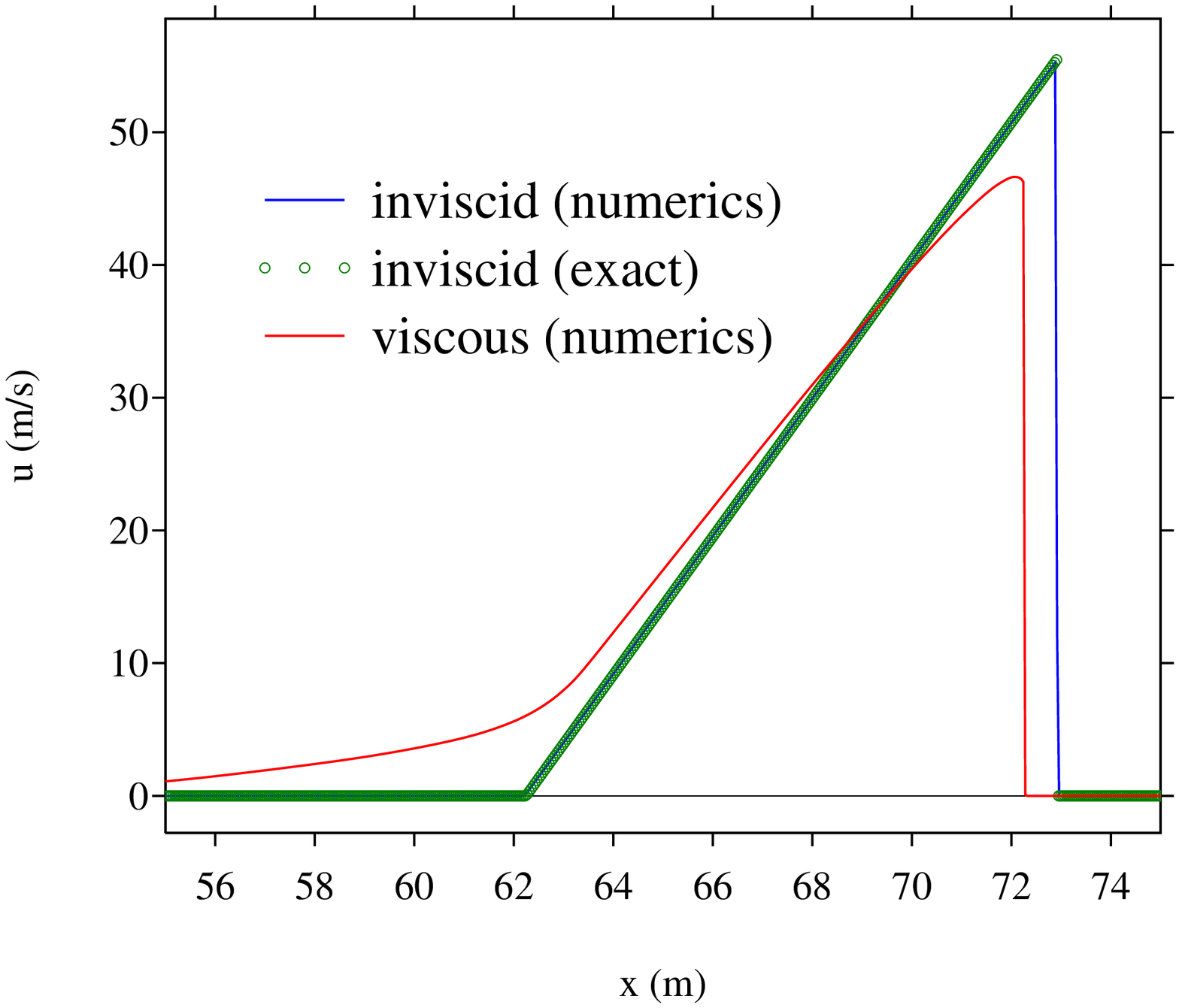} \\
(v) & (vi)\\
\includegraphics[scale=0.31]{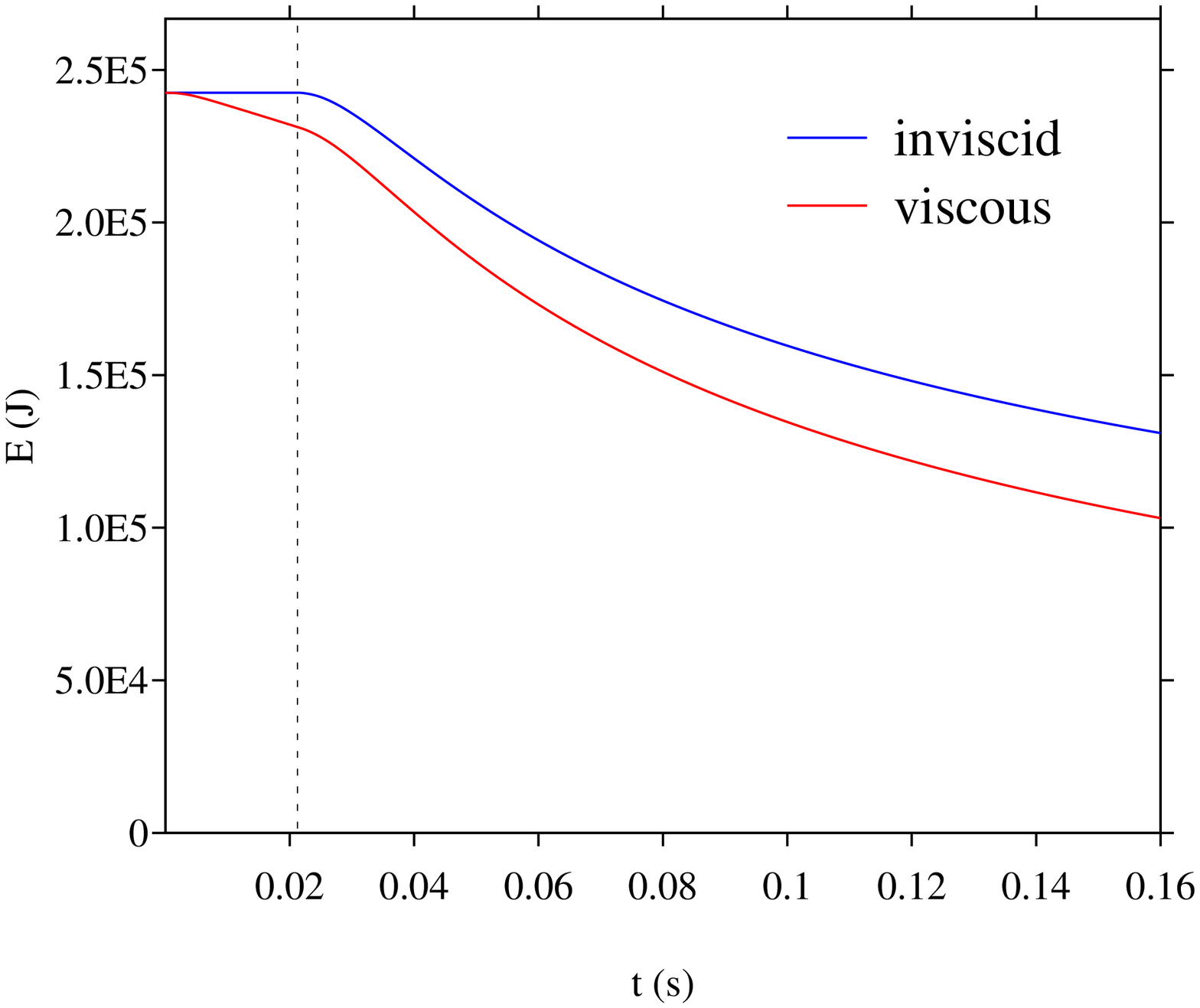} &
\includegraphics[scale=0.31]{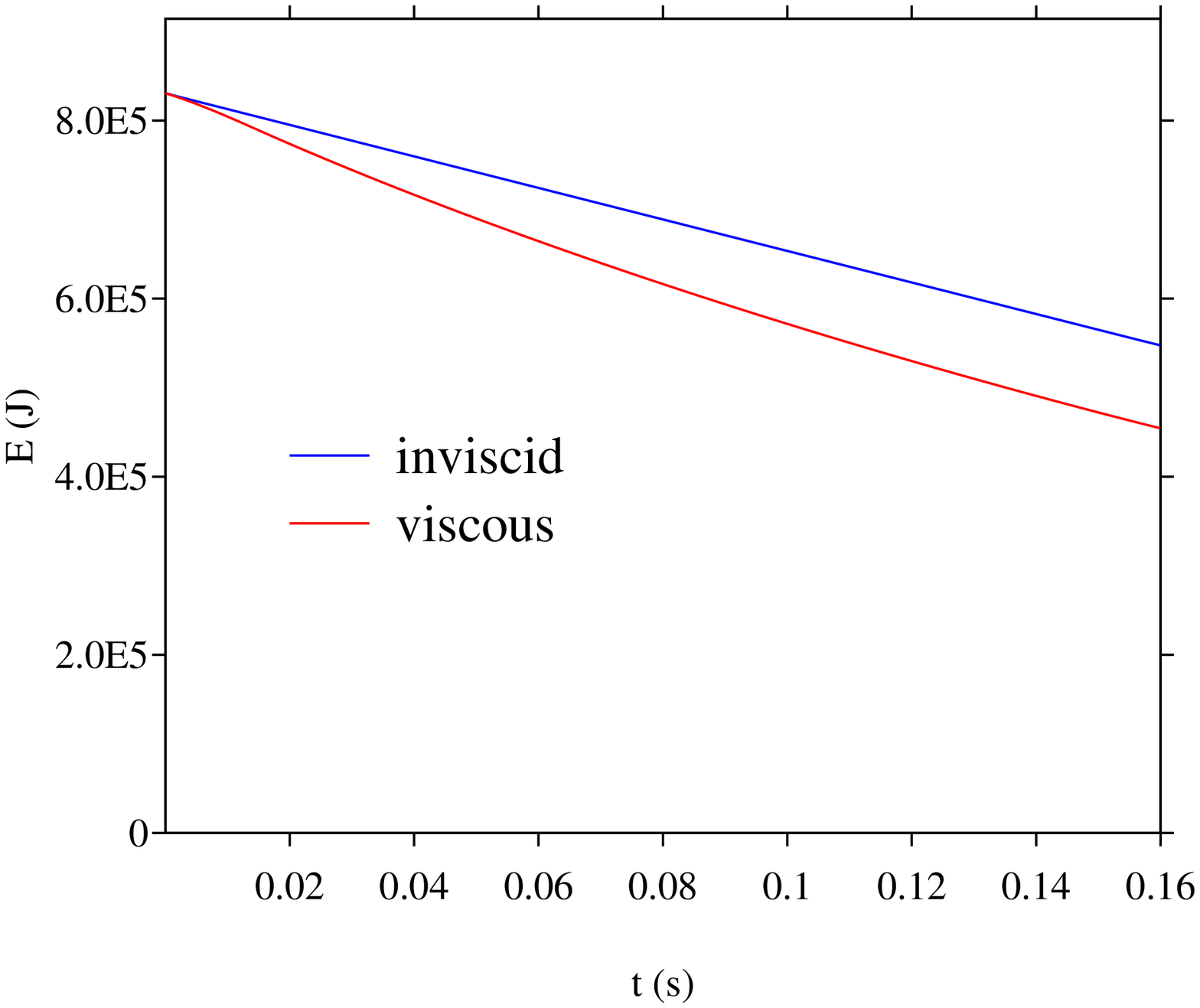} 
\end{tabular}
\vspace{-0.5cm}
\caption{test 1. Nonlinear acoustic waves in the absence of oscillators, for a Gaussian pulse (left row) and a rectangular force pulse (right row). (i-ii): initial pulse. (iii-iv): numerical and exact solution at $t=0.16$ s. (v-vi): time evolution of the energy; in (v), the vertical dotted line denote the time of shock $t^*$.} 
\end{center}
\label{FigT1} 
\end{figure}

In a first test, the height of the cavity was $H=0$, hence $e=0$ and no coupling occured with the Helmholtz resonators. The dispersion relation was therefore (\ref{DispersionGuide}), and the key adimentionalized parameters in section \ref{SecPhysRegime} were $K=0$ and $\Omega=+\infty$. The number of diffusive variables was $N=12$; since no memory variables $\xi$ are required to model dissipative effects in the resonators, only 15 variables were involved in (\ref{VecU}). Optimisation of the coefficients $\mu_{\ell}$ was performed between $\omega_{min}=185$ rad/s and $\omega_{max}=8228$ rad/s (section \ref{SecCoeff4}).

In the inviscid case, where only the coefficients $a$ and $b$ are non-null in (\ref{EDP}), smooth initial data develop shocks in finite time, yielding a decrease in energy. The Gaussian initial pulse breaks at time 
\begin{equation}
t^*=\sqrt{\frac{\textstyle \exp(1)}{2}}\,\frac{\textstyle \sigma}{\textstyle u_m\,b}.
\label{TstarGauss}
\end{equation}
It gives the time break $t^*=0.021$ s (\ref{TstarGauss}).

Figure 8 shows the initial values of the solution at the initial instant (i-ii) and at $t=0.16\,\mbox{ s}>t^*$ (iii-iv), which corresponds roughly to 7000 time steps. Both the inviscid case and the viscous case are displayed, where the viscous boundary layer in the tube and the diffusivity of sound are accounted for. In the inviscid case, typical nonlinear phenomena are observed: shock on the right part of the Gaussian pulse (iii), rarefaction waves and right-going shock for the rectangular force pulse (iv). In this latter case, good agreement is obtained with the exact solution. In the viscous case, these phenomena are qualitatively maintained. A small decrease in amplitude is observed, together with a tail on the left part of the waves. The key issue is that viscous effects are not sufficient to prevent from the occurrence of a shock (iii) or to smooth an existing discontinuity (iv), which confirms the theoretical analysis performed in \cite{Sugimoto91}. 

Lastly, time evolution of the energy $E^{(n)}=\sum_j (u_j^n)^2$ is displayed in (v-vi). For the Gaussian pulse in the inviscid case (v), energy is conserved as long as the wave is smooth; at the scale of the figure, numerical diffusion is not seen. From $t=t^*$ where the shock appears, energy decreases. For the rectangular force pulse in the inviscid case (v), energy decreases linearly with time. The results are qualitatively the same in the viscous cases, with a greater decrease in energy.


\subsection{Test 2: fractional oscillations in Helmholtz resonators}\label{SecExpTest2}

\begin{figure}[htbp]
\begin{center}
\begin{tabular}{cc}
\includegraphics[scale=0.31]{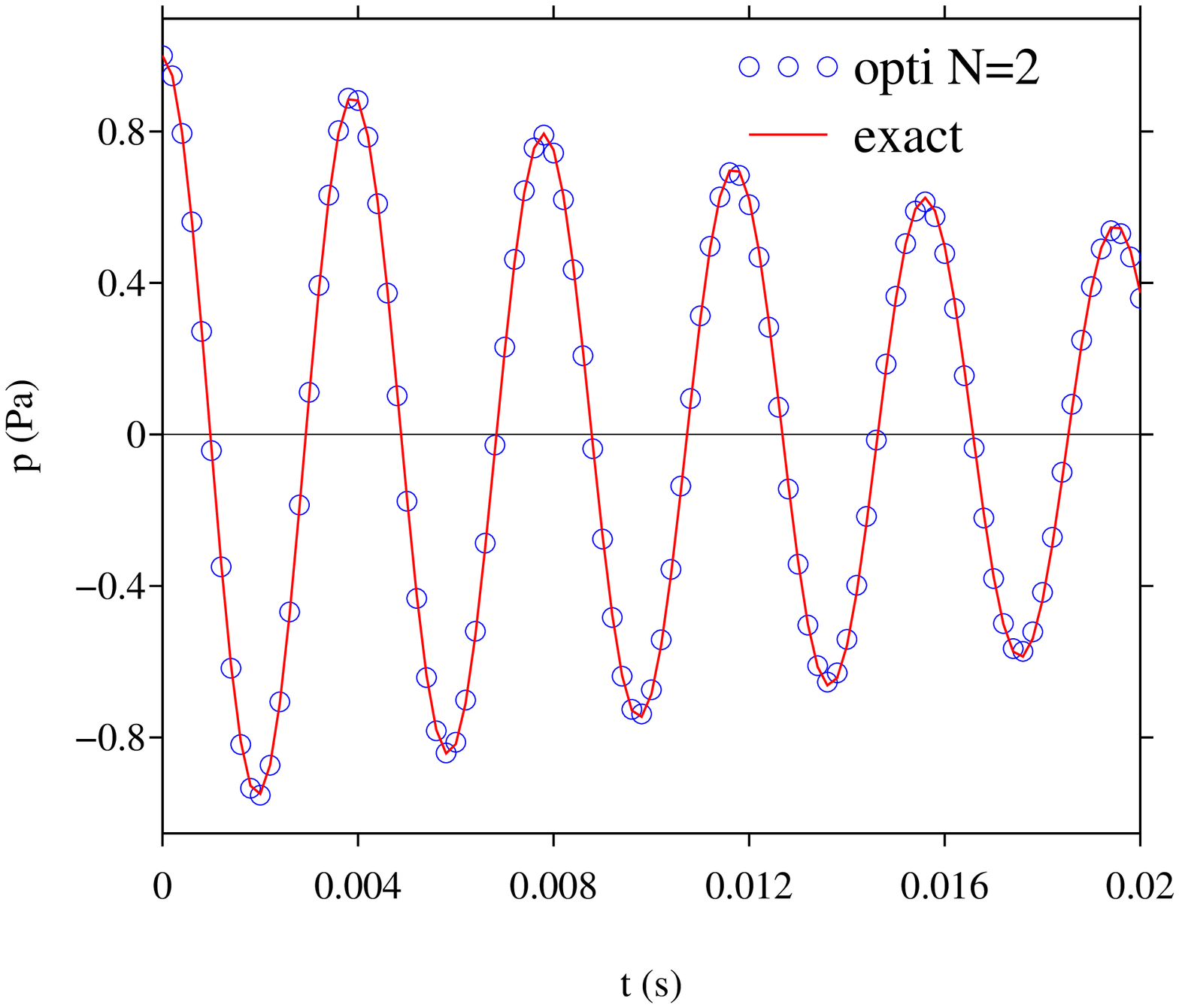}&
\includegraphics[scale=0.31]{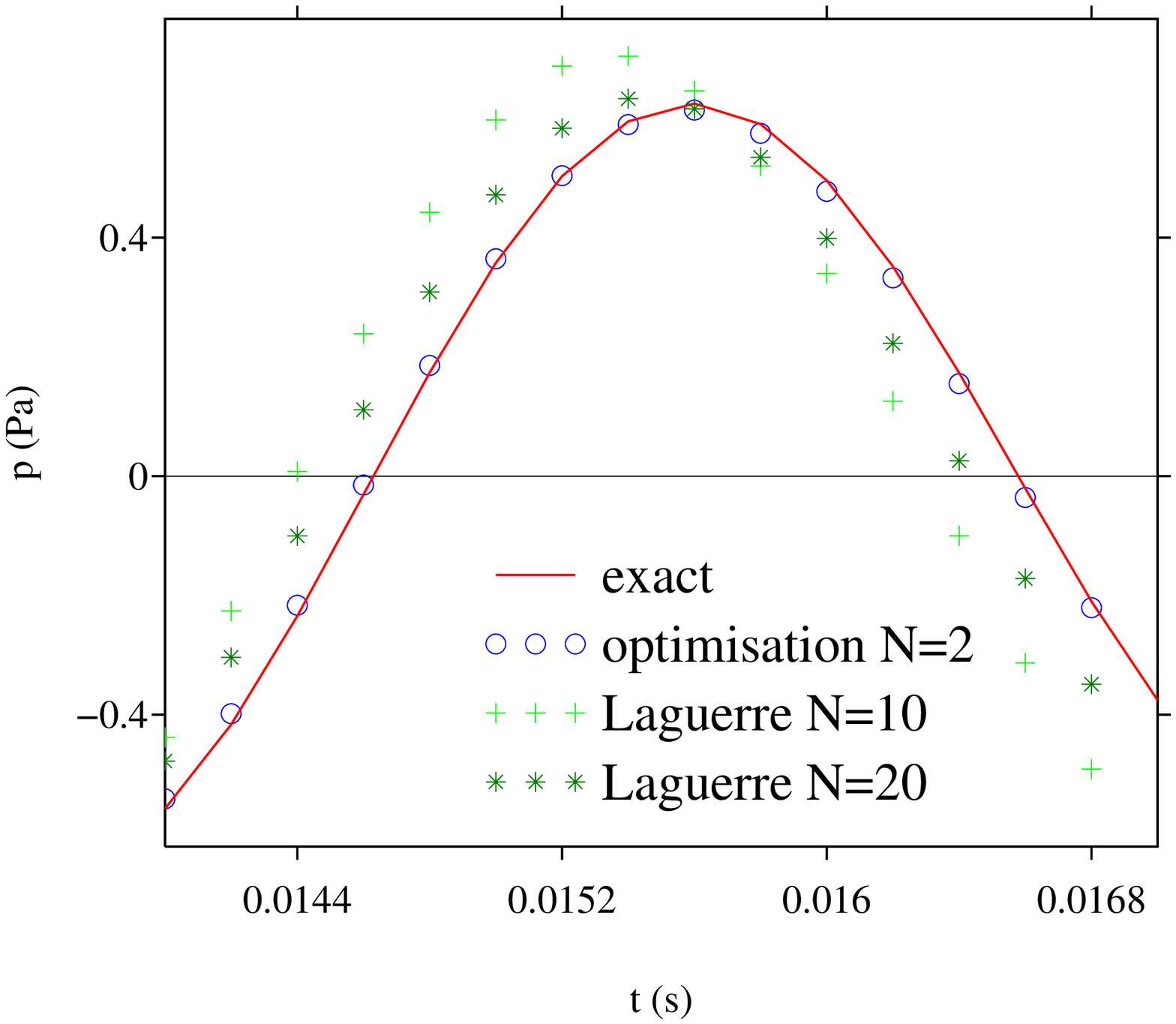}
\end{tabular}
\end{center}
\vspace{-0.8cm}
\caption{test2. Fractional oscillator of order 3/2. Comparisons between the numerical solution and the exact solution (right row: zoom).}
\label{FigTest2}
\end{figure}

In this second validation test, we focused on the fractional oscillator of order 3/2. The coupling with nonlinear acoustics in the tube was neglected, and (\ref{EDP2})  was solved with no coupling $h=0$. The initial value of pressure was $p_0(x)=1$ and $p_1(x)=0$, leading to oscillations with damping. The analytical solution is obtained in terms of fractional power series on 200 modes: see equation (10) of \cite{Deu10}. The numerical solution is obtained by solving the following problem with $N+2$ unknows
\begin{equation}
\left\{
\begin{array}{l}
\displaystyle
\frac{\textstyle \partial p}{\textstyle \partial t}=q,\\
[8pt]
\displaystyle
\frac{\textstyle \partial q}{\textstyle \partial t}=-g\,p-f\,\sum_{\ell=1}^N\mu_{\ell}\left(-\theta_{\ell}^2\,\xi_{\ell}+\frac{\textstyle 2}{\textstyle \pi}\,q\right),\\
[12pt]
\displaystyle
\frac{\textstyle \partial \xi_{\ell}}{\textstyle \partial t}=-\theta_{\ell}^2\,\xi_{\ell}+\frac{\textstyle 2}{\textstyle \pi}\,q,\hspace{1.5cm} \ell=1\cdots N.
\end{array}
\right.
\end{equation}
Therefore the numerical solution is the exact solution of the system (\ref{SplitDiffuExp}) with $N$ memory variables; consequently, the only error is the error of model $\varepsilon_m$, due to the quadrature of the diffusive representation. 

Figure \ref{FigTest2} shows the influence of the quadrature rule on the accuracy of the model. When optimization is used, only $N=2$ memory variables are required to obtain an excellent agreement with the exact solution (left row). On the contrary, large errors are observed when Laguerre quadrature is used, even with $N=20$.


\subsection{Test 3: linear dispersive waves}\label{SecExpTest3}

\begin{figure}[htbp]
\begin{center}
\begin{tabular}{cc}
$\nu =0$ & $\nu \neq 0$\\
\includegraphics[scale=0.31]{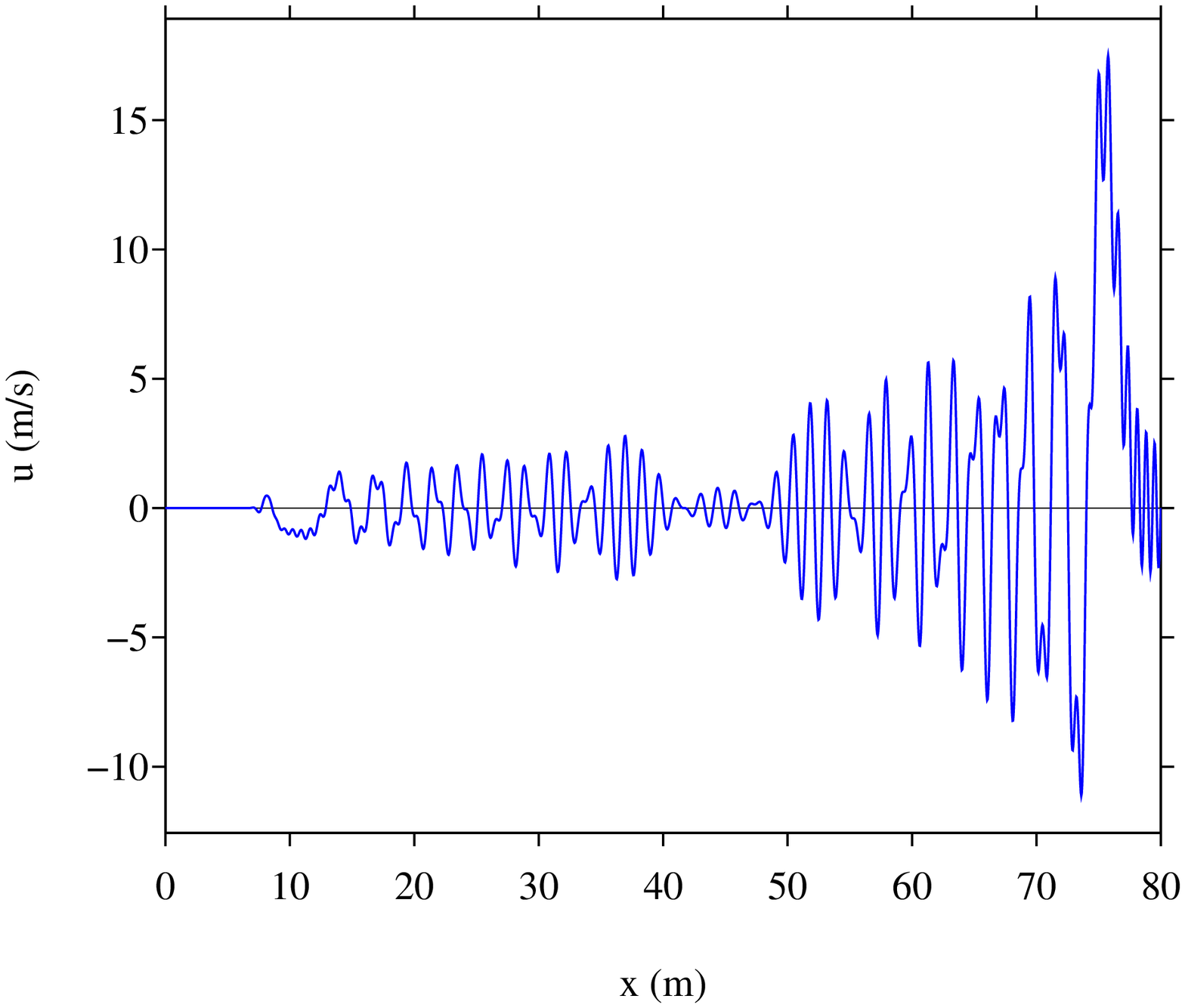}&
\includegraphics[scale=0.31]{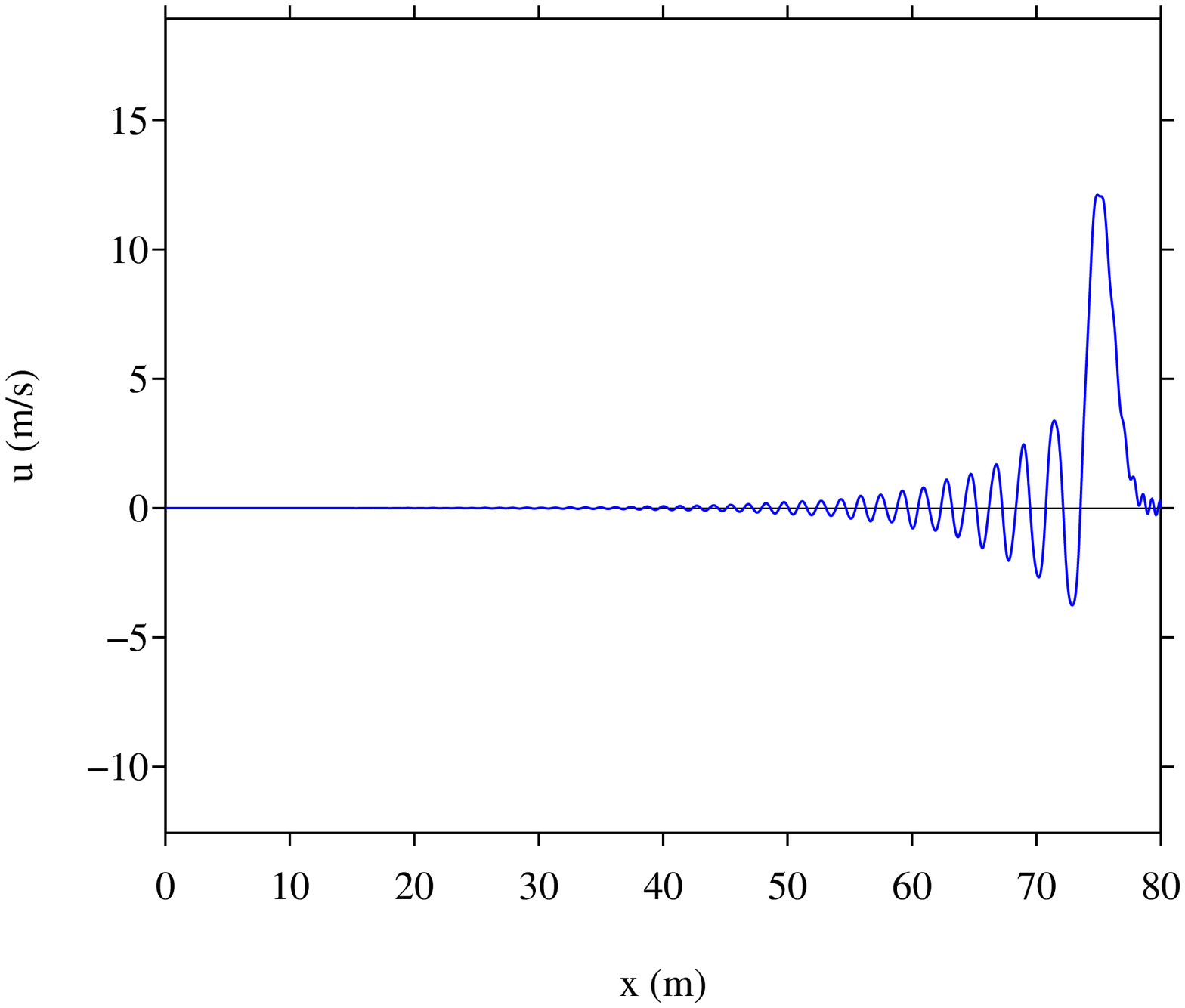}
\end{tabular}
\end{center}
\vspace{-0.8cm}
\caption{test3. Linear coupled system, with a Gaussian pulse and $\Omega=1$. Snapshots of $u$ after 8000 time steps. Left: invicid case; right: viscous case.}
\label{FigT3Om1}
\end{figure}

In a third test, we considered the coupled system with resonators, in the linear regime: $b=0$ in (\ref{EDP}). The simulations were initialized by a Gaussian pulse, and they were performed over 8000 time steps. Computations were done with and without the viscous effects of boundary layers and  diffusivity of sound. The number of diffusive variables is $N=6$, involving 15 variables in (\ref{VecU}). Optimisation of the diffusive coefficients $\mu_{\ell}$ is performed between $\omega_{min}=822$ rad/s and $\omega_{max}=2468$ rad/s if $\Omega=1$, and values 4 times smaller if $\Omega=16$ (section \ref{SecCoeff}). 

\begin{figure}[htbp]
\begin{center}
\begin{tabular}{cc}
$\nu =0$ & $\nu \neq 0$\\
\includegraphics[scale=0.31]{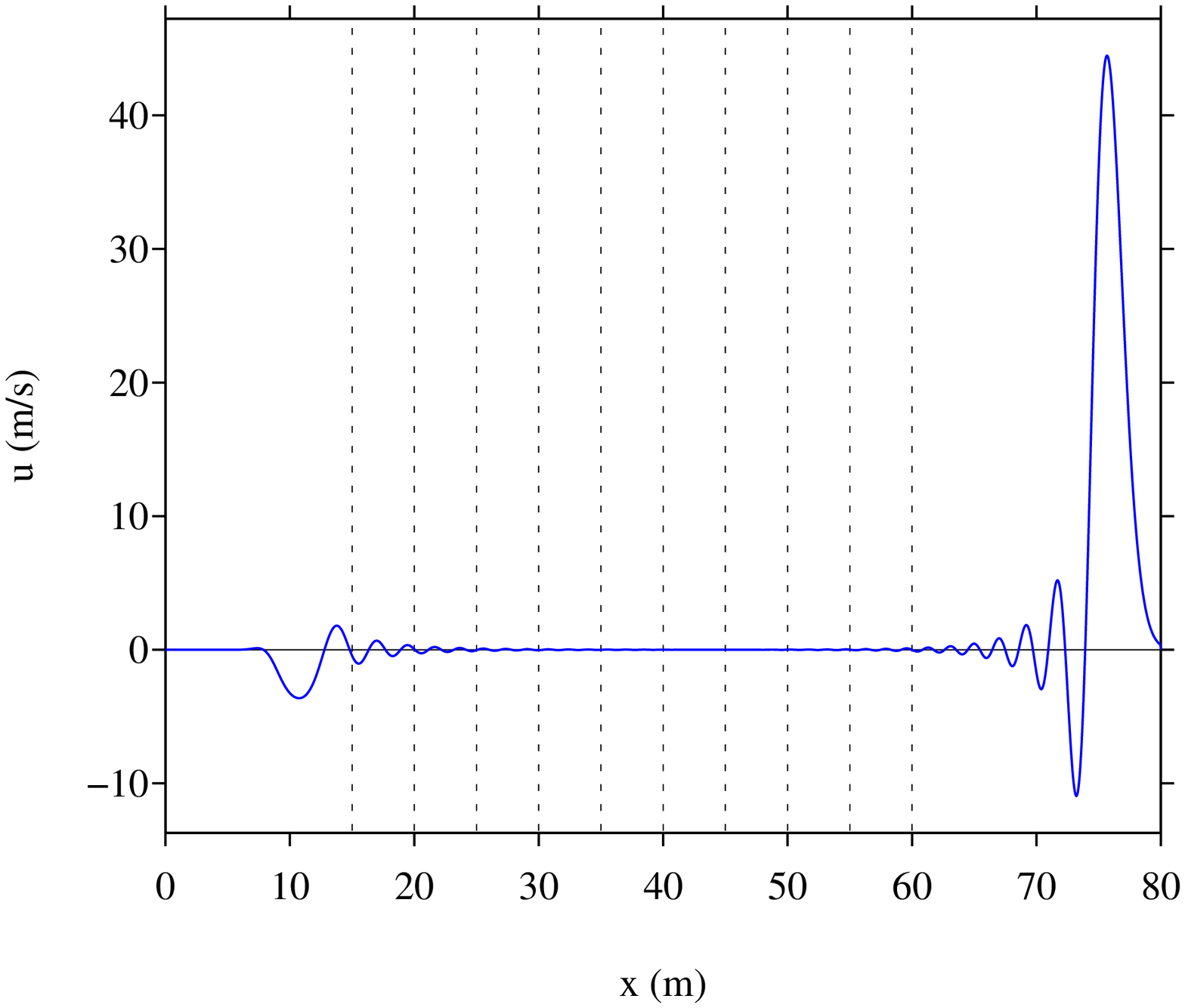}&
\includegraphics[scale=0.31]{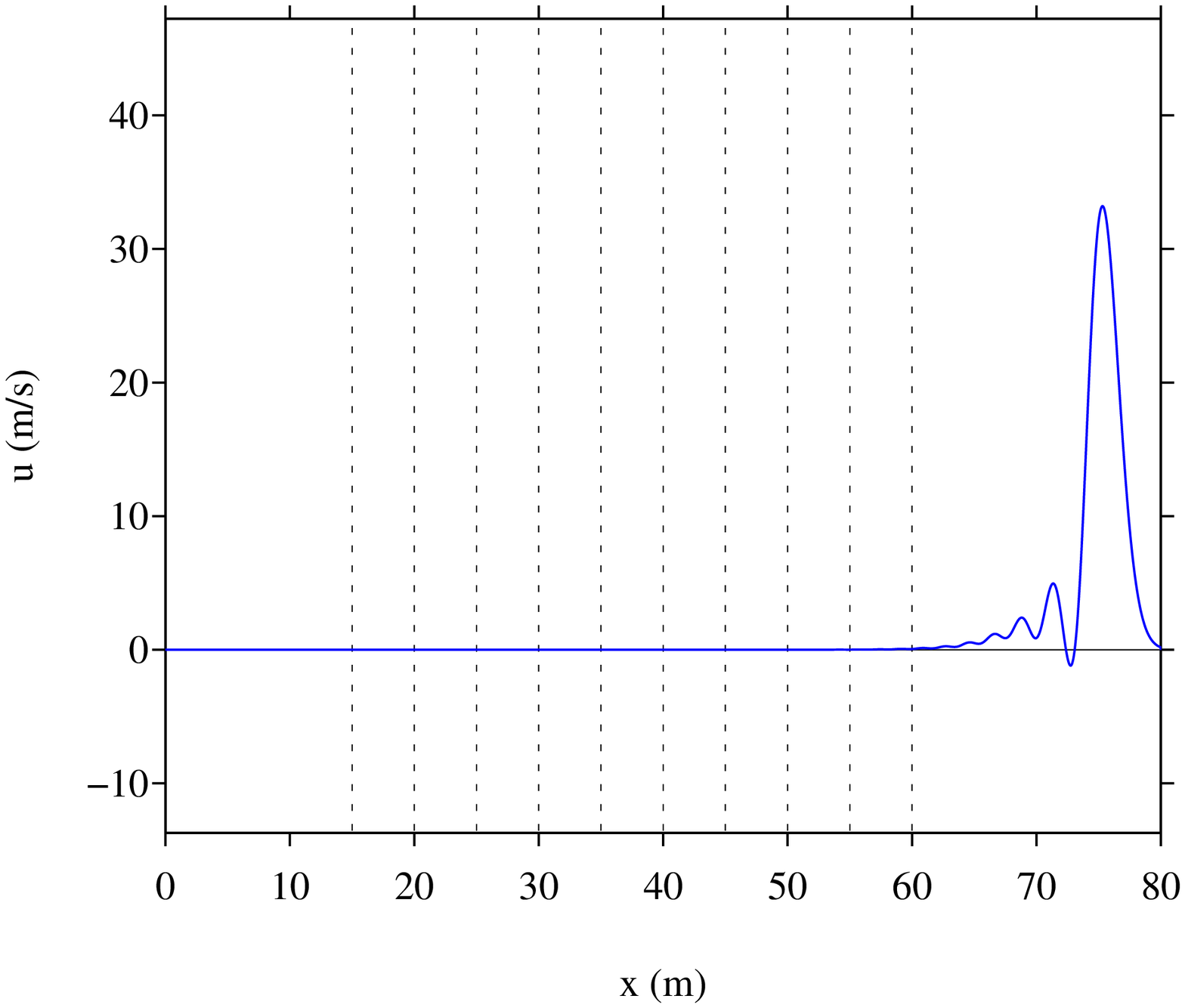}\\
\includegraphics[scale=0.31]{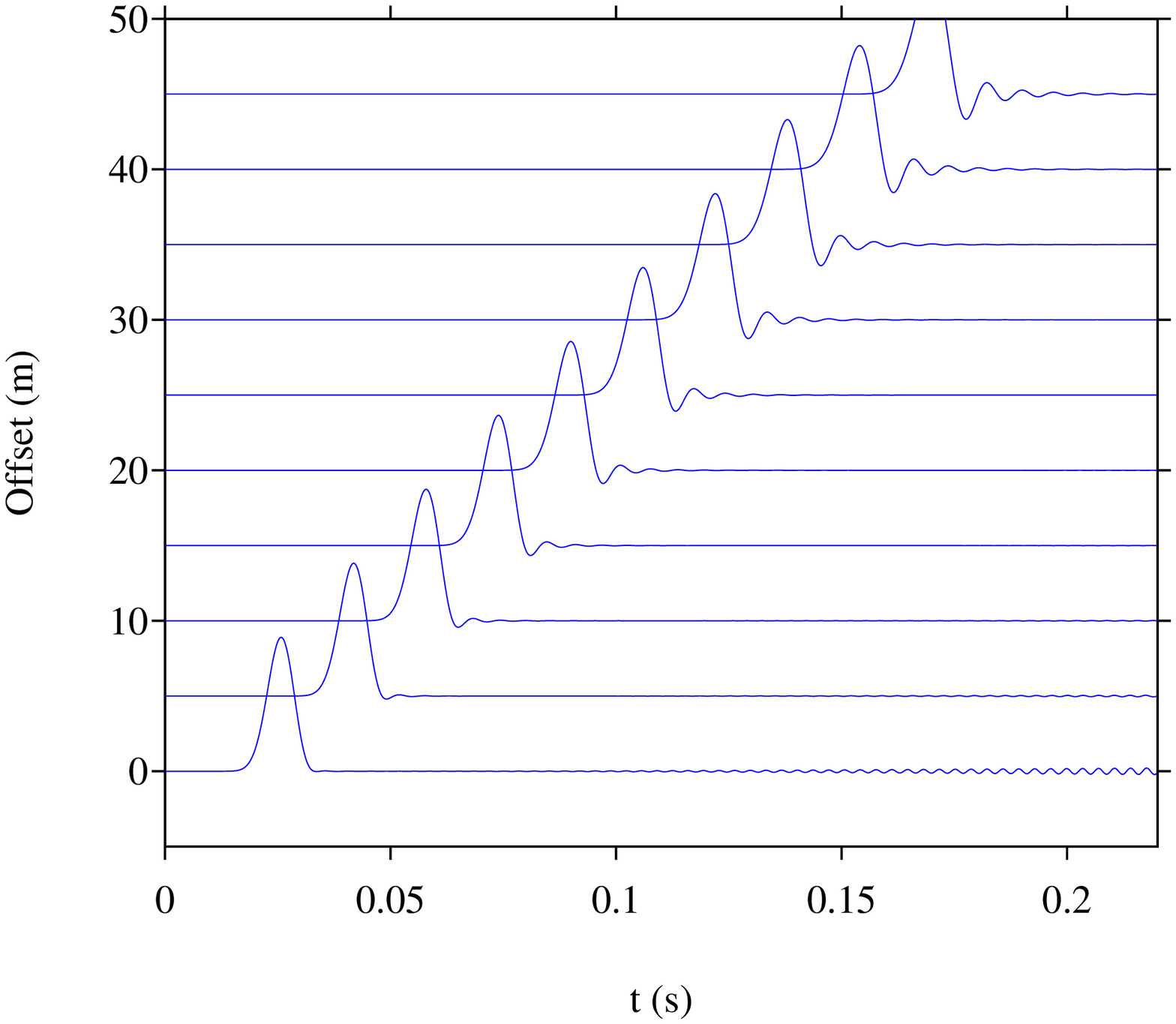}&
\includegraphics[scale=0.31]{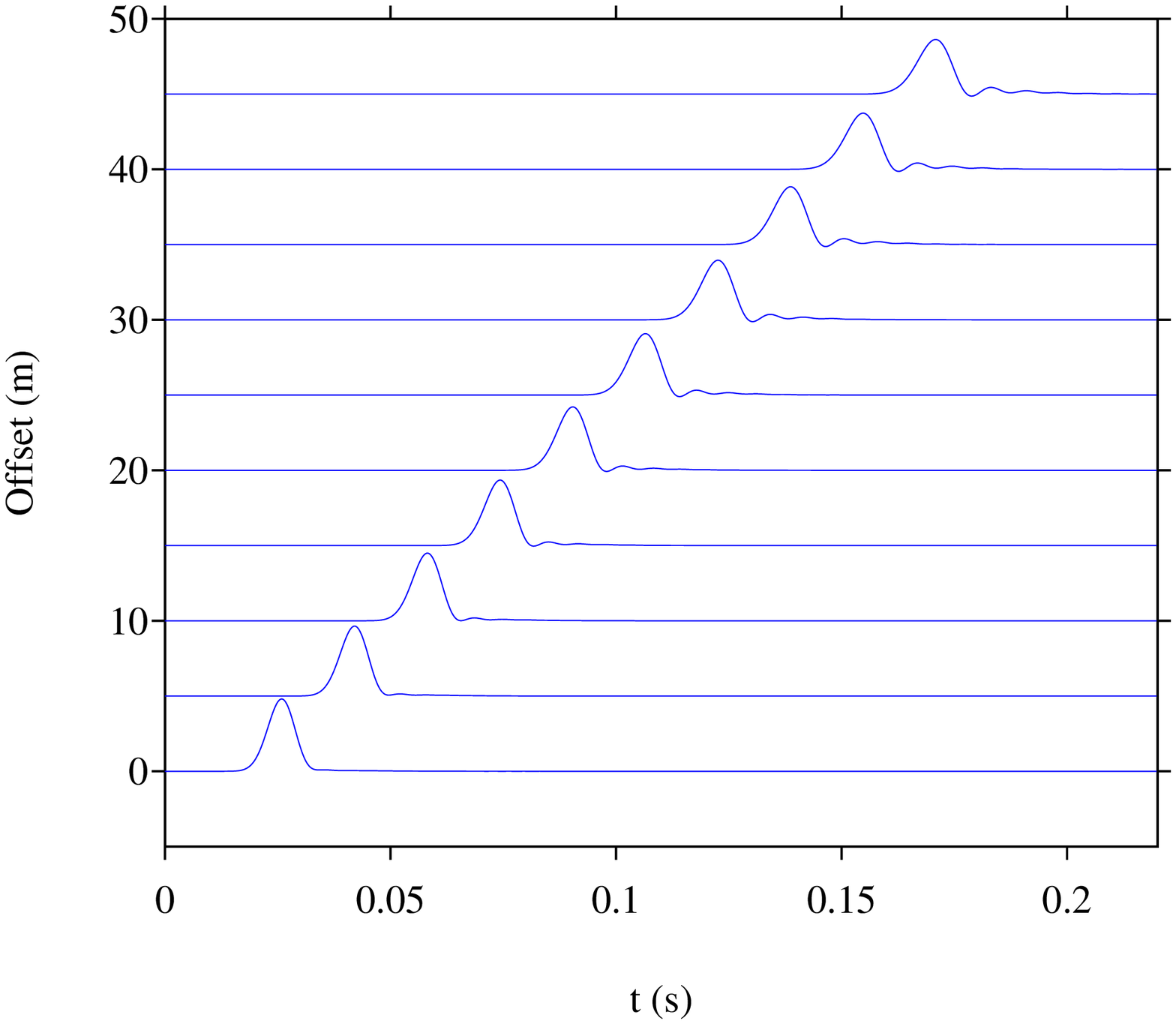}
\end{tabular}
\end{center}
\vspace{-0.8cm}
\caption{test3. Linear coupled system, with a Gaussian pulse and $\Omega=16$. Left: invicid case; right: viscous case. Top: snapshot of $u$ after 8000 time steps (the vertical dotted lines denote the receivers); bottom: seismograms.}
\label{FigT3Om16}
\end{figure}

The case $\Omega=1$ is shown in figure \ref{FigT3Om1}. High dispersion is observed in the inviscid case, which confirms the dispersion analysis performed in section \ref{SecPhysDisp} (see figure 3, near the vertical dotted lines). The oscillations are highly damped in the viscous case, due to the large value of attenuation (see figure 4).

The case $\Omega=16$ is displayed in figure \ref{FigT3Om16}. Compared with figure \ref{FigT3Om1}, the dispersion is greatly reduced. In the inviscid case, an oscillating mode remains at the place of initialization; moreover, the energy is conserved (not shown here). The static mode is damped in the viscous case. Seismograms are built from the time signals stored at the receivers. Then, the celerity ${\cal V}$ of the highest amplitude is numerically measured. One obtains ${\cal V}=312.02$ m/s (if $\nu =0$) and ${\cal V}=310.39$ m/s (if $\nu \neq0$). These values are close to the zero-frequency limit $\overline{\upsilon}=314.53$. The slight difference is due to the large-band of the intial pulse. 


\subsection{Test 4: acoustic solitary waves}\label{SecExpTest4}

\begin{figure}[htbp]
\begin{center}
\begin{tabular}{cc}
$\nu =0$ & $\nu \neq 0$\\
\includegraphics[scale=0.31]{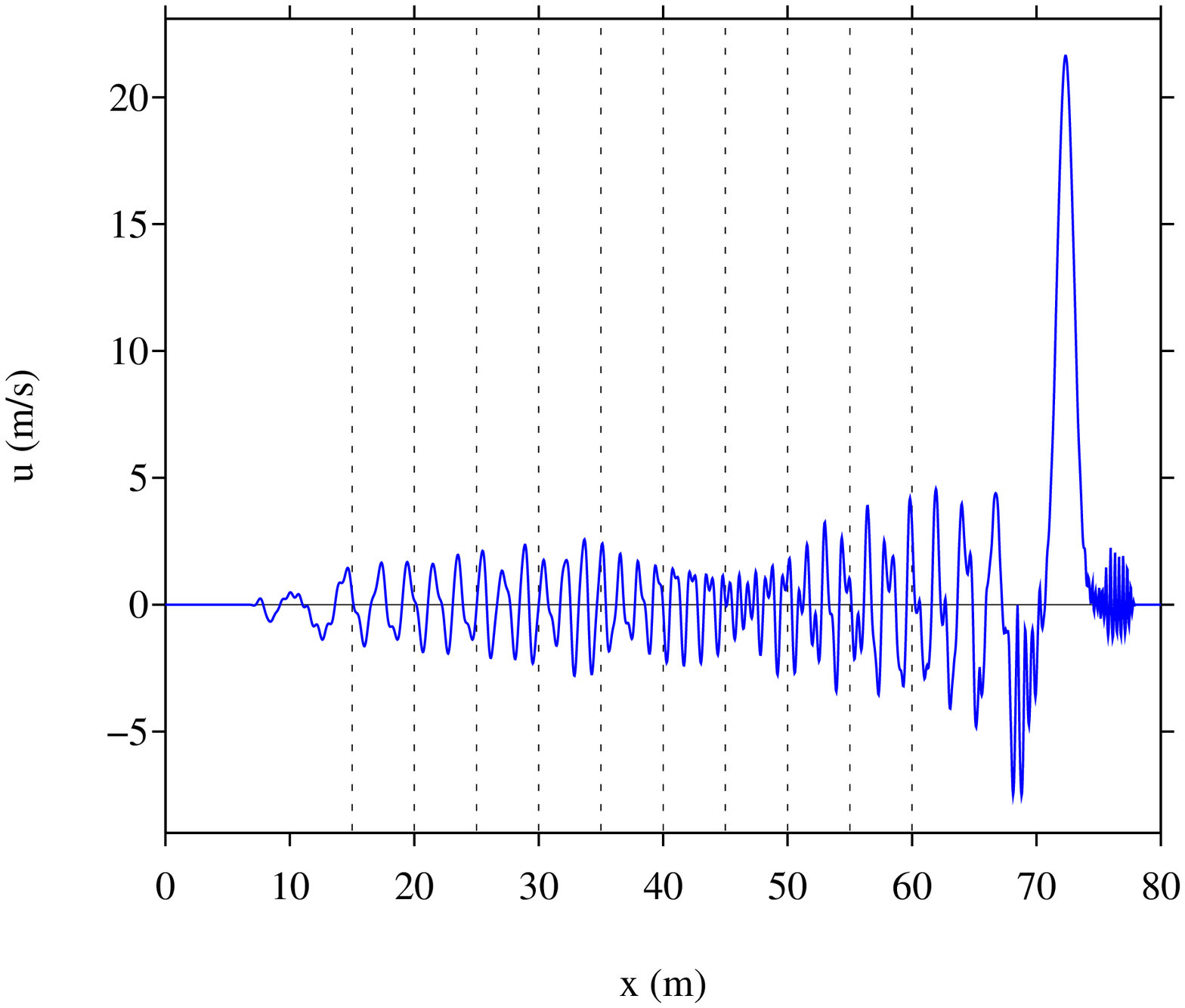}&
\includegraphics[scale=0.31]{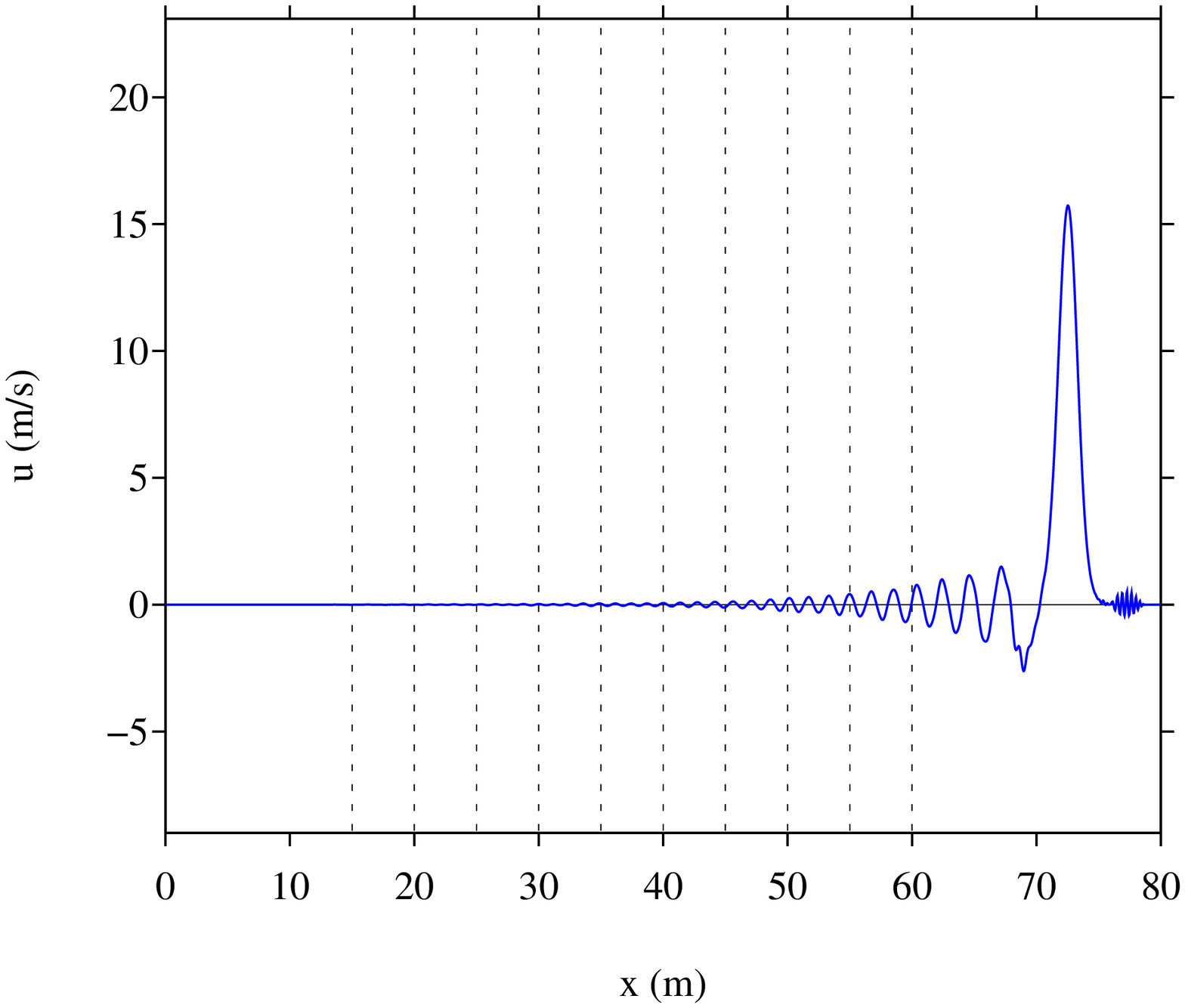}\\
\includegraphics[scale=0.31]{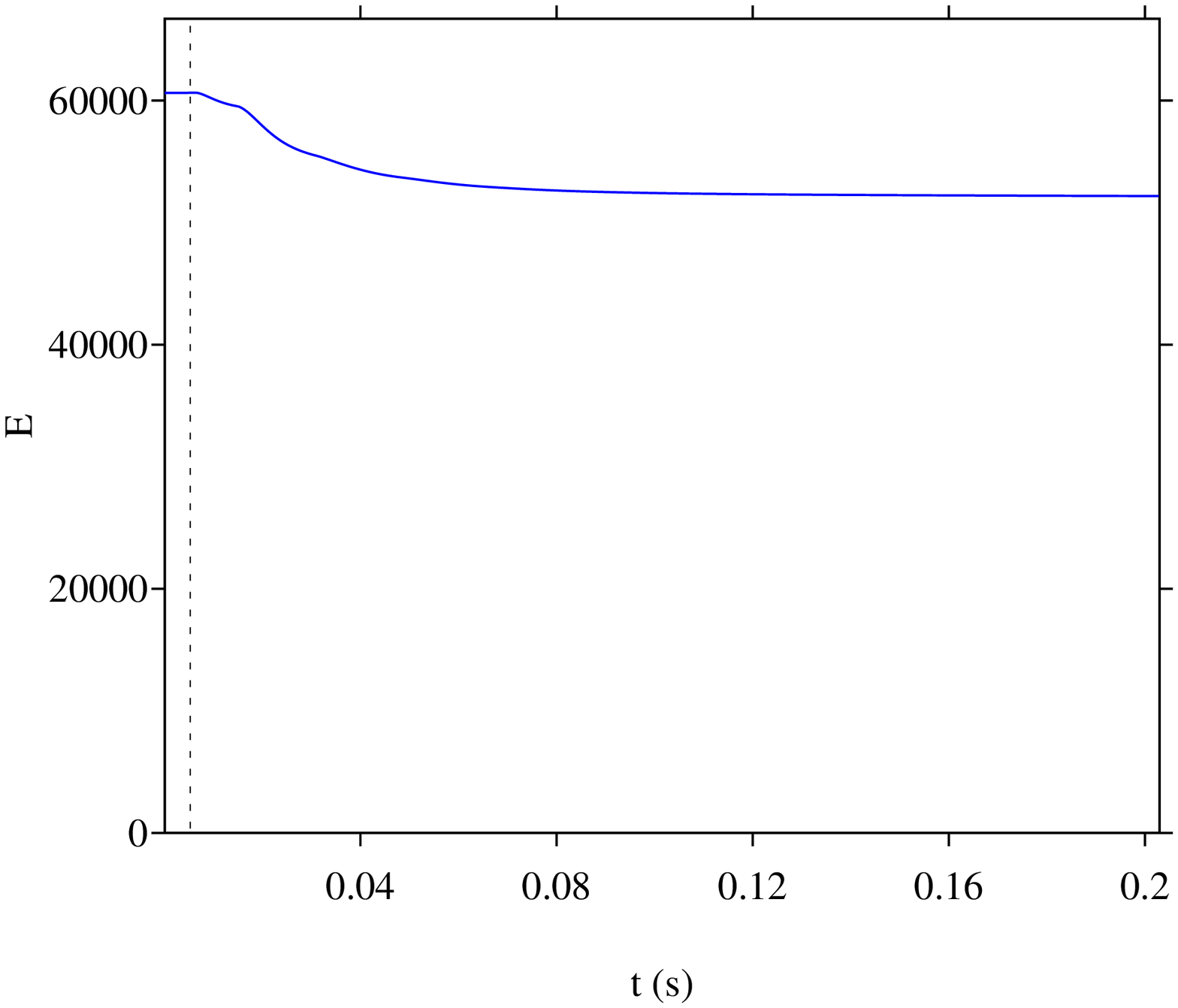}&
\includegraphics[scale=0.31]{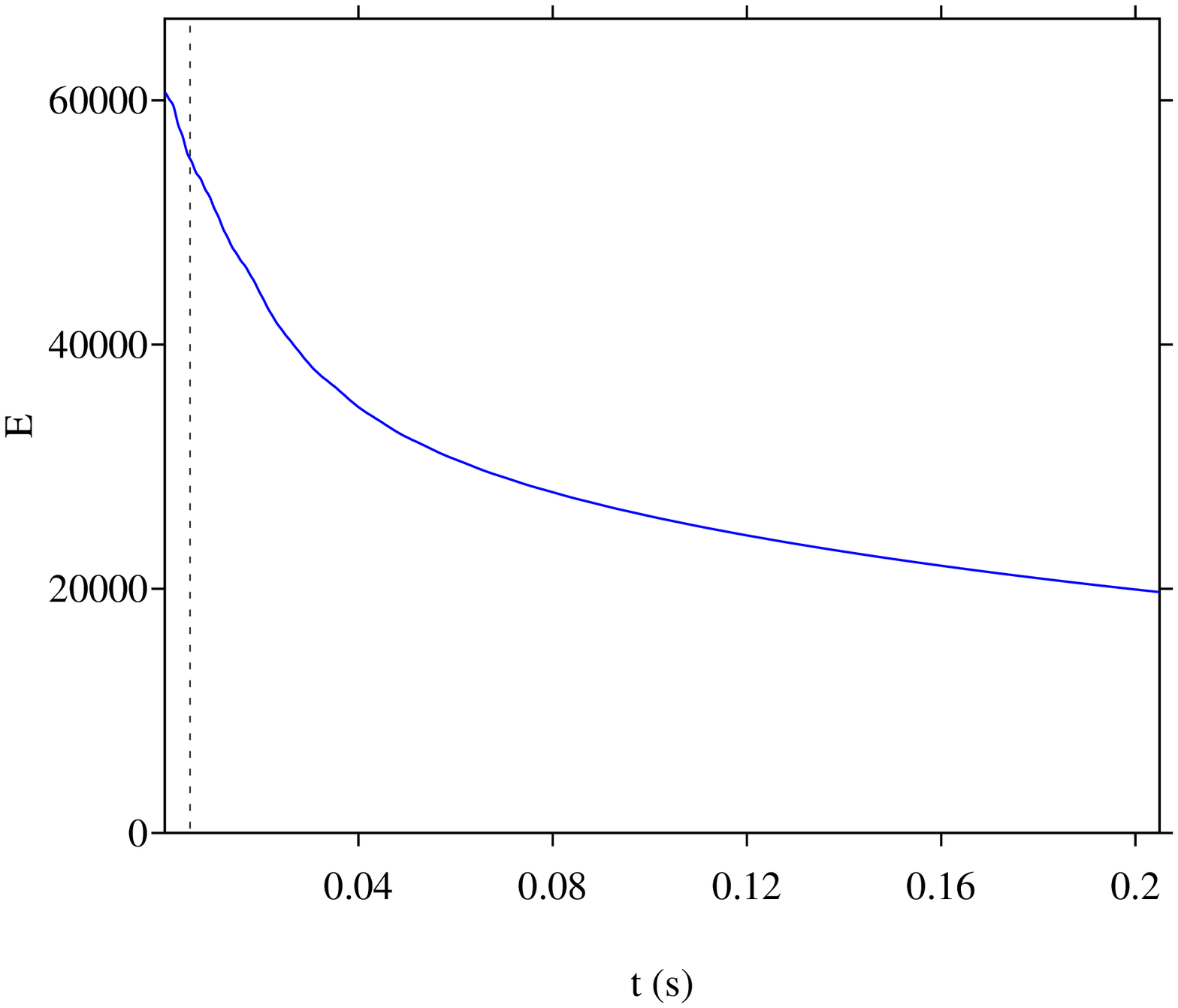}
\end{tabular}
\end{center}
\vspace{-0.8cm}
\caption{test4. Nonlinear coupled system, with a Gaussian pulse, $K=0.5$ and $\Omega=1$. Left: viscous case; right: inviscid case. Top: snapshots of $u$ after 8000 time steps (the vertical dotted lines denote the receivers); bottom: time evolution of the energy (the vertical dotted line denotes $t^*$ (\ref{TstarGauss}).}
\label{FigT4Om1}
\end{figure}

\begin{figure}[htbp]
\begin{center}
\begin{tabular}{cc}
$\nu =0$ & $\nu \neq 0$\\
\includegraphics[scale=0.31]{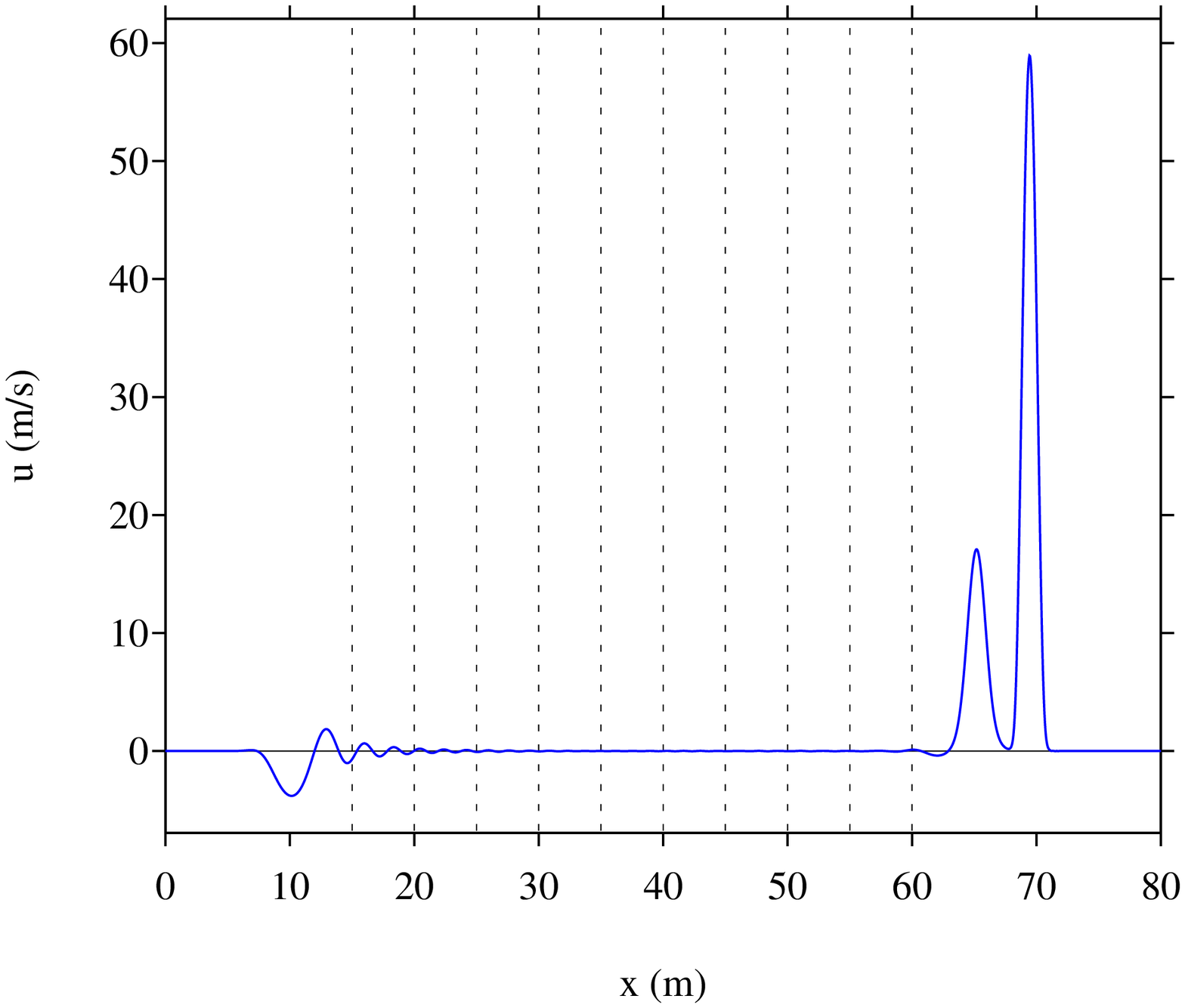}&
\includegraphics[scale=0.31]{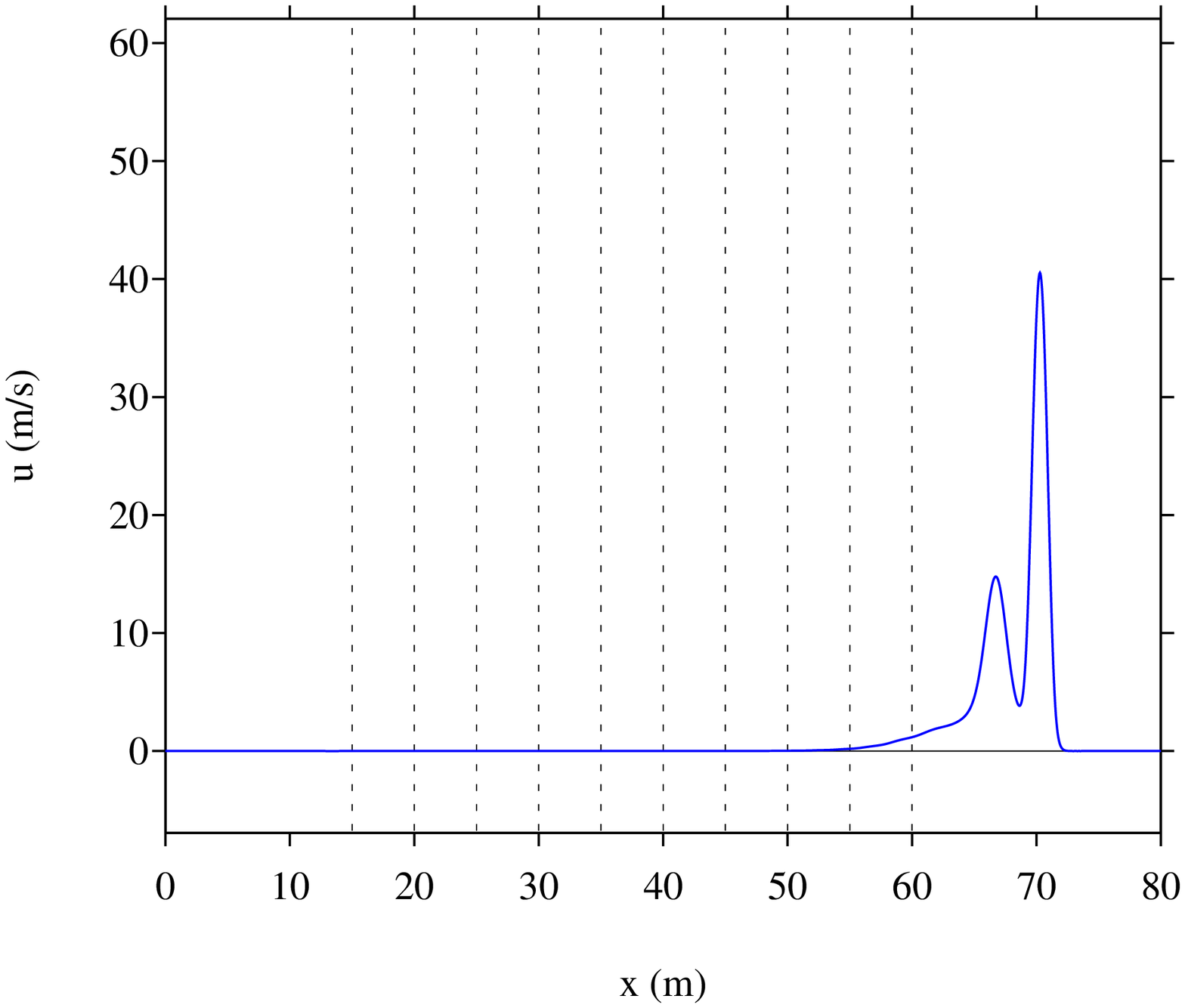}\\
\includegraphics[scale=0.31]{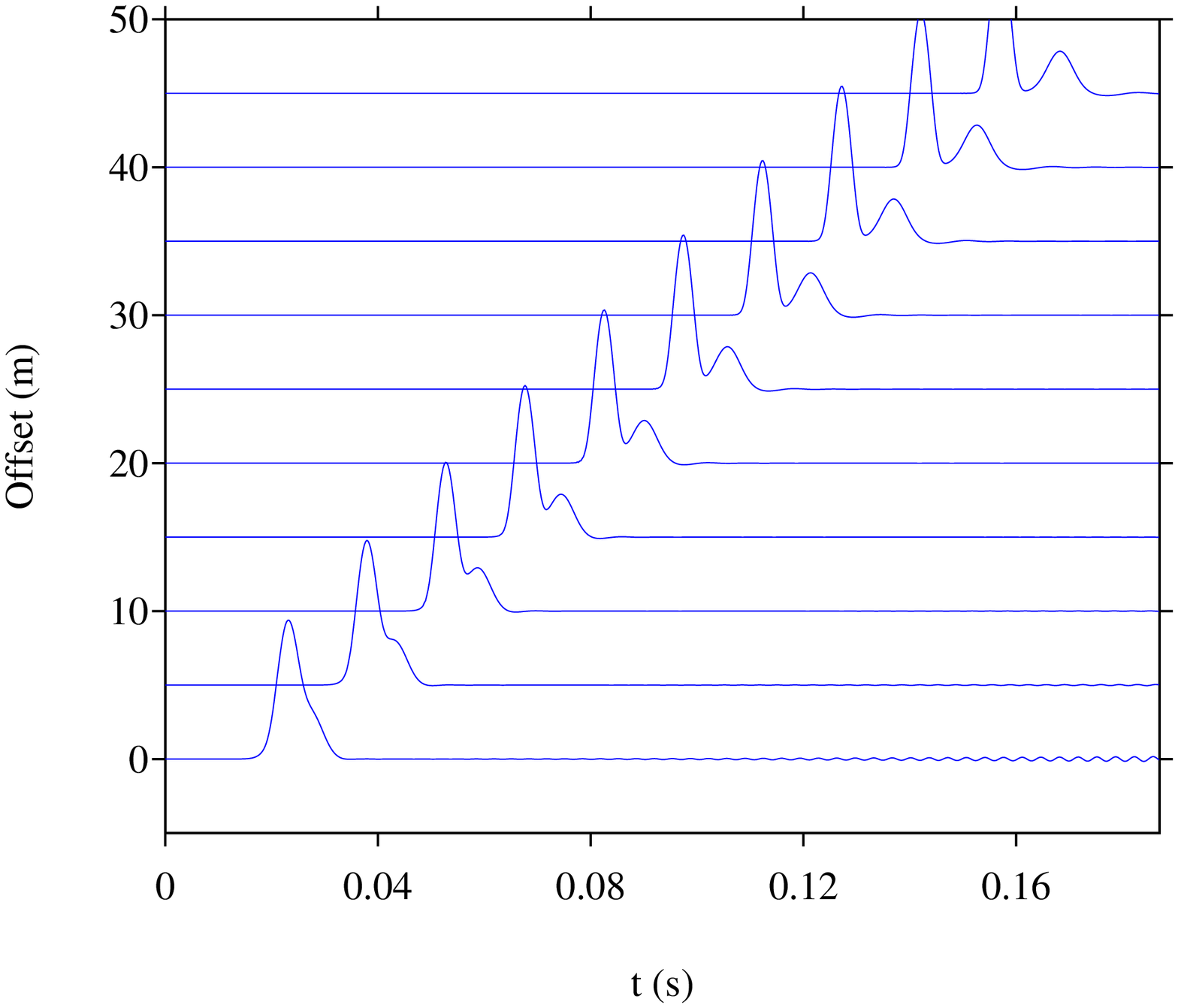}&
\includegraphics[scale=0.31]{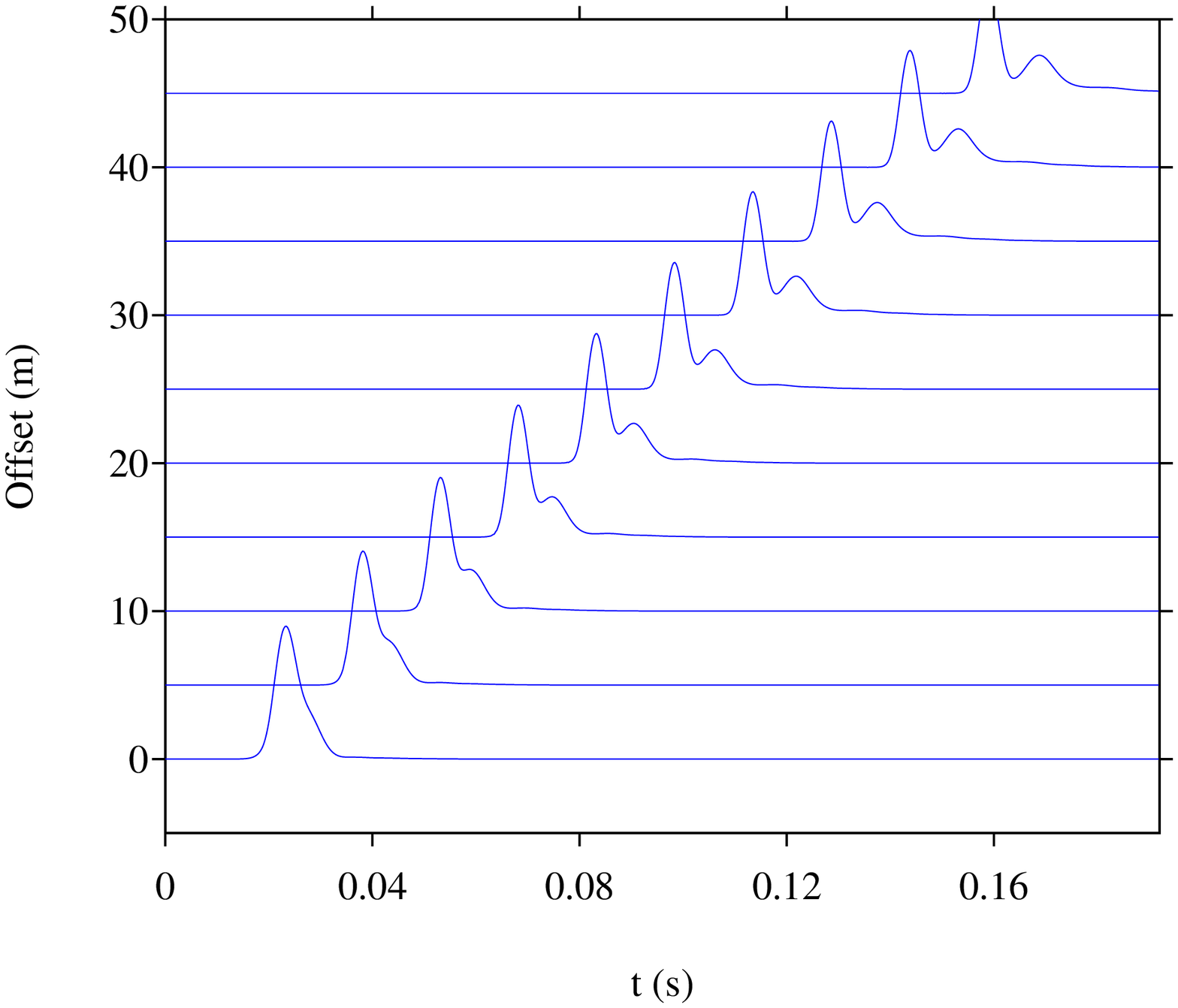}\\
\includegraphics[scale=0.31]{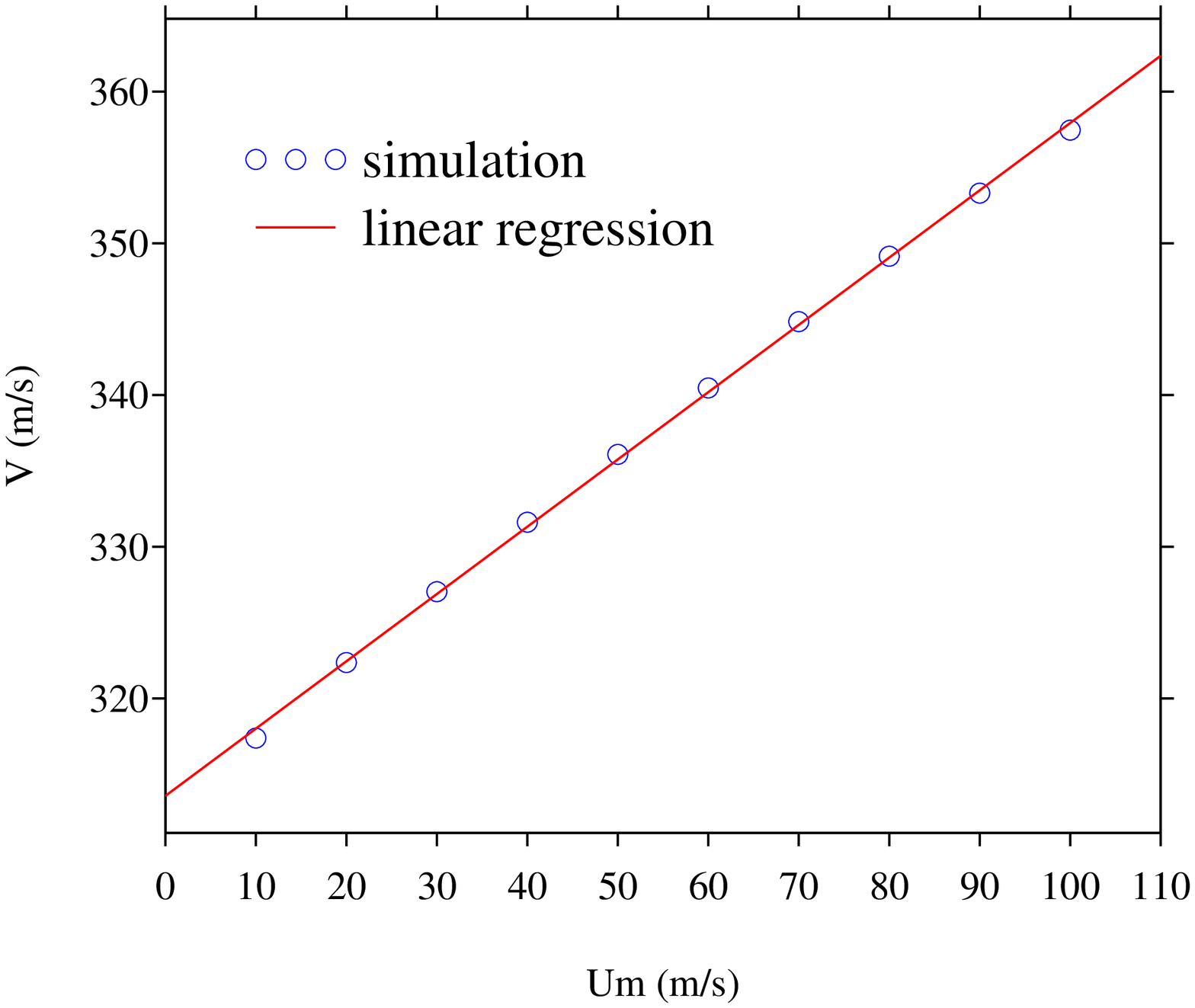}&
\includegraphics[scale=0.31]{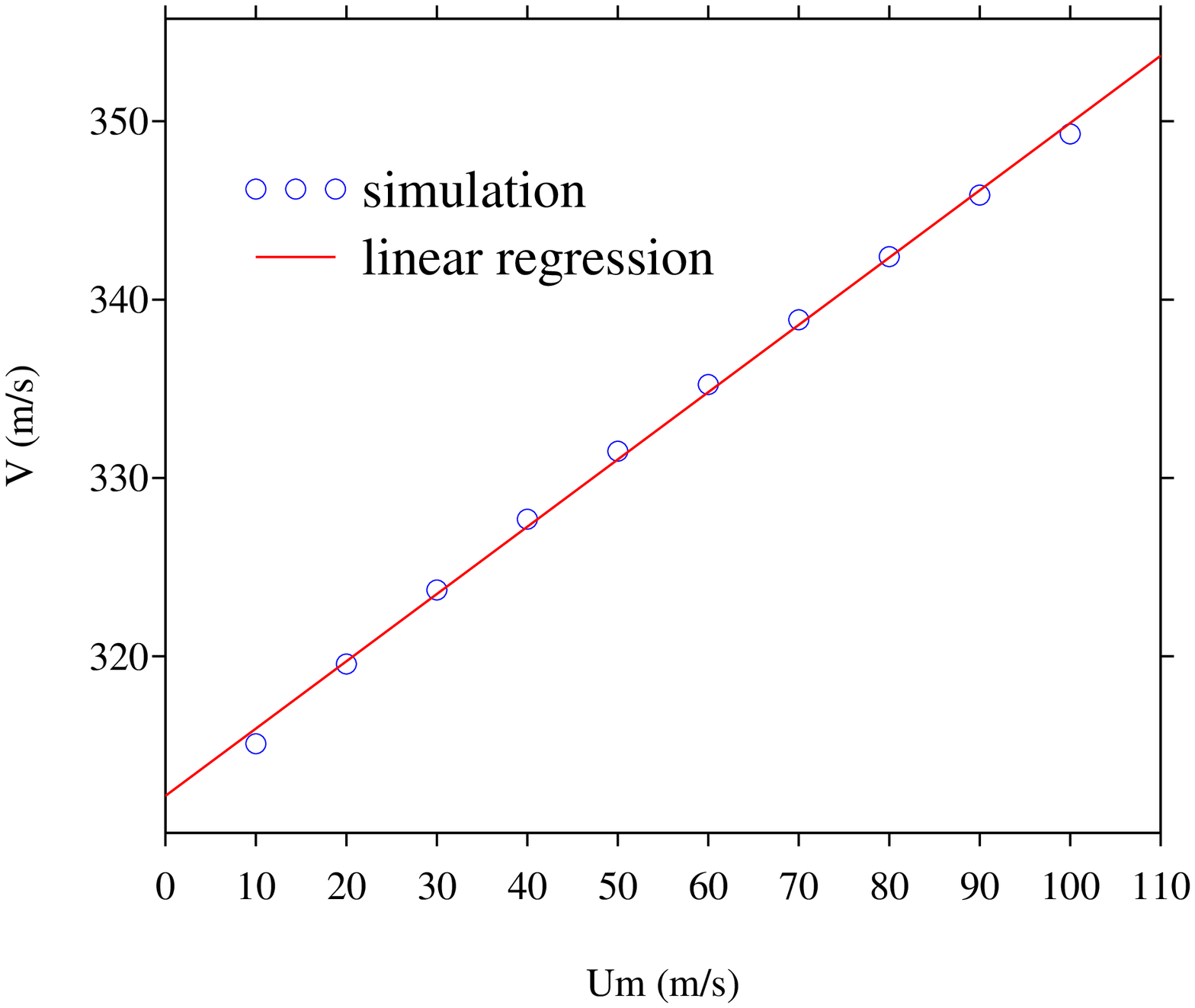}
\end{tabular}
\end{center}
\vspace{-0.8cm}
\caption{test4. Nonlinear coupled system, with a Gaussian pulse, $K=0.5$ and $\Omega=16$. Top: snapshot of $u$ after 8000 time steps. Middle: seismograms. Bottom: celerity ${\cal V}$ vs amplitude $u_m$ (from 10 m/s to 100 m/s).}
\label{FigT4Om16}
\end{figure}

Lastly, we considered the coupled system in the nonlinear regime. Computations were initialized by a Gaussian pulse, and simulations were performed during 8000 time steps, which corresponds roughly to 0.20 s of propagation. The case $K=0.5$ and $\Omega=1$ is shown in figure \ref{FigT4Om1}, to be compared with the case presented in figure \ref{FigT3Om1} obtained in the linear regime. A ballistic signal was observed, followed by a highly dispersive coda: no solitary wave emerged. If $\nu \neq 0$, a large amount of attenuation was also introduced, which damped this coda. In the inviscid case, the energy began to decrease and then was almost conserved. This means that the initial smooth pulse had led to a shock, and then an equilibrium with dispersion had led to the emergence of a smooth structure. Similar conclusion is reached in the viscous case, except that energy always decreased. 

The case $K=0.5$ and $\Omega=16$ is displayed in figure \ref{FigT4Om16}, to be compared with the case $\Omega=16$ in figure \ref{FigT3Om16} obtained in the linear regime. We recall that the theoretical analysis predicts the existence of solitary waves (section \ref{SecPhysRegime}). Compared with what can be seen in figure \ref{FigT4Om1}, the coda has disappeared. In the inviscid case, an oscillating mode remains at the place of initialization; moreover, the energy is conserved (not shown here), which indicates that no shock has been created. Two smooth structures are observed. Longer simulations show that these two components separate and propagate at different speeds. In the sequel, we will examine whether these solitary waves have the classical properties of solitons. 

In the case $K=0.5$ and $\Omega=16$, seismograms are built from the time signals stored at the receivers. The celerity ${\cal V}$ of the nonlinear wave with the highest amplitude is measured numerically. Similar measures are done for various amplitudes $u_m$ of the incident pulse, from 10 m/s to 100 m/s, or equivalently from $K=2.80$ to $K=0.28$. It is observed that ${\cal V}$ increases linearly with $u_m$: a linear regression estimation yields ${\cal V}=313.58+0.4435\,u_m$ (if $\nu =0$) and ${\cal V}=312.17+0.3773\,u_m$ (if $\nu \neq 0$). Waves propagate slightly faster in the inviscid case, because attenuation decreases the amplitude and consequently the celerity. The limit for $u_m=0$ is close to the value obtained in the linear case (test 3). 

\begin{figure}[htbp]
\begin{center}
\begin{tabular}{cc}
$\nu =0$ & $\nu \neq 0$\\
\includegraphics[scale=0.31]{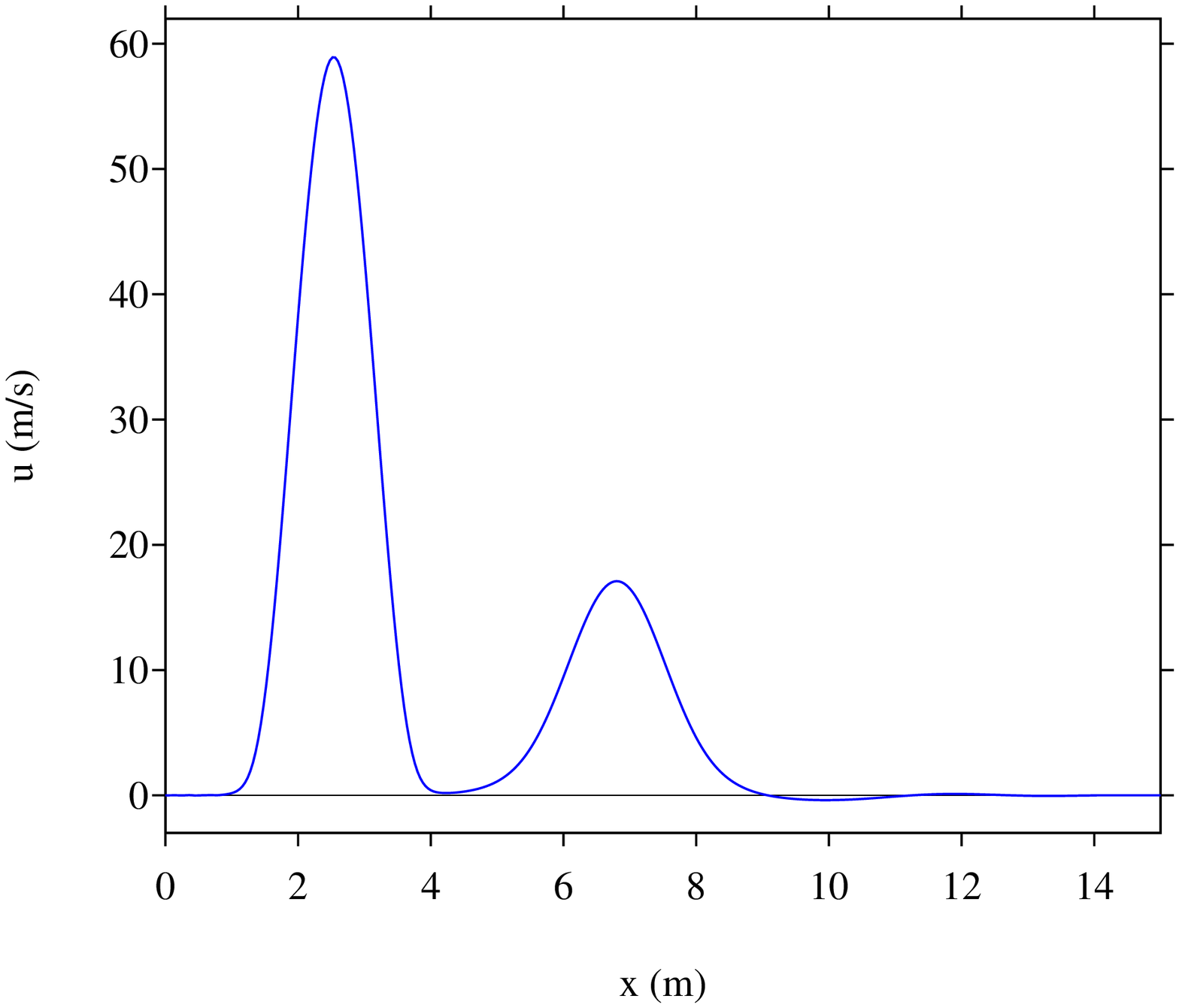}&
\includegraphics[scale=0.31]{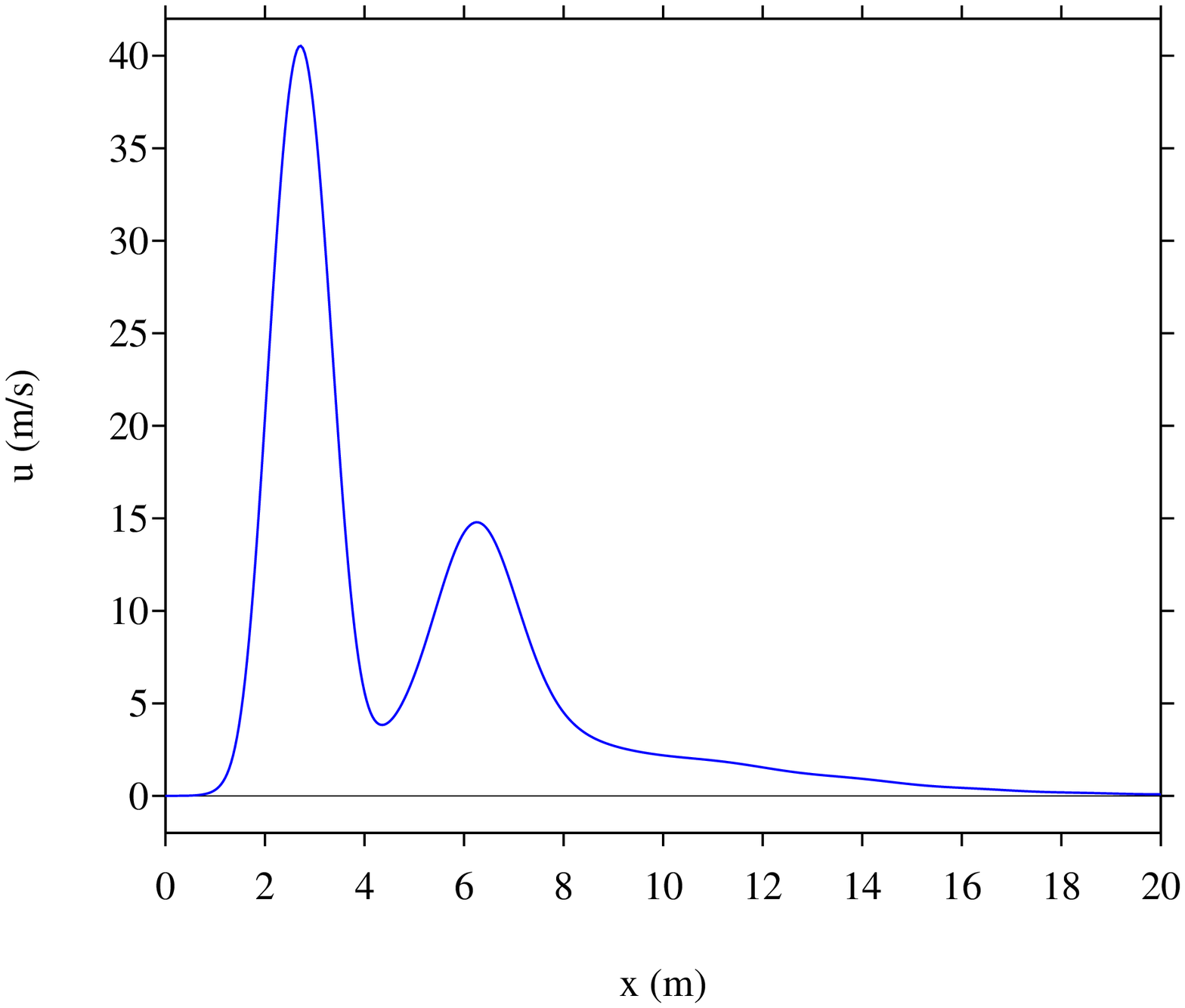}\\
\includegraphics[scale=0.31]{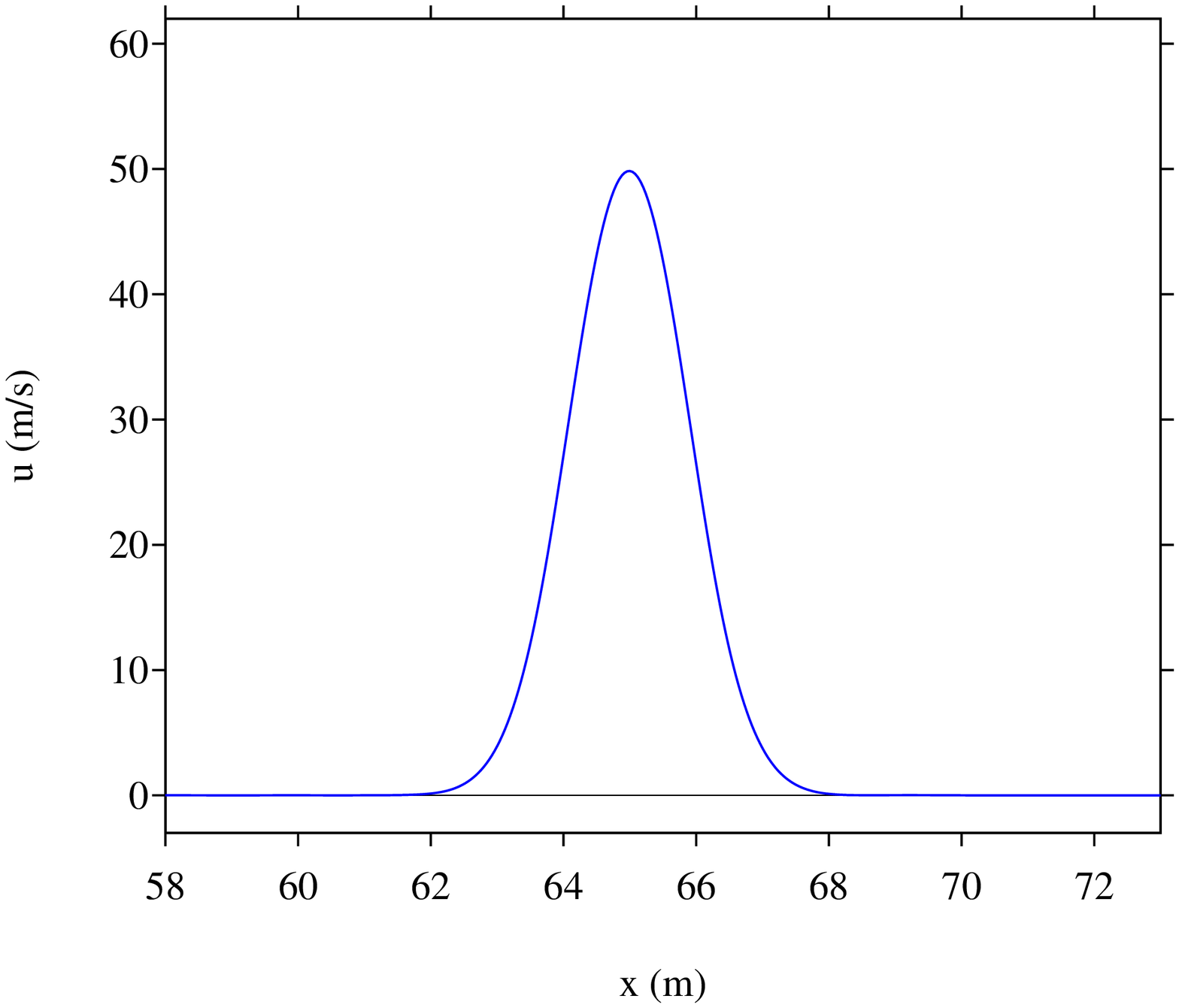}&
\includegraphics[scale=0.31]{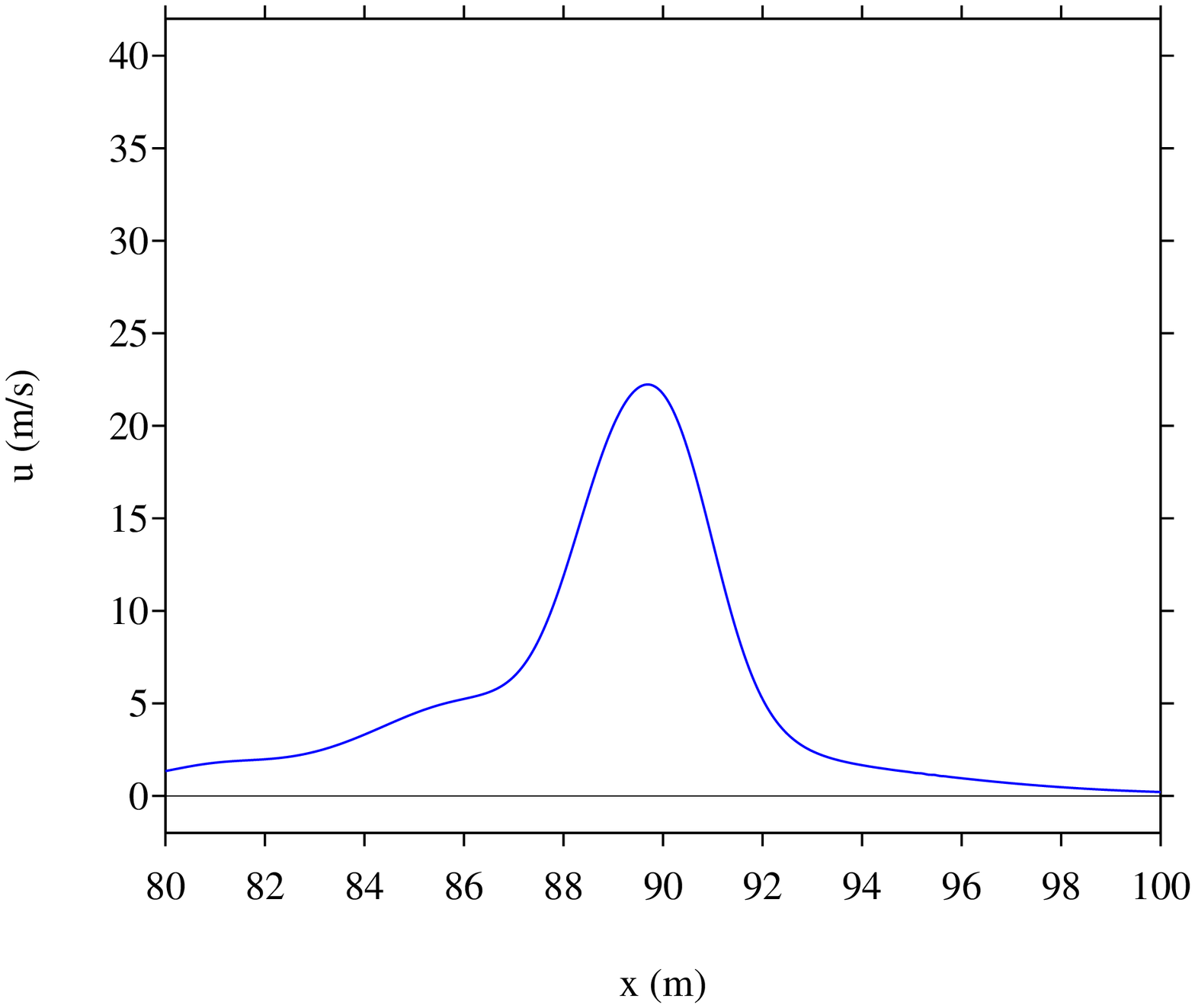}\\
\includegraphics[scale=0.31]{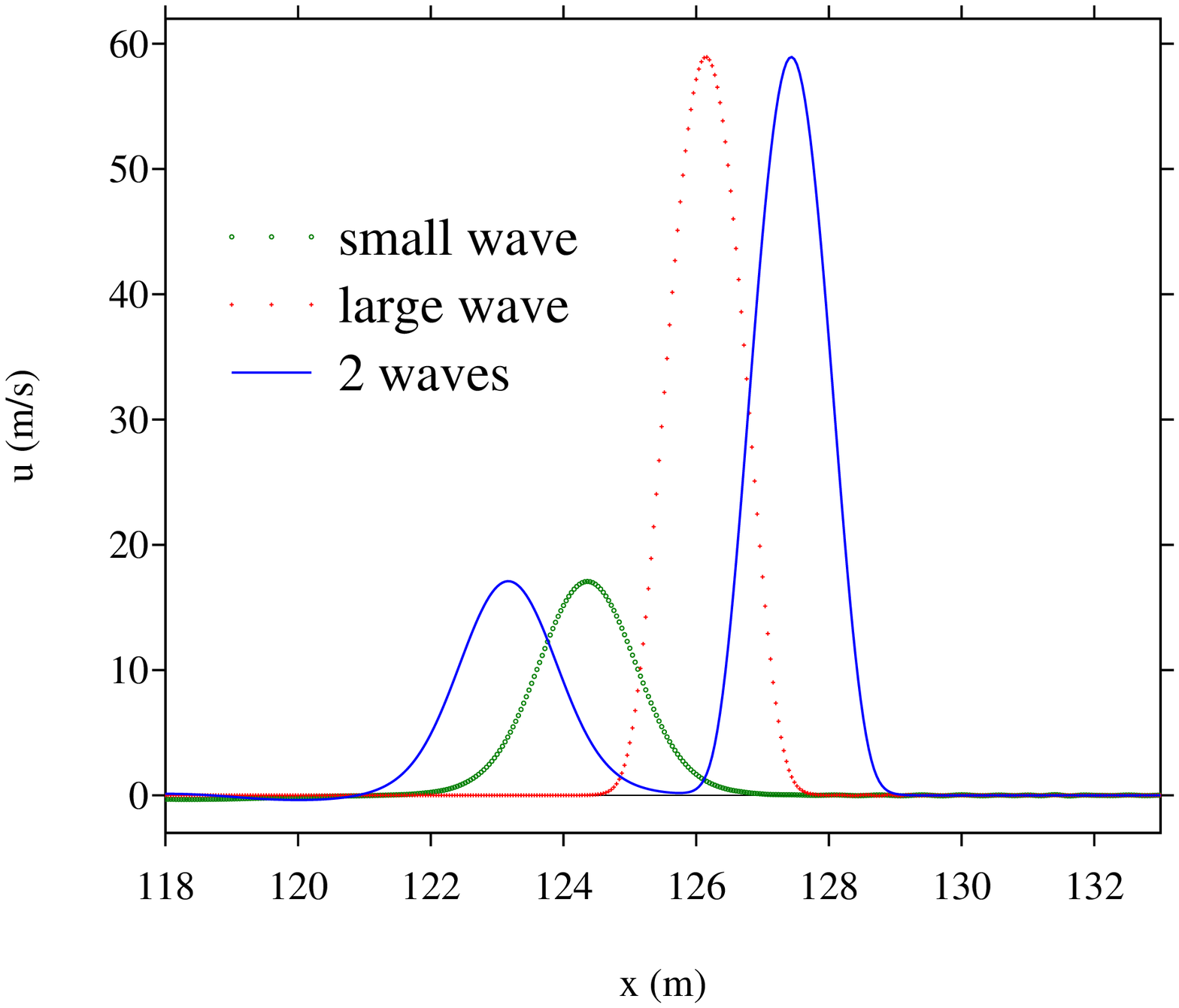}&
\includegraphics[scale=0.31]{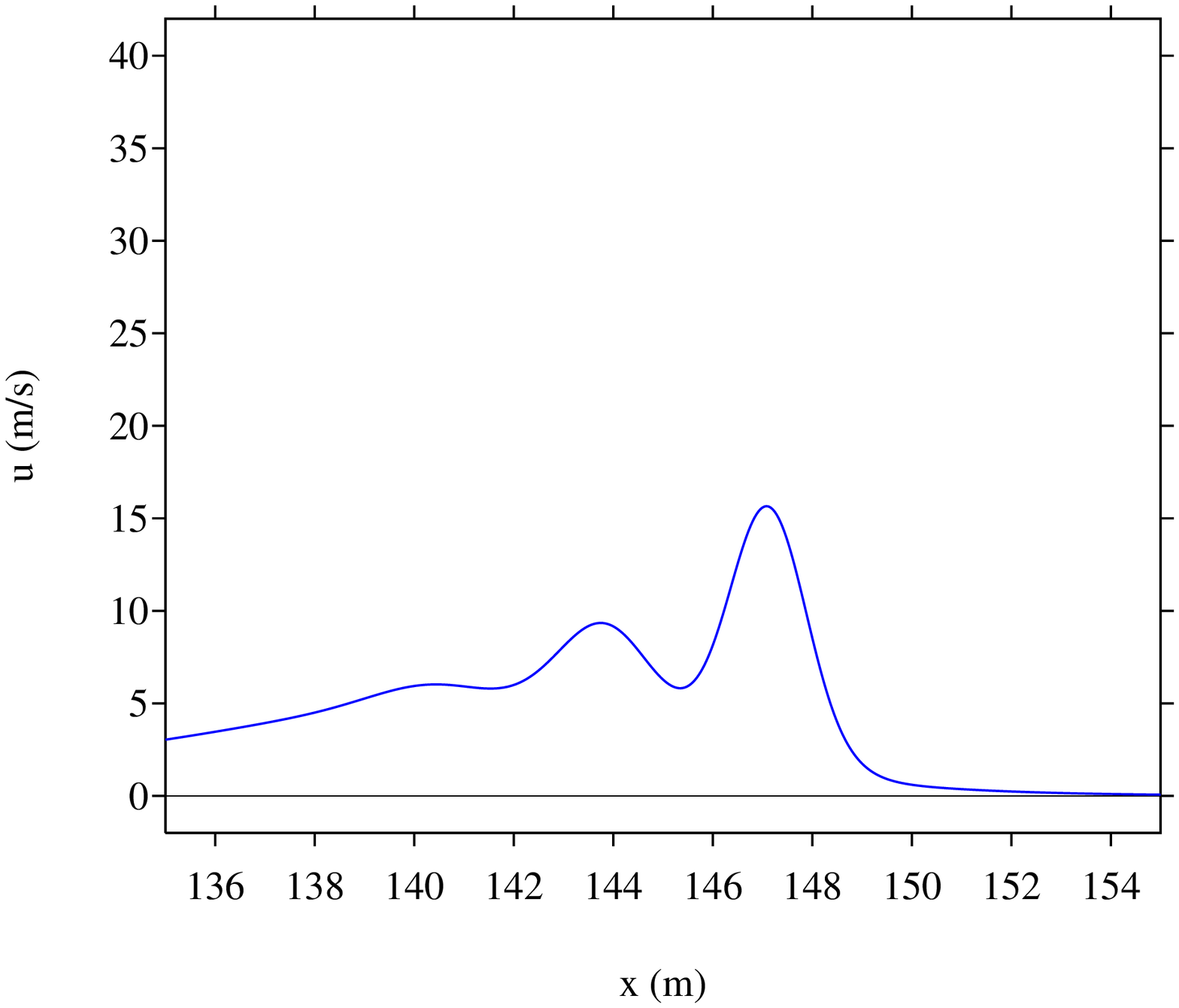}
\end{tabular}
\end{center}
\vspace{-0.8cm}
\caption{test4. Nonlinear coupled system, with a Gaussian pulse, $K=0.5$ and $\Omega=16$. Collision between two solitary waves. In green and red: location of the waves if they were alone.}
\label{FigT4LH}
\end{figure}

In figure \ref{FigT4Om16}, it can be seen that the original Gaussian pulse of $u$ separated into two smooth structures. The taller one was thinner and traveled faster than the shorter one. At the last instant of simulation, we inverted these waves to initialize a new computation (top of figure \ref{FigT4LH}). In the inviscid case, we observe that the two waves interact like classical solitons, exchanging their shape \cite{LeVeque02,LeVeque03}. After separation, the shape of the wave is the same as that of the original wave, though it was shifted in location from where it would be without interaction. These behaviors were qualitatively maintained in the viscous case, even if it is less clear due to the attenuation of these waves.


\section{Conclusion}\label{SecConclusion}

We have considered nonlinear acoustic waves in a tube connected with Helmholtz resonators. Various challenging physical features are involved: nonlinearity due to the amplitude of the waves, dispersion induced by the resonators, and fractional derivatives of order -1/2 and 3/2 due to viscous losses.

Our original contribution was to propose an efficient and accurate numerical modeling of this configuration. Some tools are conventional (TVD scheme for the nonlinear hyperbolic part), some others are novel (diffusive representation of fractional derivatives). To our knowledge, it is the first time that a diffusive representation has been considered together with the advection-Burgers equation. Eventually, a splitting strategy has ensured an optimal CFL condition for an explicit scheme. The proposed approach is computationaly efficient: the CFL stability condition is only governed by the nonlinearity (as in the usual advection-Burgers equation), and a minimum number of supplementary arrays is required to discretize the fractional derivatives.

This work was motivated by the experimental configuration shown in figure \ref{FigPhoto}, previously used in the linear propagation regime \cite{Richoux06,Richoux07}, and currently investigated in the nonlinear propagation regime. Our objective was to provide an efficient and accurate numerical modeling, validating (or not) the model (\ref{EDP}) and the underlying hypotheses. The numerical experiments showed that the viscous effects do not modify qualitatively the wave phenomena. Consequently, the theoretical predictions made in the inviscid case about the existence of acoustic solitons were also obtained in the viscous case \cite{Sugimoto92}.

A first extension of this work concerns the coefficients of the diffusive representation. In some linear problems, is is possible to determine the time evolution of the energy and to prove that this energy decreases as soon as all the coefficients $\mu_l>0$ ($l=1 \cdots N$) of the diffusive representations of the fractional derivatives are positive \cite{Haddar10,Deu10}. Therefore methods leading to positive $\mu_l>0$ are usually prefered. As already pointed out, it is the case for the Laguerre method but not for the optimization method we used. One perspective is to develop an alternative method ensuring the positivity of the coefficients $\mu_l$. A first possibility is to use an analytical method consisting in appoximating the function $\chi(\omega)$ by rational fractions. A second approach consists in using an optimization process with a positivity constraint, for instance a Shor algorithm \cite{Rekik11} initialized with the results from the Laguerre method. In the case of propoelasticity \cite{Blanc14}, this method led to 10 up to 100 times more accurate results.

From a physical point of view, the dissipation effects also require further investigations. To get solitons, it is necessary to have $\Omega \gg 1$ (\ref{NotationsEDPscale}), which implies $\omega\ll \omega_0$. In this regime, the dispersion analysis indicates that the attenuation is quite low (left part of figure 3). On the contrary, a high attenuation of waves is observed experimentally. A possible explanation of this mismatch is that some mechanisms of attenuation are not incorporated in the model. A good candidate is given by turbulence and nonlinear losses in the resonators. To account for these losses, a nonlinear theory for the response of the resonators has been proposed in the appendix of \cite{Sugimoto92}. Equation (\ref{EDP2}) with notations (\ref{NotationsEDP}) should be replaced by the nonlinear fractional ODE
\begin{equation}
\begin{array}{l}
\displaystyle
\frac{\textstyle \partial^2 p}{\textstyle \partial t^2}+\frac{\textstyle 2\,\sqrt{\nu}}{\textstyle r}\,\frac{\textstyle L^{'}}{\textstyle L_e}\frac{\textstyle \partial^{3/2} p}{\textstyle \partial t^{3/2}}+\omega_e^2\,p-\frac{\textstyle \gamma-1}{\textstyle 2\,\gamma}\frac{\textstyle 1}{\textstyle p_0}\frac{\textstyle \partial^2 \left(p\right)^2}{\textstyle \partial t^2}\\
\\
\displaystyle
\hspace{2cm}
+\frac{\textstyle V}{\textstyle B\,L_e\,\rho_0\,a_0^2}\left|\frac{\textstyle \partial p}{\textstyle \partial t}\right|\,\frac{\textstyle \partial p}{\textstyle \partial t}
=\omega_e^2\,\frac{\textstyle \gamma\,p_0}{\textstyle a_0}\,u,
\end{array}
\label{ODETurbulence}
\end{equation}
with the new parameters 
\begin{equation}
L^{'}=L+2\,r,\qquad L_e=L+\eta,\qquad \omega_e^2=\frac{\textstyle L}{\textstyle L_e}\,\omega_0^2,
\label{ParamTurbu}
\end{equation}
where $\eta$ is determined experimentally ($\eta\approx 0.82\,r$). The term $\partial^2 \left(p\right)^2/\partial t^2$ models the nonlinearity due to the adiabatic process in the cavity, whereas the semi-empirical term depending on the sign of $\partial p/\partial t$ accounts for the jet loss resulting from the difference in inflow and outflow patterns \cite{Sugimoto92}. A more sophisticated numerical method must be developed to integrate (\ref{ODETurbulence}).

A last extension of our work concerns the case where the height $H$ of each resonator may vary with position, leading to variable coefficients in (\ref{EDP}). Numerically, this requires smooth functions $e(x)$, $g(x)$ and $h(x)$, for instance with cubic splines, to be built. The exponential of ${\bf S}$ in (\ref{MatS}) and (\ref{SplitDiffuExp}) needs to be computed at each grid node and at each time step, which increases the computational cost, but no other modifications are required. It will make it possible to investigate numerically the propagation of acoustic solitons in random media \cite{Garnier98}. This topic is a subject of intense research in various fields of physics, with possible applications in the transport of information.

\vspace{0.5cm}
\noindent
{\bf Acknowledgments}. This study has been initiated with Agn\`es Maurel (ESPCI, France), Olivier Richoux and Vincent Pagneux (LAUM, France), and has received financial support from the Agence Nationale de la Recherche through the grant ANR ProCoMedia, project ANR-10-INTB-0914. Pierre Haldenwang (AMU, France) is thanked for his insights into hyperbolic equations.


\end{document}